\DeclareSymbolFont{UPM}{U}{eur}{m}{n}
\DeclareMathSymbol{\umu}{0}{UPM}{"16}
\let\oldumu=\umu
\renewcommand\umu{\ifmmode\oldumu\else\math{\oldumu}\fi}

\newcounter{magicrownumbers}

\documentclass[twocolumn]{aastex63}

\usepackage{longtable}
\usepackage{chngcntr}

\usepackage{hyperref}

\expandafter\def\expandafter\UrlBreaks\expandafter{\UrlBreaks
  \do\*\do\-\do\~\do\'\do\"\do\-}%

\received{2022 November 22}
\revised{2023 March 7}
\accepted{2023 March 23}
\submitjournal{the Planetary Science Journal}

\usepackage{todonotes}

\shorttitle{Assessing Thermochemical Equilibrium Estimation Methods}
\shortauthors{Himes et al.}
\graphicspath{{./}{figures/}}

\begin{document}

\title{Towards 3D Retrieval of Exoplanet Atmospheres: Assessing Thermochemical Equilibrium Estimation Methods}

\correspondingauthor{Michael D. Himes}
\email{mhimes@knights.ucf.edu}

\author[0000-0002-9338-8600]{Michael D. Himes}
\affiliation{Planetary Sciences Group, Department of Physics, University of Central Florida}

\author[0000-0002-8955-8531]{Joseph Harrington}
\affiliation{Planetary Sciences Group, Department of Physics and Florida Space Institute, University of Central Florida}

\author[0000-0001-9854-8100]{At{\i}l{\i}m G\"{u}ne\c{s} Bayd{\rlap{\.}\i}n}
\affiliation{Department of Computer Science, University of Oxford}

\begin{abstract}

Characterizing exoplanetary atmospheres via Bayesian retrievals requires assuming some chemistry model, such as thermochemical equilibrium or parameterized abundances.  
The higher-resolution data offered by upcoming telescopes enables more complex chemistry models within retrieval frameworks.  
Yet, many chemistry codes that model more complex processes like photochemistry and vertical transport are computationally expensive, and directly incorporating them into a 1D retrieval model can result in prohibitively long execution times.  
Additionally, phase-curve observations with upcoming telescopes motivate 2D and 3D retrieval models, further exacerbating the lengthy runtime for retrieval frameworks with complex chemistry models. 
Here, we compare thermochemical equilibrium approximation methods based on their speed and accuracy with respect to a Gibbs energy-minimization code.  
We find that, while all methods offer orders of magnitude reductions in computational cost, neural network surrogate models perform more accurately than the other approaches considered, achieving a median absolute dex error $<0.03$ for the phase space considered.  
While our results are based on a 1D chemistry model, our study suggests that higher dimensional chemistry models could be incorporated into retrieval models via this surrogate modeling approach.


\end{abstract}

\keywords{techniques: machine learning --- planets and satellites: atmospheres --- planets and satellites: composition}


\section{Introduction}
\label{sec:intro}

Improvements in signal-to-noise ratio and spectral resolution of exoplanet observations, such as those offered by the James Webb Space Telescope (JWST), motivate and enable the use of more sophisticated models to characterize exoplanetary atmospheres via retrieval \citep[see review by][]{Madhusudhan2018bookRetrieval}.  
Early retrieval studies made simplifying assumptions about atmospheric chemistry, such as gas abundances that were constant with altitude or by scaling equilibrium profiles calculated assuming solar metallicity \citep[e.g.,][]{MadhusudhanSeager2009apjRetrieval, LineEtal2014apjRetrieval2}.  
Later, as new data began to offer a more detailed picture of exoplanet atmospheres, groups incorporated more complex chemistry models into retrieval frameworks, such as (dis)equilibrium abundances for sub- and super-solar metallicites \citep[e.g.,][]{OreshenkoEtal2017apjlWASP12b, MolliereEtal2020aapDisequilibriumHR8799e, ChubbMin2022aap3DPhaseCurveRetrieval}. 

Many exoplanet atmospheres are likely in disequilibrium \citep{MosesEtal2011apjDisequilibrium, LineYung2013apjRetrieval3, MosesEtal2013apjGJ436b, VenotEtal2015aapCarbonRichAtmospheres, RoudierEtal2021ajHubbleDisequilibrium}. 
\citet{StevensonEtal2010natDisequilibriumGJ436b} presented the first detection of disequilibrium chemistry in an exoplanet atmosphere, finding CH$_4$ depleted relative to thermochemical equilibrium calculations.  
More recent studies considering disequilibrium processes, such as photochemistry and vertical quenching, have found evidence of disequilibrium in multiple hot-Jupiter atmospheres \citep[e.g.,][]{MolliereEtal2020aapDisequilibriumHR8799e, KawashimaMin2021aapDisequilibriumRetrieval, RoudierEtal2021ajHubbleDisequilibrium}, while \citet{ChangeatEtal2022apjsEquilibriumHST} found that some hot Jupiter spectra measured by the Hubble Space Telescope are not well fit by equilibrium chemistry models.
While these recent studies still find some equilibrium models consistent with the data, JWST will provide sufficient precision to more definitively differentiate between equilibrium and disequilibrium atmospheres \citep{BlumenthalEtal2018apjDisequilibriumJWST}.
Furthermore, JWST's early-release science data for WASP-39b has led to the detection of SO$_2$, likely a product of photochemistry, providing additional motivation for more complex chemistry models within retrieval algorithms \citep{AldersonEtal2023natJwstErsSO2, RustamkulovEtal2023natJwstErsWASP39b}.

Both the more detailed measurements of JWST and future telescopes as well as recent results from 1D retrieval studies motivate more complex retrieval models.  
While the findings of \citet{BlecicEtal2017apjThermal3D} show that 1D retrieval models can recover a thermal profile comparable to the 3D structure's arithmetic average, \citet{CaldasEtal2019aapRetrievalBiases} and \citet{PlurielEtal2022aapRetrievalBiases} found biases in the retrieved gas abundances. 
A more complete understanding of the atmospheric chemistry thus requires a 2D or 3D model to properly capture longitudinal variations.
Community efforts are underway to develop such retrieval codes, though the computational costs associated with the increase in dimensionality require simplifying assumptions \citep[e.g.,][]{IrwinEtal2020mnrasWASP43b2D, FengEtal2020ajRetrieval2D, ChangeatEtal2021apjWASP43bRetrieval, CubillosEtal2021apjReSpect, ChallenerRauscher2022ajThERESA, ChubbMin2022aap3DPhaseCurveRetrieval}.
With 1D retrievals requiring on the order of $10^5$--$10^6$ forward model evaluations \citep{Madhusudhan2018bookRetrieval}, the difference of one second per forward model evaluation adds up to days of computing time, and higher-dimensional models multiply this further.  
While any chemistry model could be incorporated into a retrieval framework, the additional computational time of some models can be prohibitive, such as the Atmos photochemical-climate model \citep{ArneyEtal2016asbiohazyArcheanEarthATMOS} and the free molecular retrieval setup discussed in \citet{ChubbMin2022aap3DPhaseCurveRetrieval}.

Recently, \citet{HimesEtal2022psjMARGEHOMER} presented a novel machine learning approach to exoplanet atmospheric retrieval, where the radiative transfer forward model is replaced with a neural network (NN) surrogate model.  
We found that this approach achieves similar quantitative results as the classical approach, but at a fraction of the computational cost.  
Similar results have been reported for other scientific problems using similar approaches \citep[e.g.,][]{GilmerEtal2017icmlQuantumChemistry, brehmer2018guide, BaydinEtal2019neuripsQuestForPhysics, munk2019deep, kasim2021up}.

In this study, we seek to determine a generalized method that can accurately approximate a given chemistry model, such that computationally expensive chemistry models could be included within a retrieval of any dimensionality.  
To investigate this, we compare thermochemical equilibrium estimation methods and consider each approximation's applicability to retrieval based on runtime and accuracy.
In Section \ref{sec:methods6}, we describe the analytical, interpolation-based, and NN-based models considered and detail our methodology.  
In Section \ref{sec:results6}, we present and discuss the results.  
Finally, we draw conclusions in Section \ref{sec:conclusions6}. 

\section{Methods}
\label{sec:methods6}

We utilize four equilibrium estimation methods: minimization of the Gibbs free energy via the Thermochemical Equilibrium Abundances (TEA) code \citep{BlecicEtal2016apjsTEA}; the analytical approximation for equilibrium used in the Reliable Analytic Thermochemical Equilibrium (RATE) code \citep{CubillosEtal2019apjRATE}; interpolation within a grid of models produced by TEA; and a surrogate model based on an NN trained on data produced by TEA \citep[using the approach of][]{HimesEtal2022psjMARGEHOMER}.

\subsection{Equilibrium via Gibbs Energy Minimization}

\citet{BlecicEtal2016apjsTEA} presented the open-source TEA code, which calculates thermochemical equilibrium via minimizing the Gibbs free energy in a Lagrangian optimization framework.  
They demonstrate that TEA reproduces the results of other thermochemical equilibrium implementations when utilizing the same thermodynamic data.  
Here, we use TEA to calculate the ``ground truth" data to compare with the other methods as well as the data set used for interpolation and NN surrogate model training.
For more details on TEA's implementation and validation, see \citet{BlecicEtal2016apjsTEA}.

\subsection{Equilibrium via Analytical Formulae}

\citet{CubillosEtal2019apjRATE} presented the open-source RATE code, an analytical formalism to approximate thermochemical equilibrium, which built upon and resolved the instability issues present in the formalism of \citet{HengEtal2016apjAnalyticalThermochemicalEquilibrium},  \citet{HengLyons2016apjAnalyticalThermochemicalEquilibrium}, and 
\citet{HengTsai2016apjAnalyticalThermochemicalEquilibrium}.  
They determined that the RATE approximation is valid over a parameter space of roughly 200--2000 K, $10^{-8}$--$10^3$ bar, and $10^{-3}$--$10^2{\times}$ solar elemental abundances. 
These stability improvements enable broad application to arbitrary combinations of parameters within this phase space, such as those considered in a Bayesian retrieval on optical and/or infrared spectra of most observed exoplanetary atmospheres.
For more details on their approach, see \citet{CubillosEtal2019apjRATE}.

\subsection{Equilibrium via Interpolation}
\label{sec:interpmethod}

We generate a grid of 74,088,000 points based on pressure, temperature, the carbon-to-hydrogen elemental abundance ratio (C/H), nitrogen-to-hydrogen abundance ratio (N/H), and oxygen-to-hydrogen abundance ratio (O/H); the parameter minima, maxima, and number of uniformly log-spaced samples are given in Table \ref{tbl:teagrid}. 
We consider the same molecules as \citet{CubillosEtal2019apjRATE}, with the addition of helium: H$_2$O, CO, CH$_4$, CO$_2$, HCN, C$_2$H$_2$, C$_2$H$_4$, NH$_3$, N$_2$, H$_2$, H, and He.  
This enables direct applicability of the models to gas-giant atmospheres.

\begin{table*}[tb]
\caption{TEA Model Grid Parameters}\vskip -.1in
\label{tbl:teagrid}
\begin{center}
\begin{tabular}{ l  c c c}
 \toprule
Parameter & Minimum & Maximum & Number of samples\\
\hline
$\textrm{log}$ pressure                  & -8     &  3     & 100 \\
$\textrm{log}$ temperature               &  2.305 &  3.778 &  80 \\
$\textrm{log}$ C/H                       & -6.57  & -0.57  &  21 \\
$\textrm{log}$ N/H                       & -7.17  & -1.17  &  21 \\
$\textrm{log}$ O/H                       & -6.31  & -0.31  &  21 \\
\hline
\multicolumn{4}{p{0.55\linewidth}}{\textbf{Notes.}  The mean abundance values correspond to solar abundance ratios as reported by \citet{AsplundEtal2009araaSolarAbundances}. Temperatures are limited to $200-6000$ K by the available data.}
\end{tabular}
\end{center}
\end{table*}

We consider linear and inverse-distance weighting (IDW) interpolation.  
For IDW, we calculate the Euclidean distance between the logarithm of the inputs, and we vary the exponent on the inverse distance, $p$, between 1 and 40.  
To interpolate, we consider the $n$ nearest neighbors along each axis, where $n$ can vary between 1 and 4.
In the interest of the ability to scale the considered models to higher dimensionalities, we do not consider spline interpolation due to longer runtimes than TEA in this setup, likely attributable to the inability to hold this data set in cached memory.
Similarly, we do not consider radial basis function interpolation due to the amount of memory required for the necessary $N \times M \times M$ matrices, where N is the number of dimensions and M is the number of data points.  
For this problem's dimensionality, using a data set of 100,000 points --- $\sim0.13$\% of the total TEA data set considered here --- would require 372.5 GiB of memory to calculate the radial basis functions in double precision.

\subsection{Equilibrium via NN-based Surrogate Model}

\citet{HimesEtal2022psjMARGEHOMER} presented a NN-based surrogate modeling method along with a software package, MARGE, that implements the technique.  
Here, we use MARGE to approximate TEA.

We use two different data sets to train the NN models: (1) a grid of TEA models as described in the previous section but with only 40 temperatures and 11 elemental abundance ratios (5,324,000 total grid points), and (2) a set of models whose inputs were randomly drawn from log-uniform distributions spanning the ranges in Table \ref{tbl:teagrid}.  
For (1), we randomly split the grid of TEA models into training, validation, and test sets, which contain 70\%, 20\%, and 10\% of the total data, respectively.  
For (2), we make 125,000 random draws of temperatures, C/H, N/H, and O/H; each of those cases is computed over the log-uniform grid of pressures defined in Table \ref{tbl:teagrid}, producing a total of 12,500,000 combinations of the 5 free parameters.
We randomly split these cases such that approximately 64\%, 16\%, and 20\% of the cases are in the training, validation, and test sets, respectively.

Each data case is comprised of the 5 input parameters given in Table \ref{tbl:teagrid} and the corresponding 12 output gas abundances.  
We normalize the data in two subsequent steps by (1) taking the base-10 logarithm of the inputs and outputs and (2) scaling the log-data to be within the closed interval [-1, 1] based on the training set extrema.

The NN model consists of an input layer with 5 nodes corresponding to the 5 input parameters; a 1D convolutional layer with a kernel size of 3 and 256 feature maps, which uses a rectified linear unit (ReLU) activation function; 3 dense layers with 4096 nodes, each of which use ReLU activation functions; and an output dense layer with 12 nodes corresponding to the output gas abundances.  
This architecture was selected through an extensive model grid search.
We train using a batch size of 1024, the mean-squared-error loss on the validation set, the Adam optimizer, and the {\tt triangular2} learning rate policy \citep{Smith2015arxivClyclicalLearningRates, HimesEtal2022psjMARGEHOMER} with a learning rate cycle of 12 epochs and range determined from the range test \citep[see Appendix A of][]{HimesEtal2022psjMARGEHOMER}. 

Using the aforementioned data sets, we train four NN models based on the above architecture (Table \ref{tbl:models}).  
NN1 is trained for 864 epochs on the grid of TEA models.  
NN2 is trained for 750 epochs on the random data.
NN3 is trained for 500 epochs on the random data, but sized to match the gridded data set.
NN4 is trained for 500 epochs on the combination of both the grid of TEA models as well as the random data.
As in \citet{HimesEtal2022psjMARGEHOMER}, we find that the models indefinitely improve by minutia and therefore choose to stop training after the above numbers of epochs, as further training offers minimal practical improvement.

\begin{table}[tb]
\caption{Training Differences Between NN Models}\vskip -.25in
\label{tbl:models}
\begin{center}
\begin{tabular}{l l c}
\toprule
Model & Data set & Training epochs\\
\hline
NN1    & Grid of 53,240 TEA & 864  \\
        & models         & \\     
NN2    & 125,000 randomized  & 750  \\
        & TEA models         & \\     
NN3    & Randomized, sized to  & 500  \\
        & match gridded         & \\     
NN4    & Combination of gridded  & 500  \\
        & and randomized         & \\     
\hline
\multicolumn{3}{p{0.85\linewidth}}{\textbf{Notes.}  All TEA models are computed over a grid of 100 pressures.}
\end{tabular}
\end{center}
\end{table}

In training the model to predict equilibrium abundances for a given pressure, temperature, C/H, N/H, and O/H, this NN surrogate model can directly predict the molecular abundance profiles for an atmospheric model with some pressure--temperature profile and (possibly non-uniform) elemental abundance ratios.  
To do so, we normalize the inputs as described above, make the predictions, and denormalize the outputs.

\subsection{Performance Assessment}

We treat TEA as the control.
We assess each non-TEA model by computing (1) the root-mean-square error (RMSE), coefficient of determination ($R^2$), and absolute dex errors (ADEs) over a grid of temperatures, pressures, metallicities, and C/O, and (2) the speedup factor compared to TEA.  
For (1), we use a grid similar to that used in \citet{CubillosEtal2019apjRATE}, except with the temperature and metallicity grids shifted by a half cell.  
This results in metallicities spanning $-1.5$ -- $2.5$ dex and temperatures spanning $221$ -- $5485$ K.
We also consider the original grid used in \citet{CubillosEtal2019apjRATE} to test the NN performance at the edge of the phase space.

Note that we compute the ADEs as
\begin{equation}
    ADE = |log_{10}(\chi_{true}) - log_{10}(\chi_{pred})|,
\end{equation}
where $\chi_{true}$ is the molecular abundance calculated by TEA and $\chi_{pred}$ is the molecular abundance predicted by a given model.  
To reduce the multidimensional matrix of ADEs to a single interpretable value, we tabulate the median ADE for each model.  
Additionally, we compute the speedup factors by averaging over all cases within a grid of temperatures, pressures, metallicities, and C/O.

\section{Results \& Discussion}
\label{sec:results6}

\begin{figure*}[htb]
\centering
\includegraphics[width=0.9\textwidth, clip]{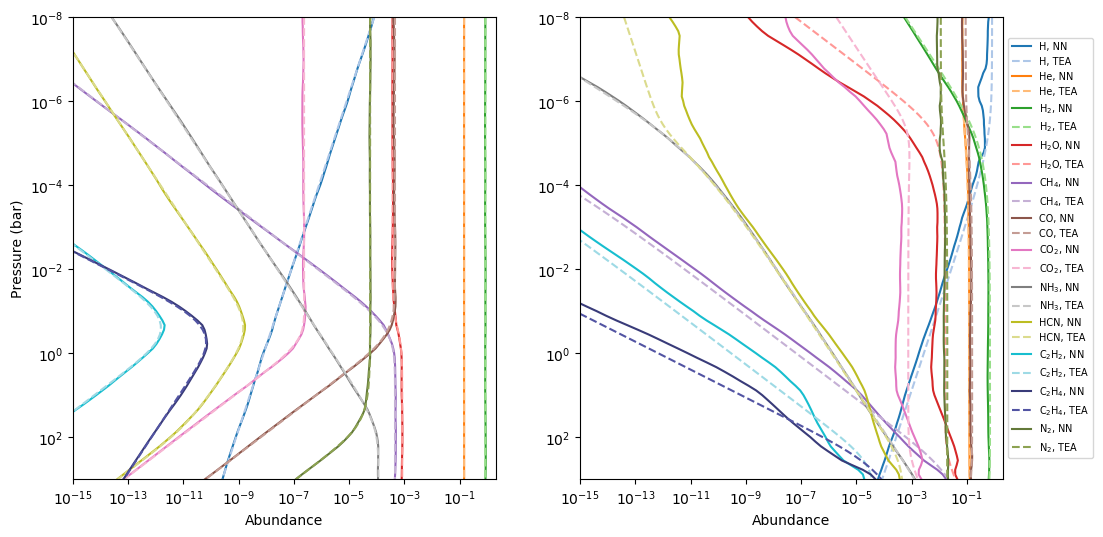}
\caption{Two example cases comparing the outputs of one of the NN models with the corresponding TEA outputs.  \textbf{Left:} Case where the NN's predictions closely match the outputs of TEA, at a temperature of $\sim$1053 K, log C/H of -3.57, log N/H of -4.17, and log O/H of -3.31 over the full range of pressures considered in Table \ref{tbl:teagrid}.  \textbf{Right:} Case where the NN performs less accurately, at a temperature of $\sim$2112 K, M/H of 2.5, and C/O of 0.9 (log C/H of -0.956, log N/H of -1.556, and log O/H of -0.910).  Despite the less accurate performance, the NN's predictions are within an order of magnitude of the true values except for H$_2$O, CO$_2$, and HCN at the lowest pressures.  Only abundances $\geq 10^{-15}$ are shown for clarity.
}
\label{fig:example}
\end{figure*}

\begin{figure*}[htb]
\centering
\includegraphics[width=0.49\textwidth, clip]{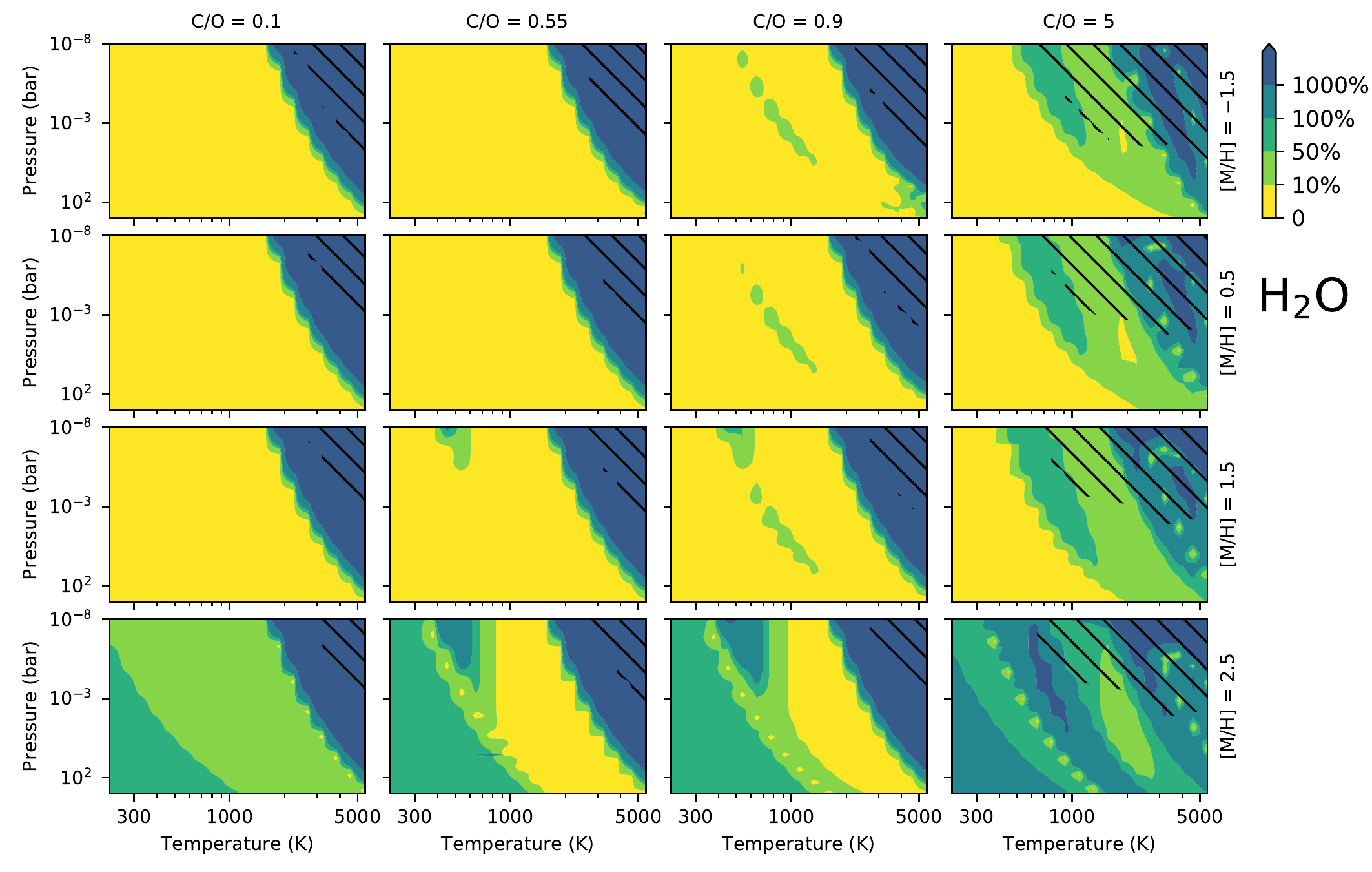}\hfill
\includegraphics[width=0.49\textwidth, clip]{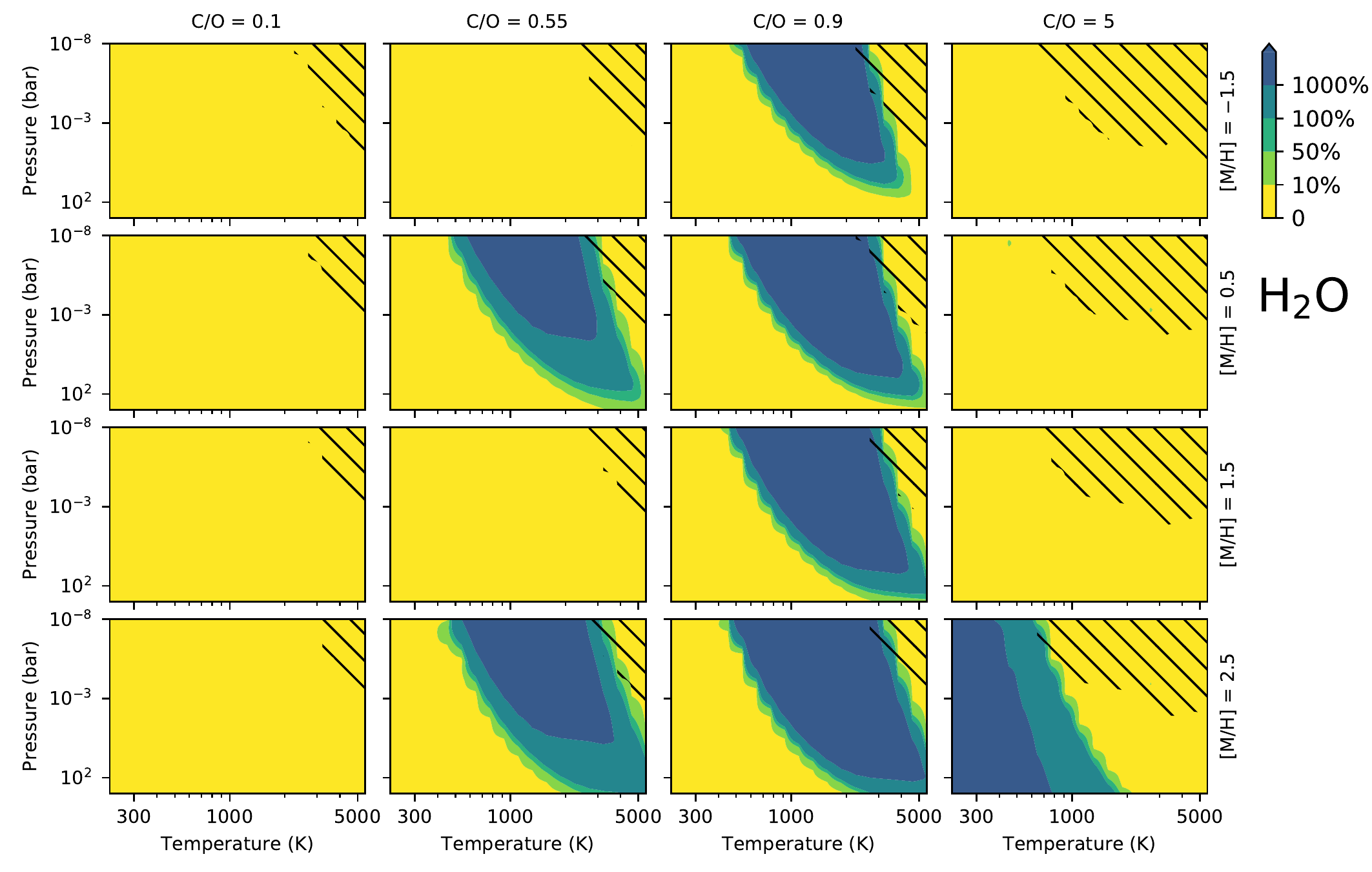}\\
\includegraphics[width=0.49\textwidth, clip]{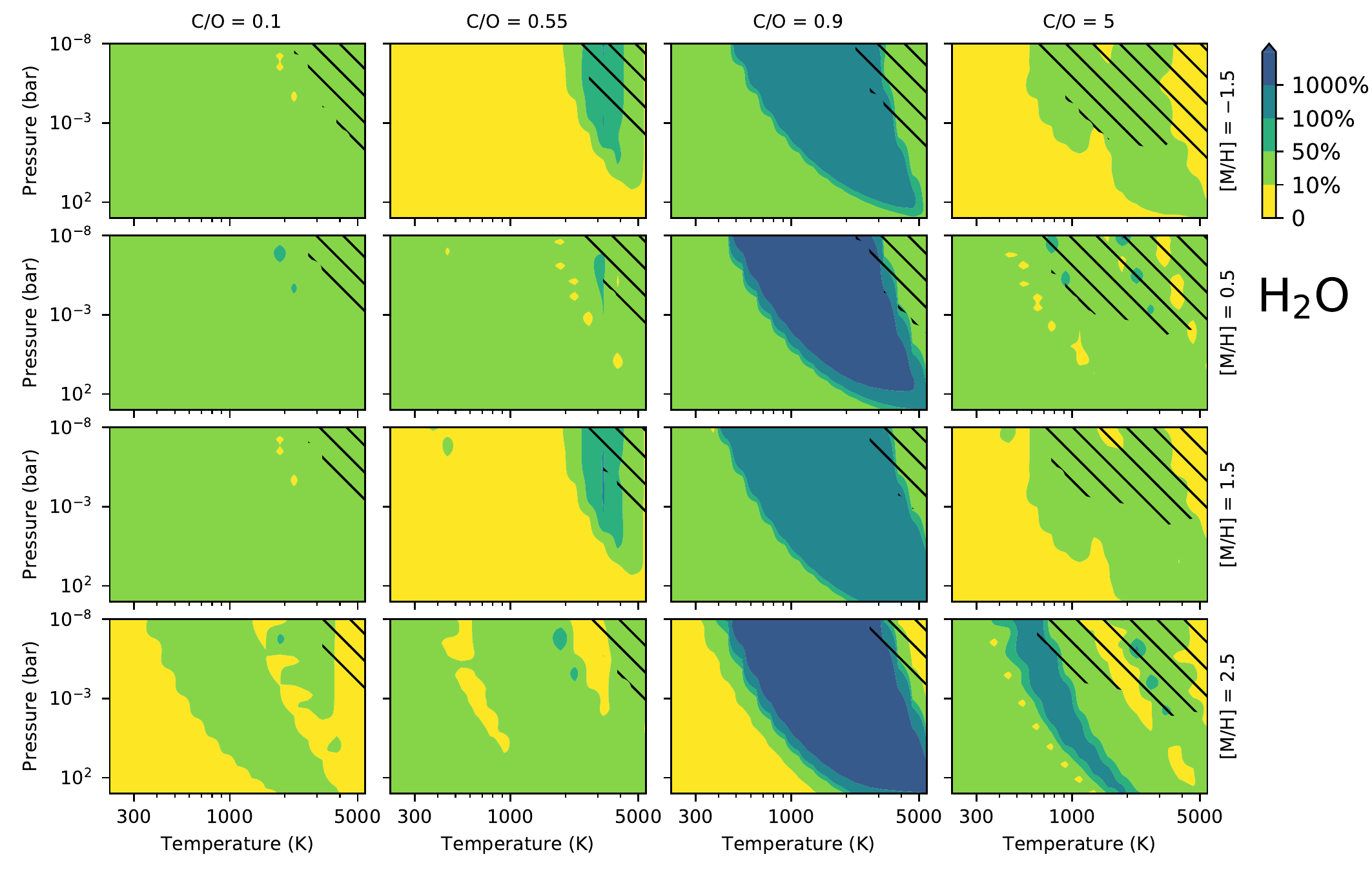}\hfill
\includegraphics[width=0.49\textwidth, clip]{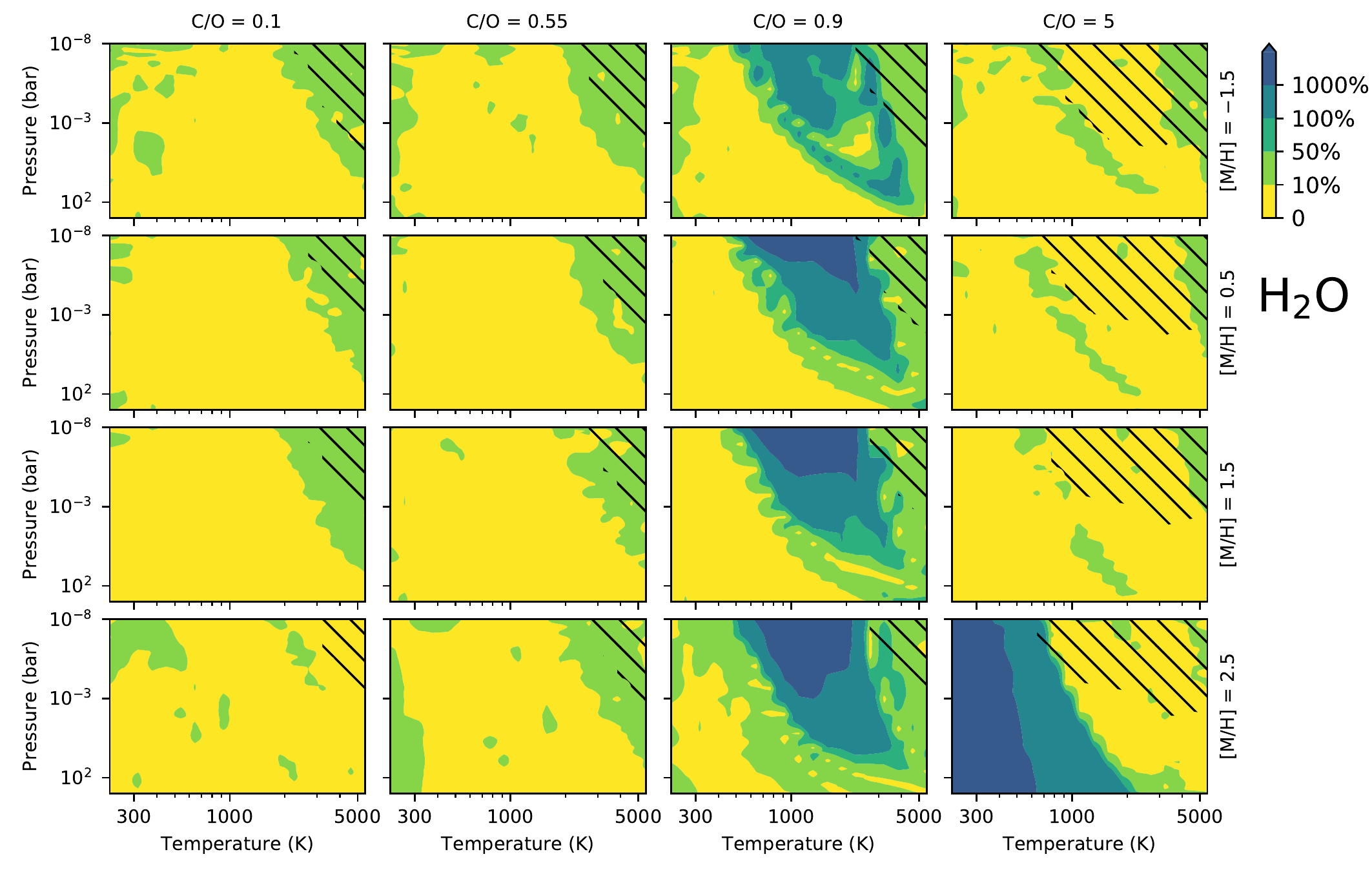}\\
\includegraphics[width=0.49\textwidth, clip]{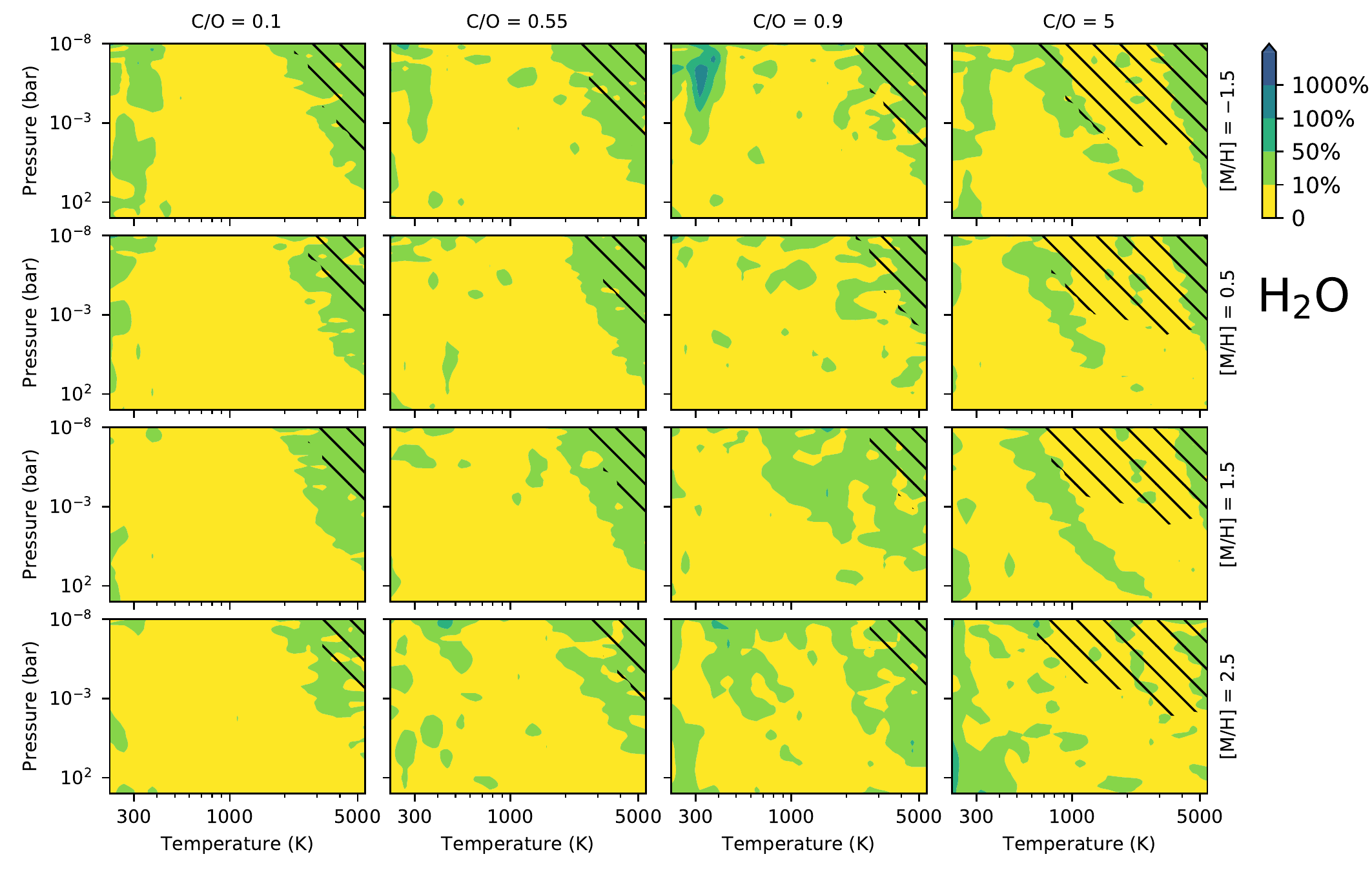}\hfill
\includegraphics[width=0.49\textwidth, clip]{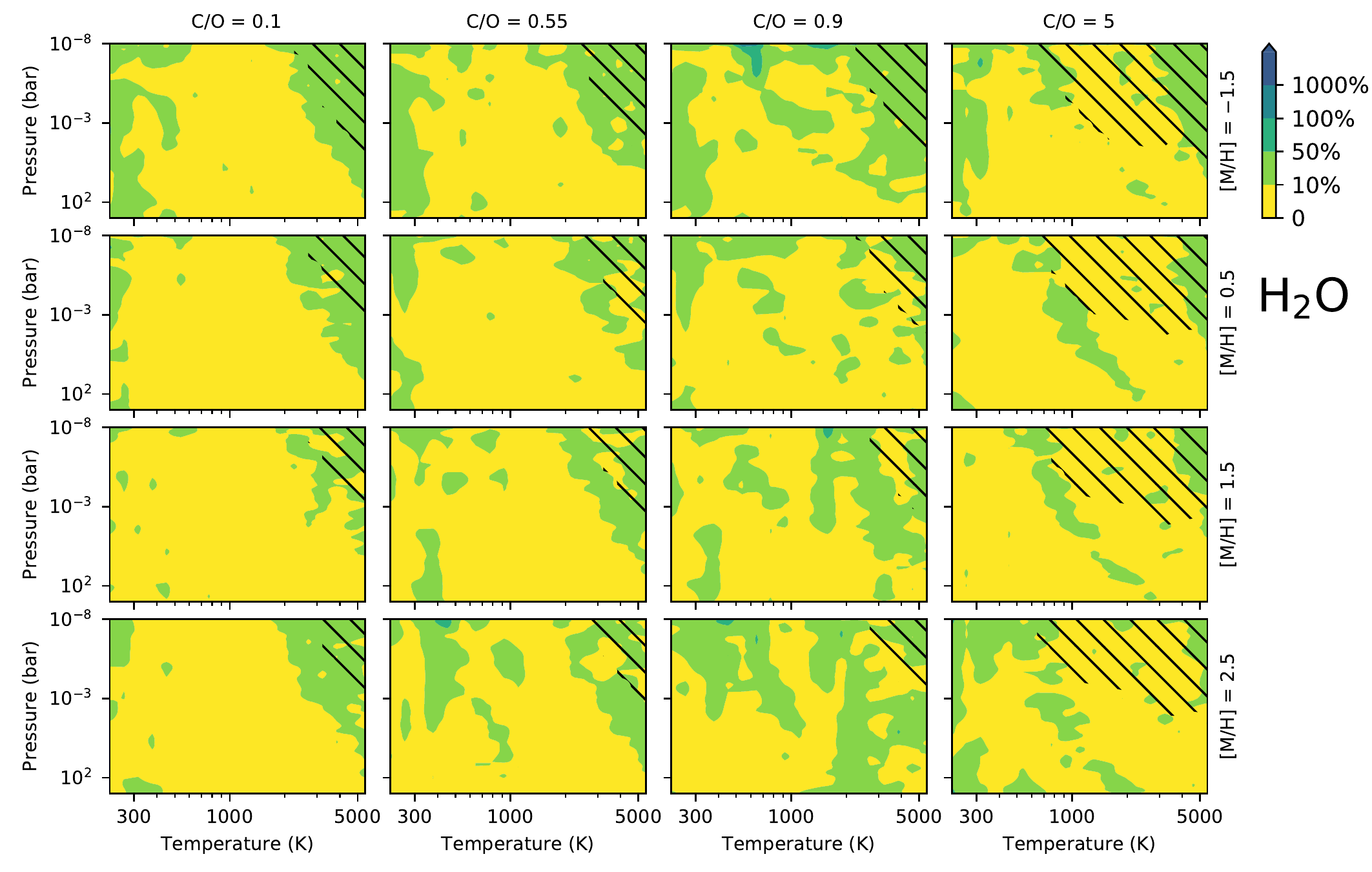}\\
\caption{Absolute percent differences between the abundances predicted by the various approximation models and TEA for H$_2$O.  Cross-hatched regions indicate where the molecular abundance is less than 10$^{-10}$.
\textbf{Top left:} RATE.  
\textbf{Top right:} Best-performing linear interpolation model.  
\textbf{Middle left:} Best-performing IDW interpolation model.
\textbf{Middle right:} NN1.
\textbf{Bottom left:} NN2.
\textbf{Bottom right:} NN3.
}
\label{fig:H2O}
\end{figure*}

\begin{figure*}[htb]
\centering
\includegraphics[width=0.49\textwidth, clip]{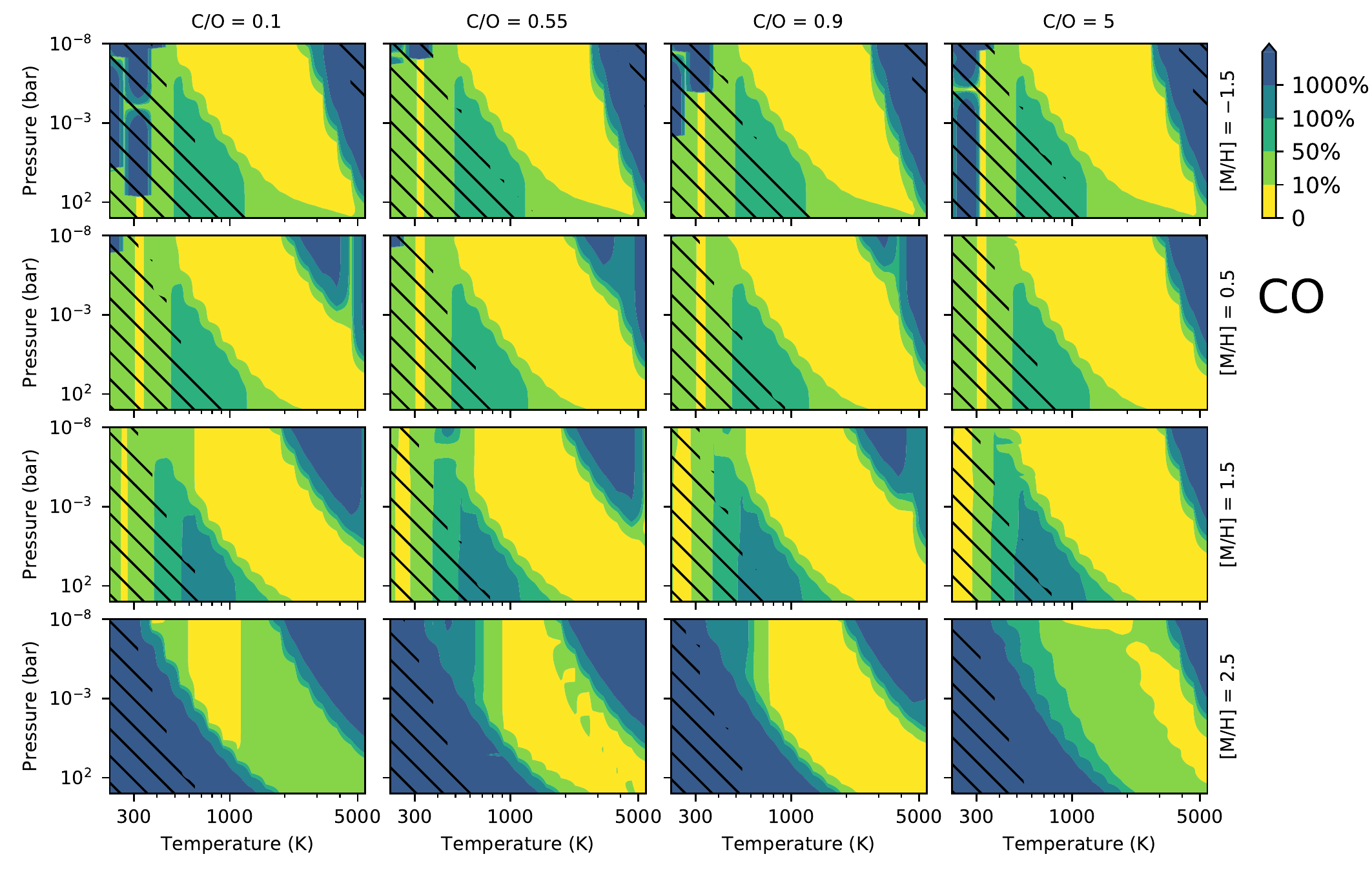}\hfill
\includegraphics[width=0.49\textwidth, clip]{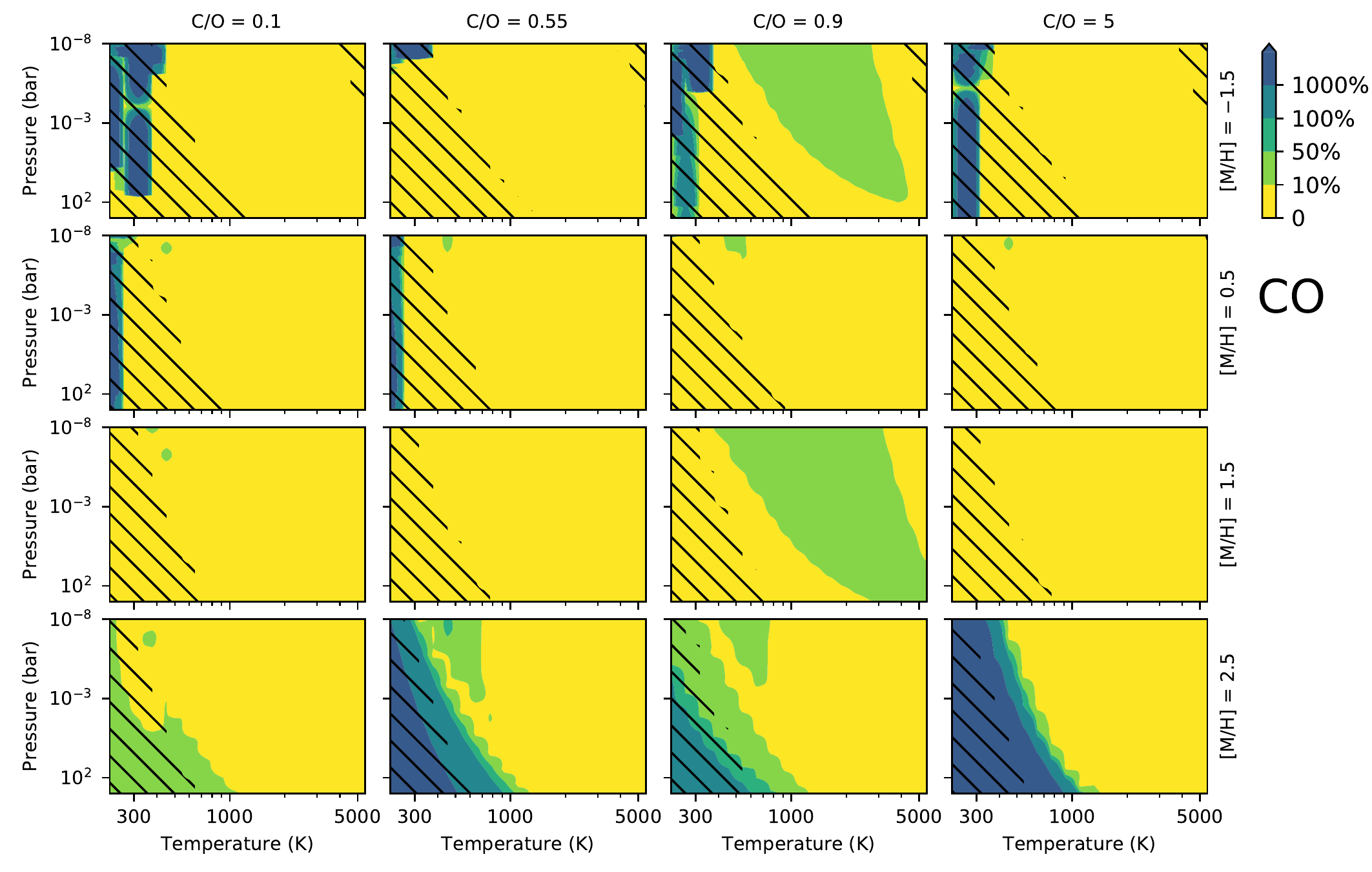}\\
\includegraphics[width=0.49\textwidth, clip]{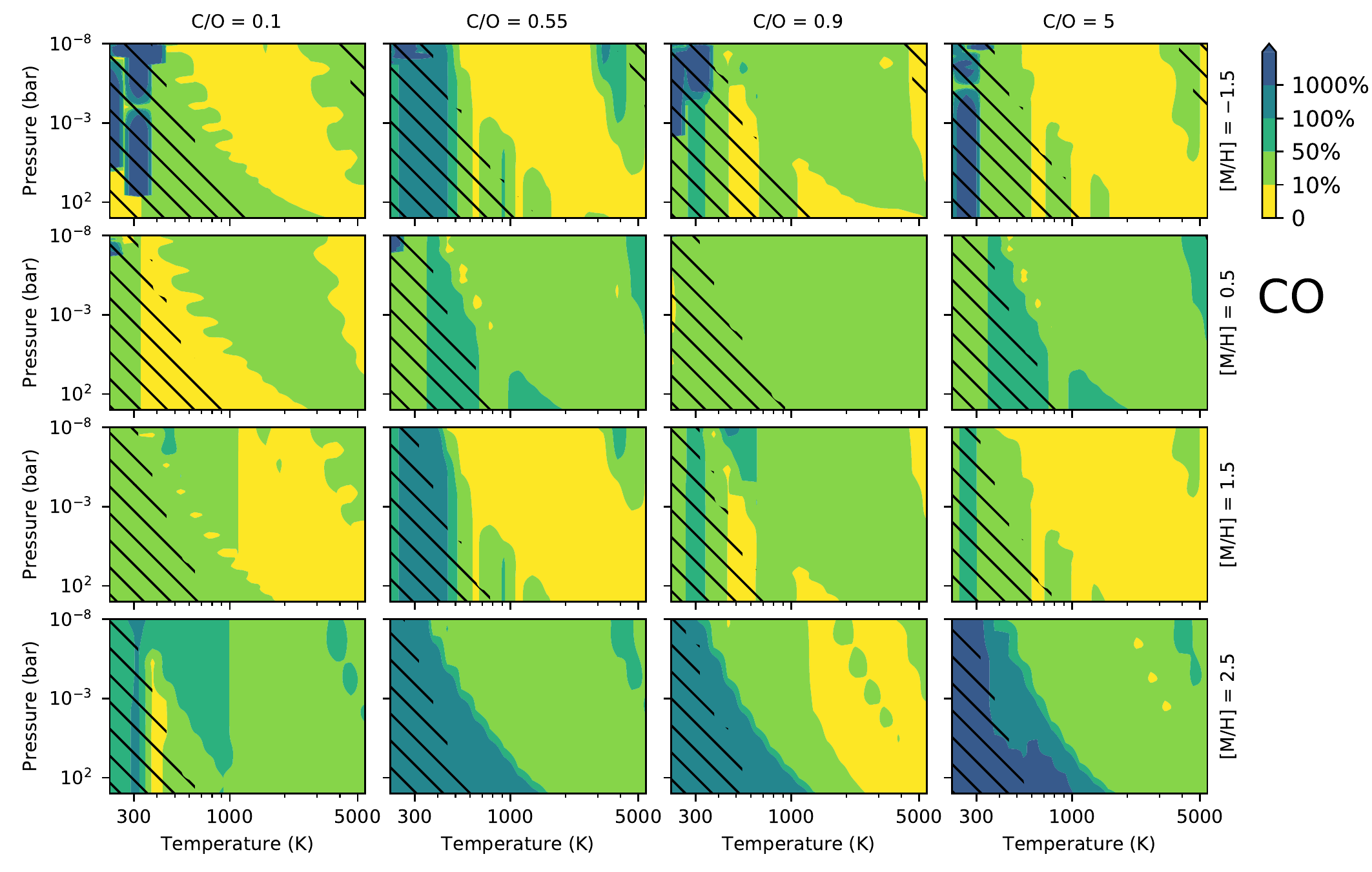}\hfill
\includegraphics[width=0.49\textwidth, clip]{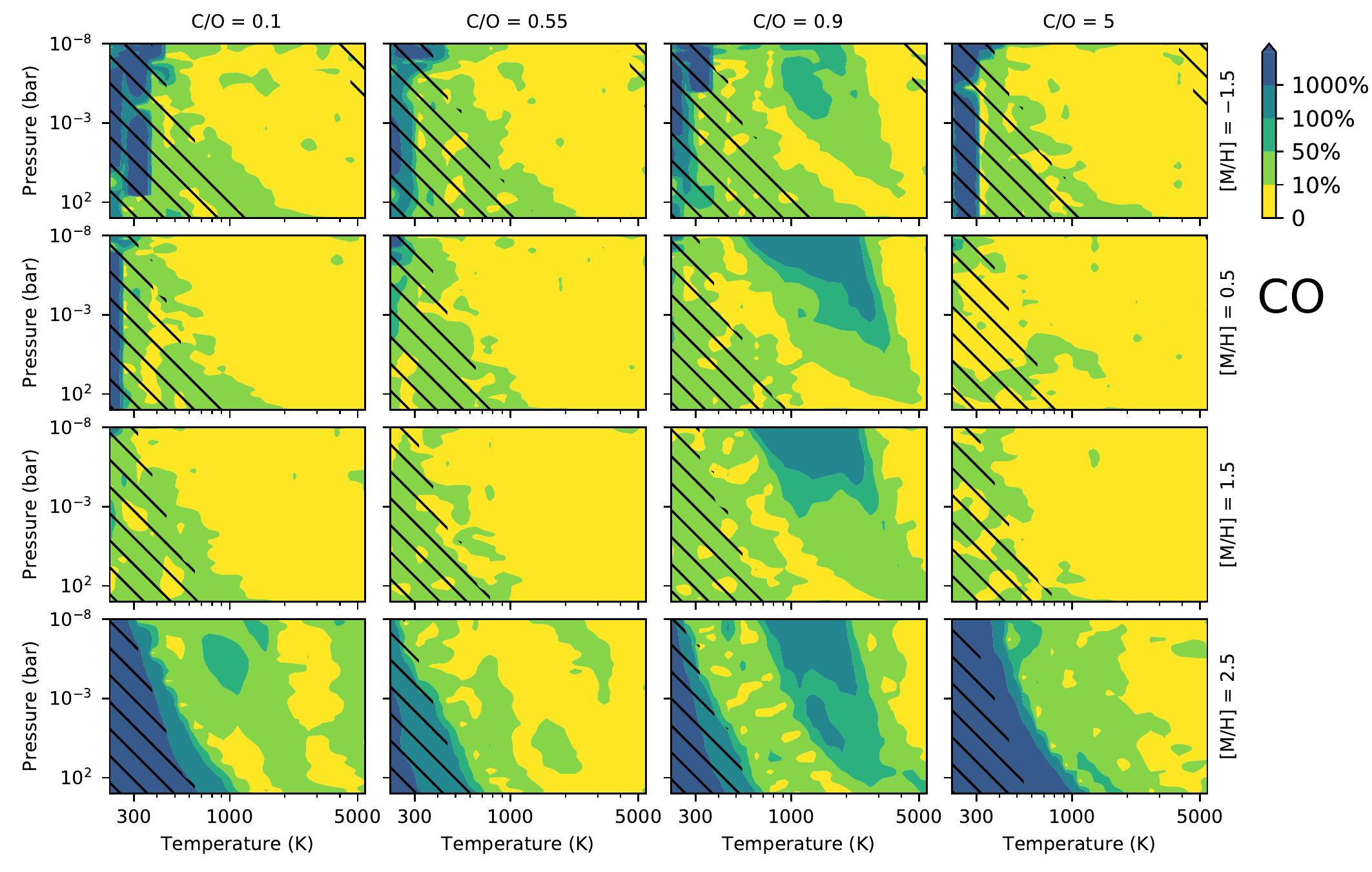}\\
\includegraphics[width=0.49\textwidth, clip]{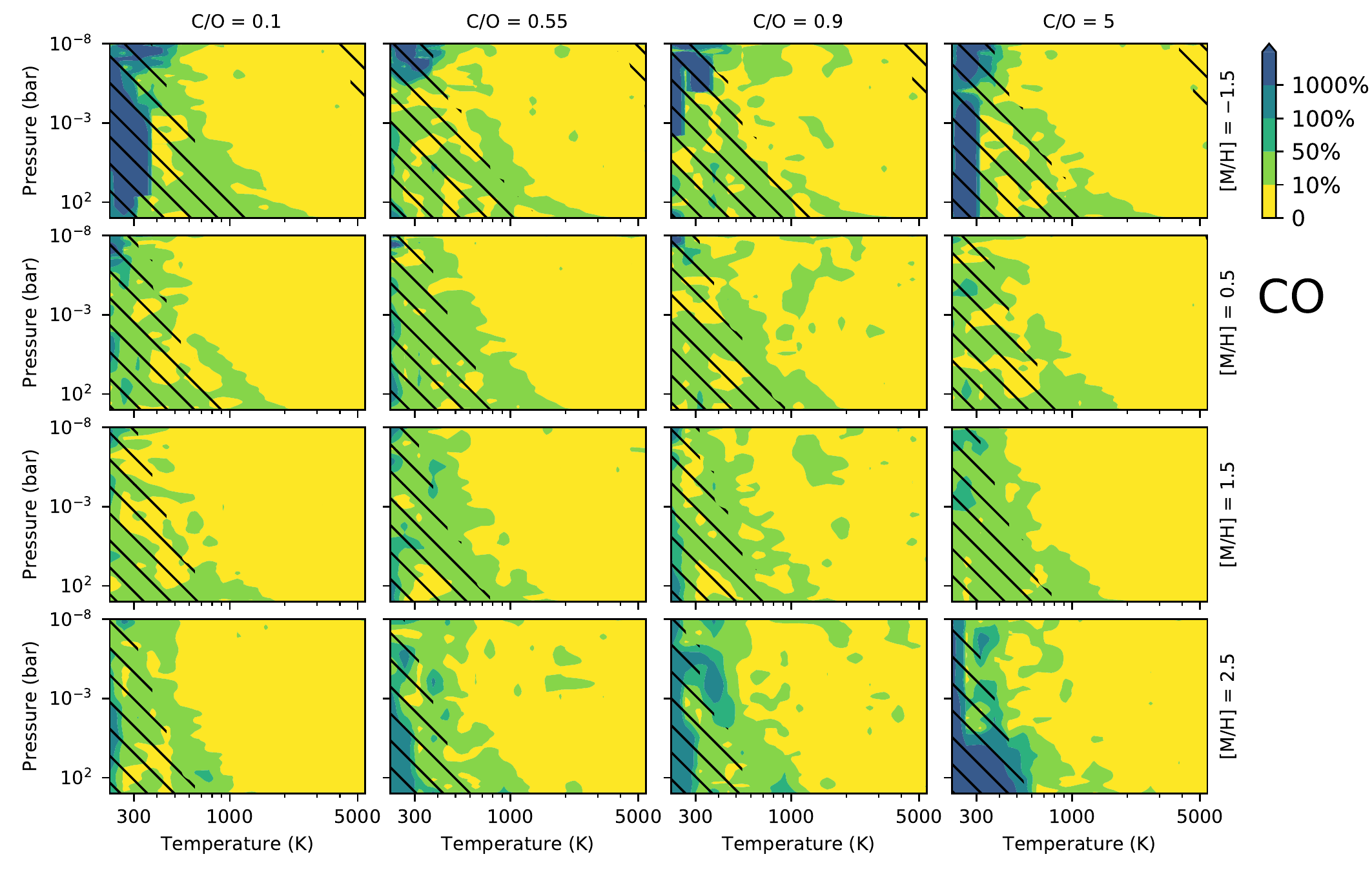}\hfill
\includegraphics[width=0.49\textwidth, clip]{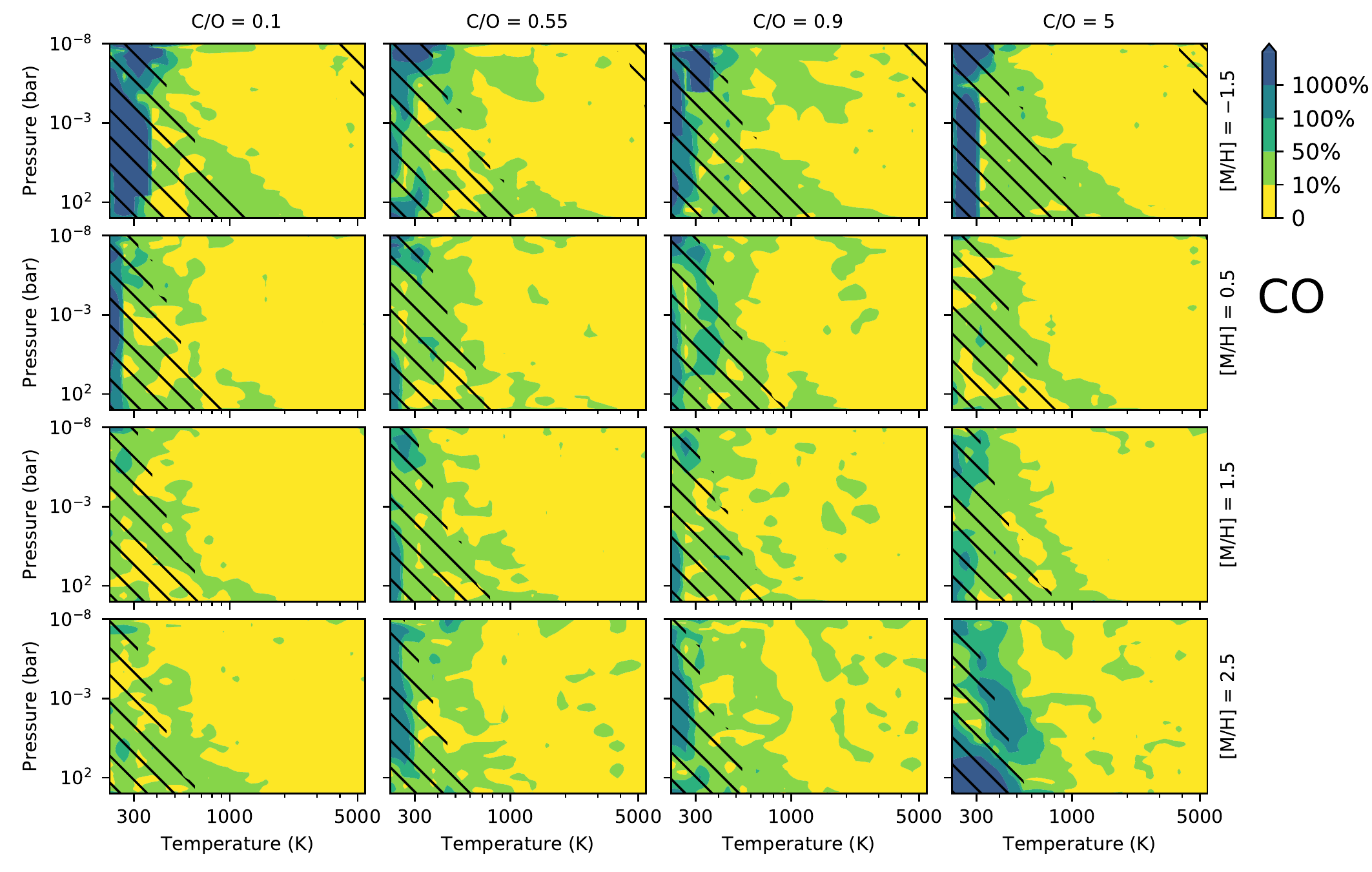}\\
\caption{As in Figure \ref{fig:H2O}, but for CO.
}
\label{fig:CO}
\end{figure*}

\begin{figure*}[htb]
\centering
\includegraphics[width=0.49\textwidth, clip]{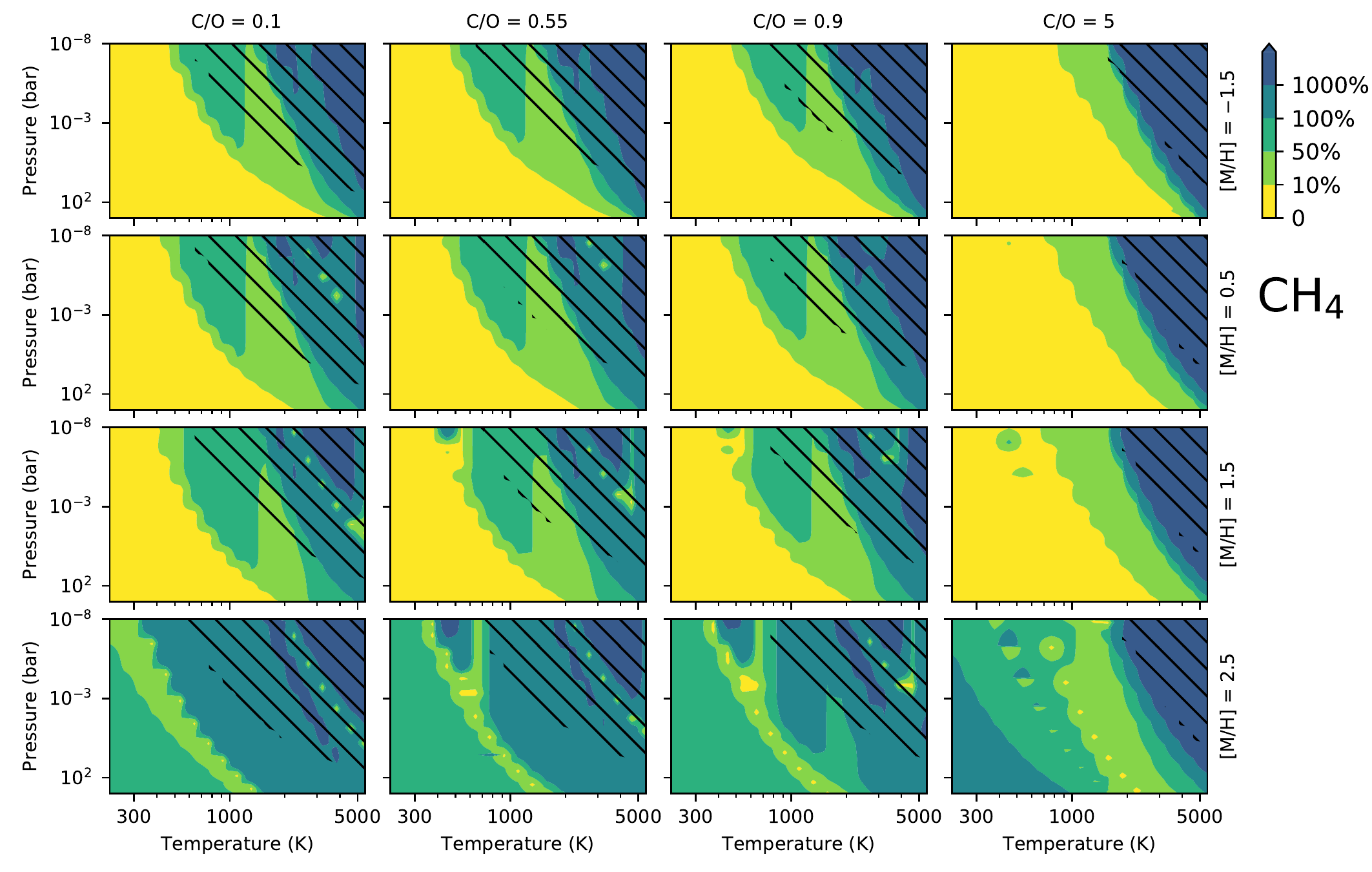}\hfill
\includegraphics[width=0.49\textwidth, clip]{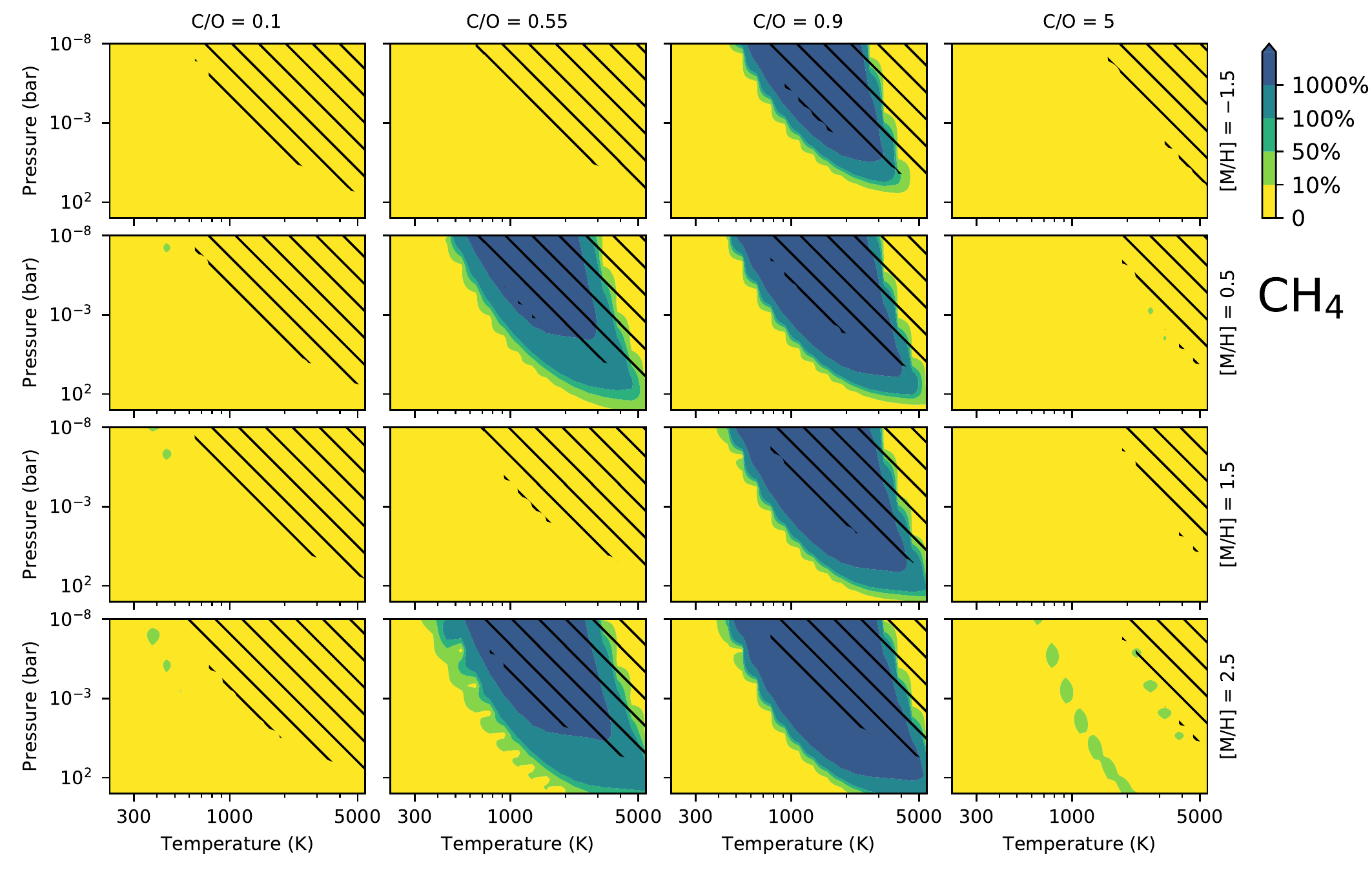}\\
\includegraphics[width=0.49\textwidth, clip]{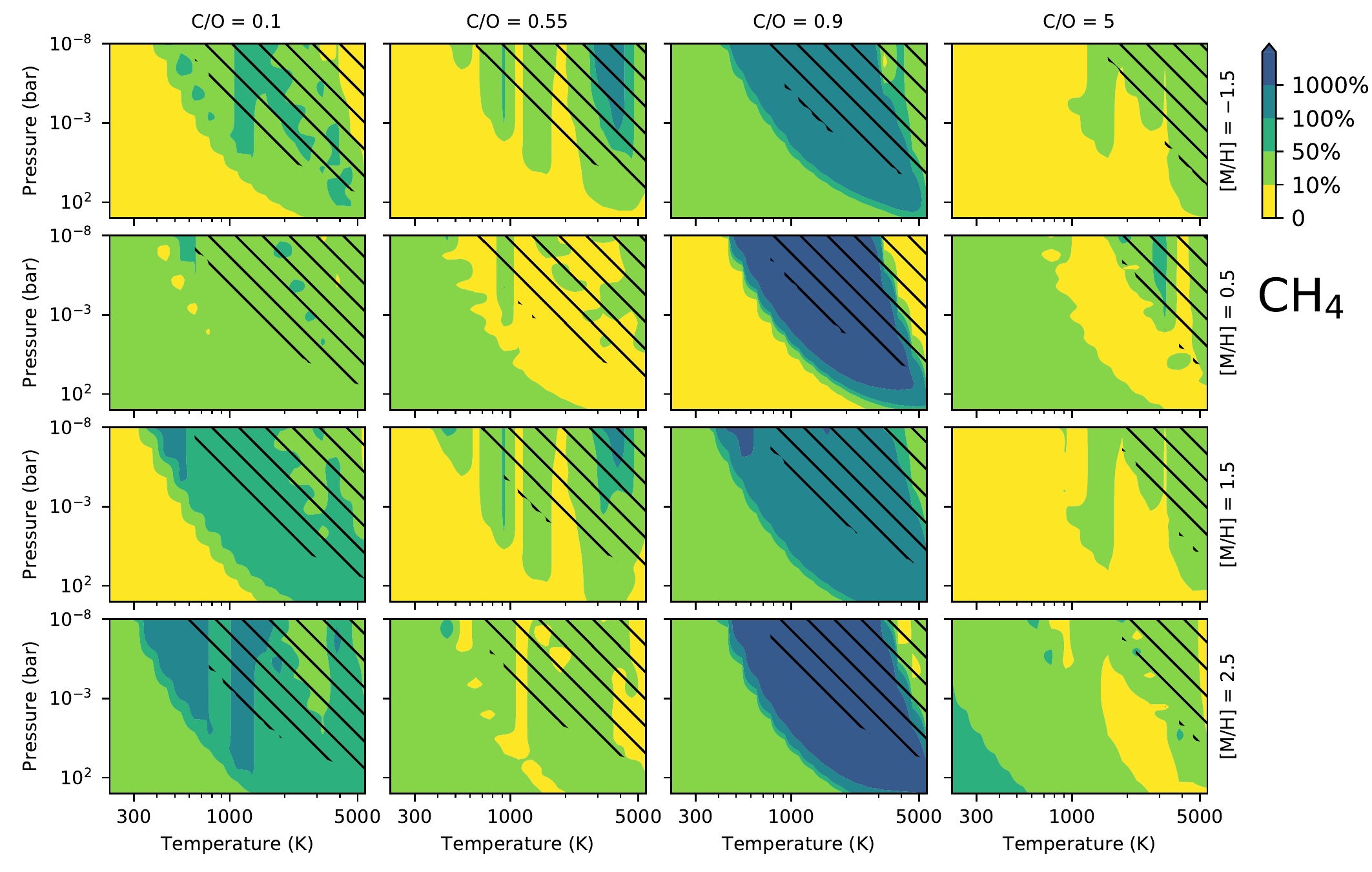}\hfill
\includegraphics[width=0.49\textwidth, clip]{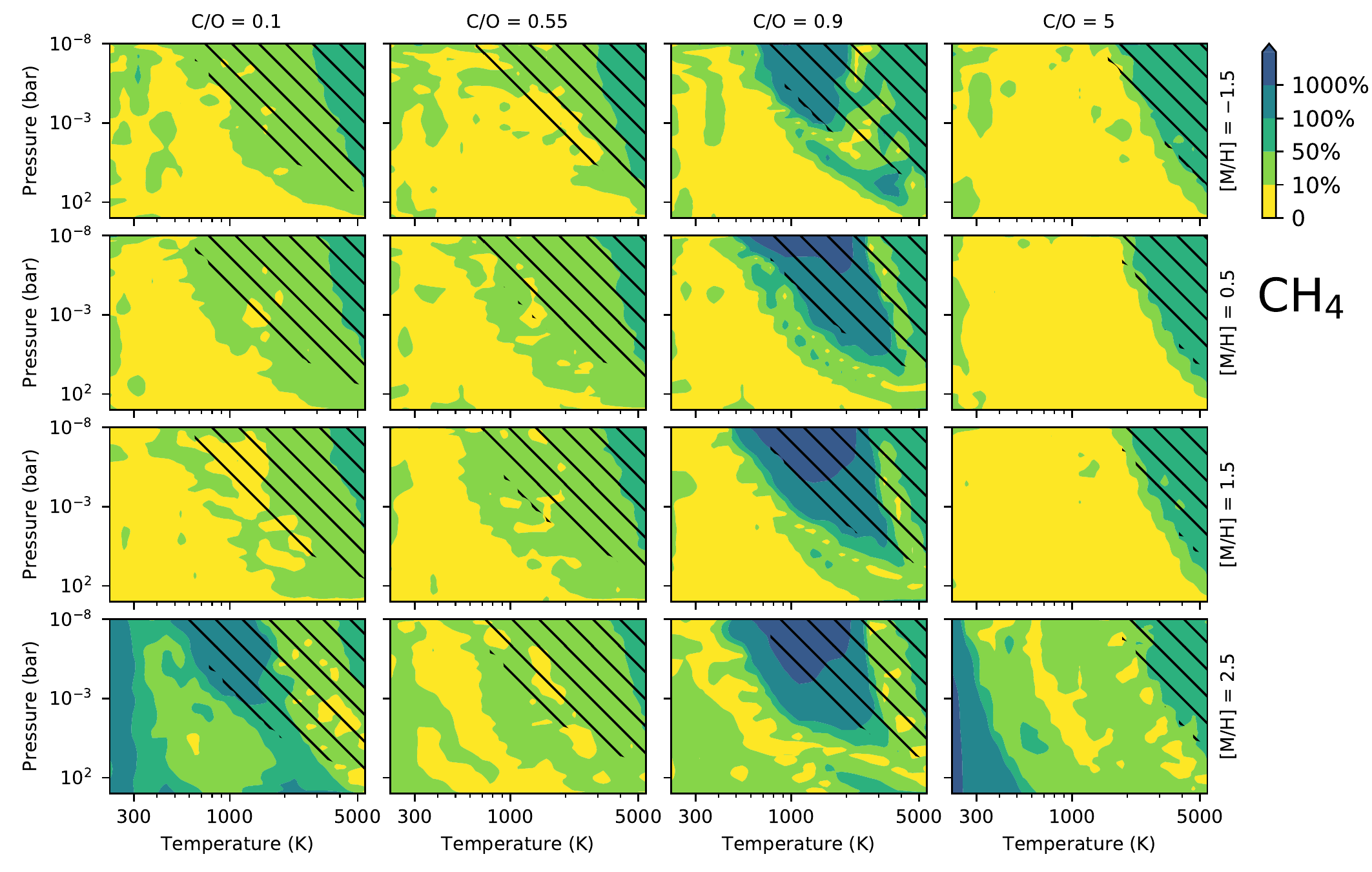}\\
\includegraphics[width=0.49\textwidth, clip]{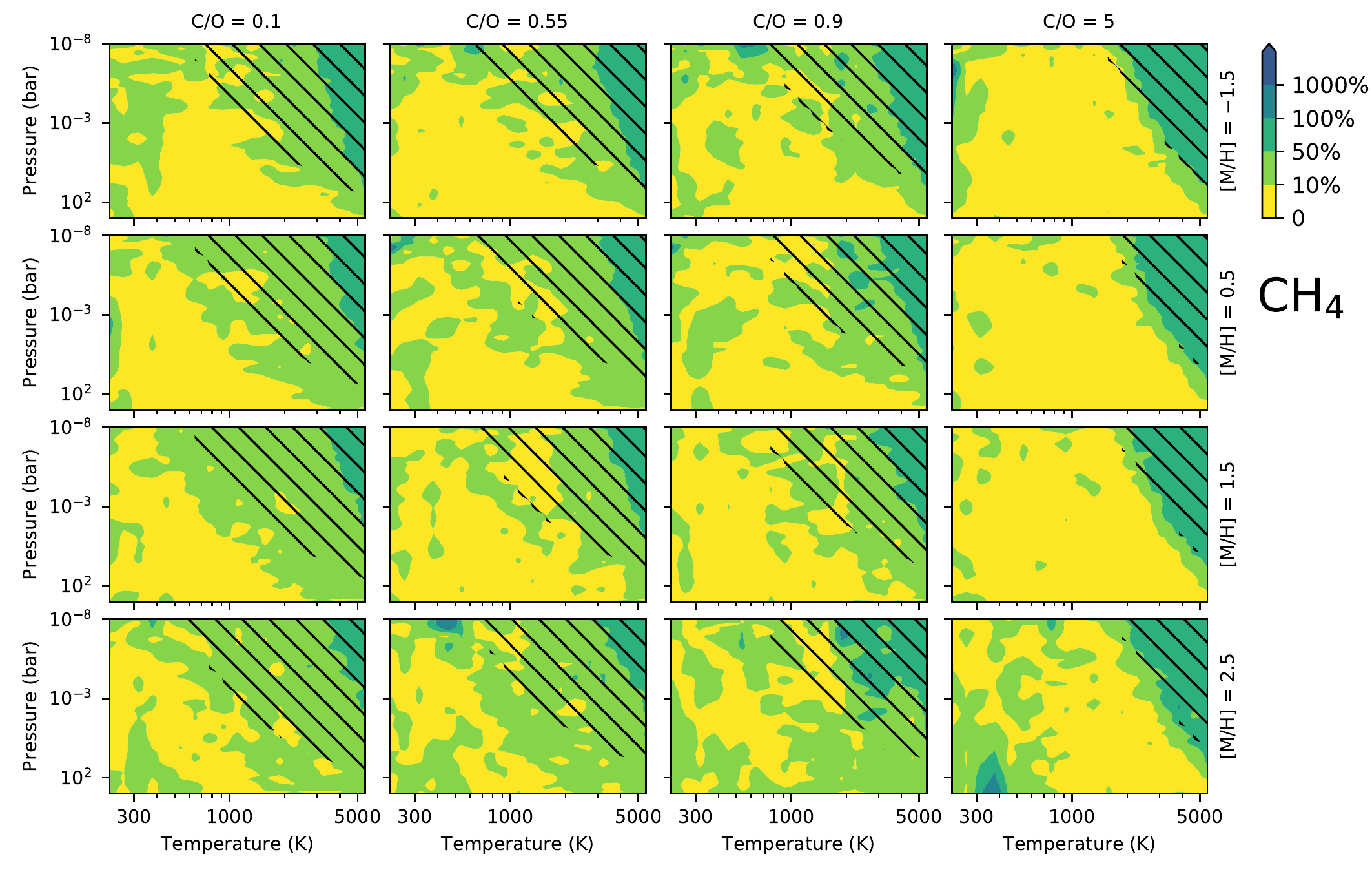}\hfill
\includegraphics[width=0.49\textwidth, clip]{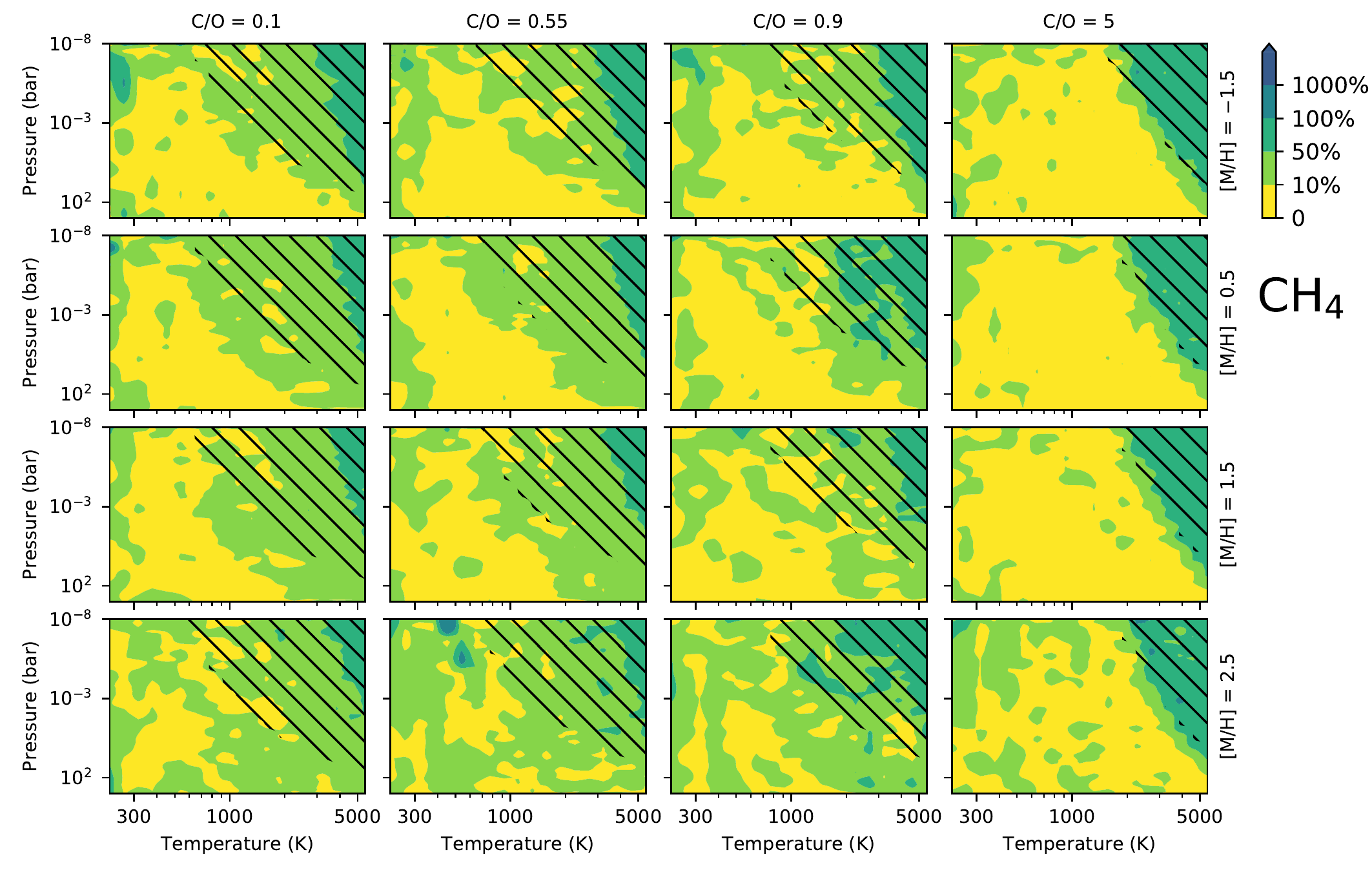}\\
\caption{As in Figure \ref{fig:H2O}, but for CH$_4$.
}
\label{fig:CH4}
\end{figure*}

\begin{figure*}[htb]
\centering
\includegraphics[width=0.49\textwidth, clip]{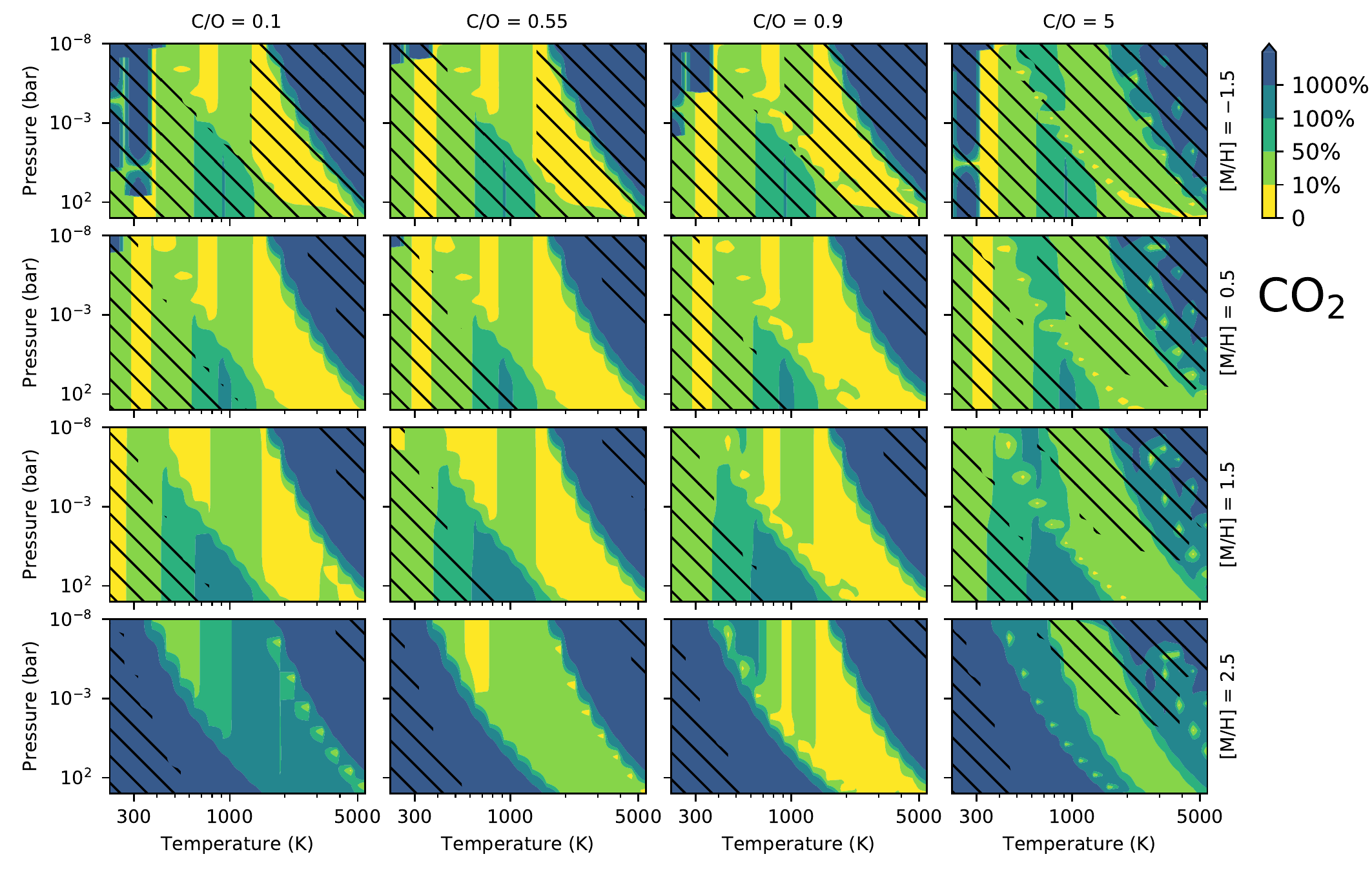}\hfill
\includegraphics[width=0.49\textwidth, clip]{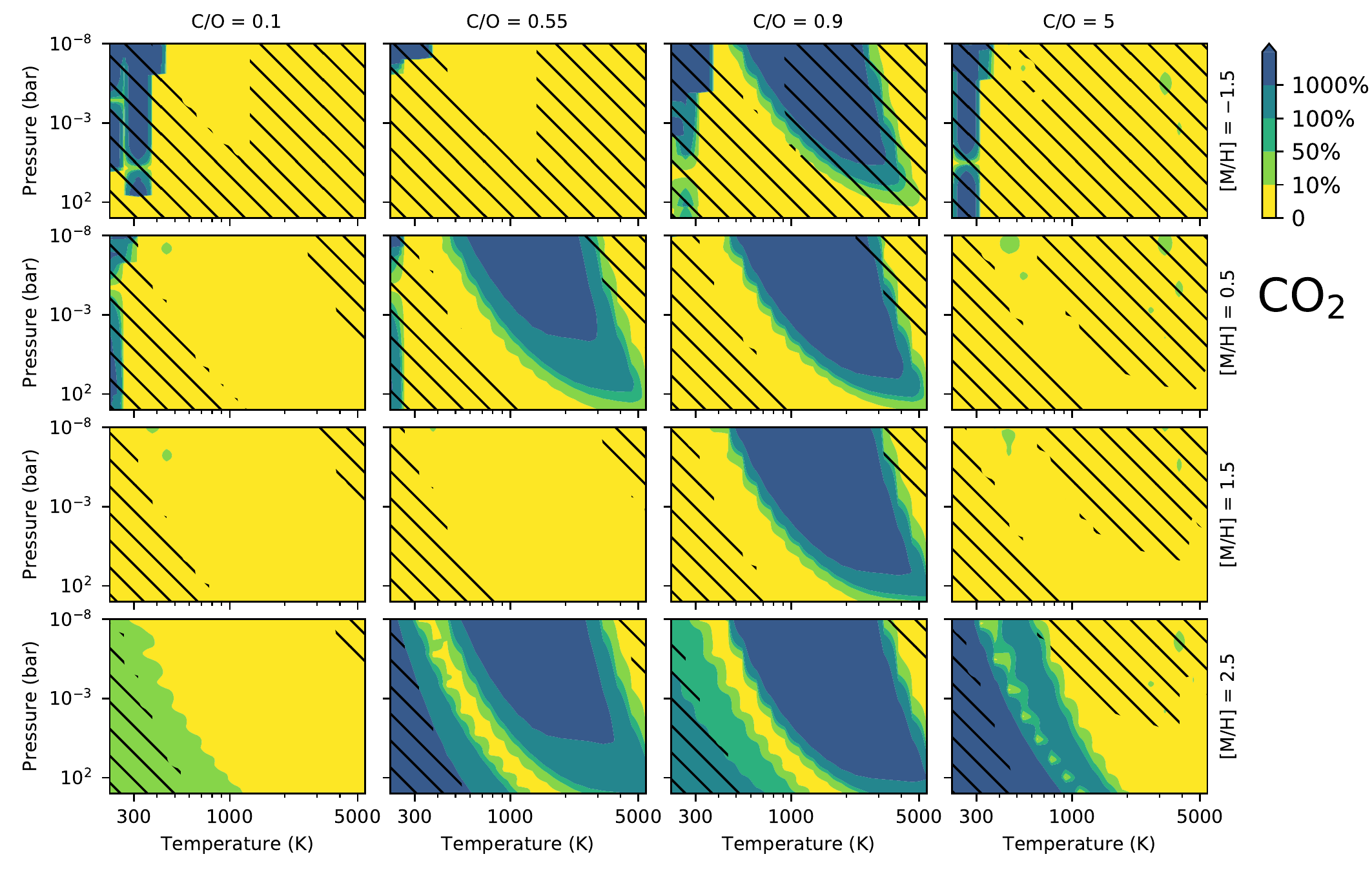}\\
\includegraphics[width=0.49\textwidth, clip]{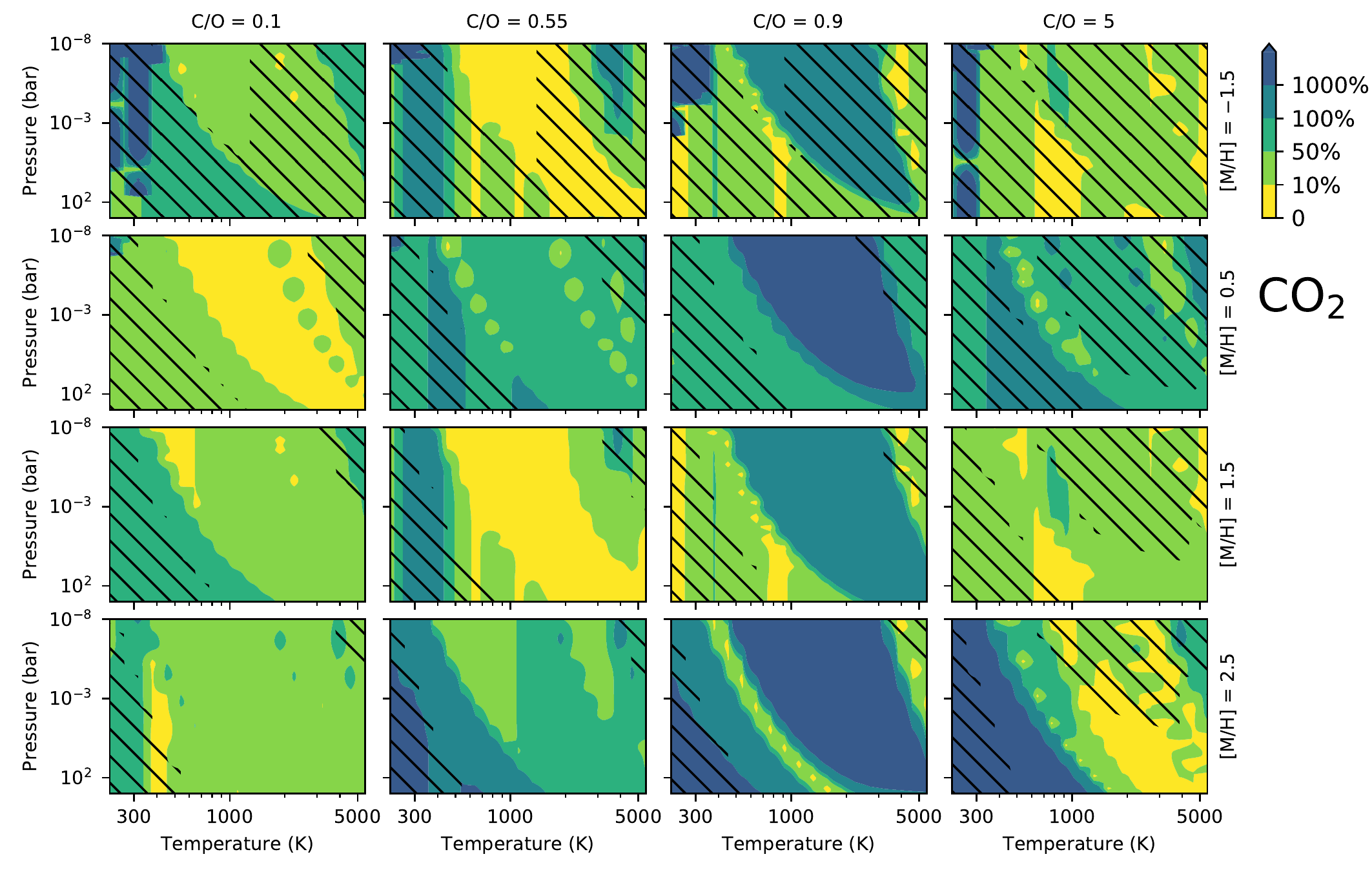}\hfill
\includegraphics[width=0.49\textwidth, clip]{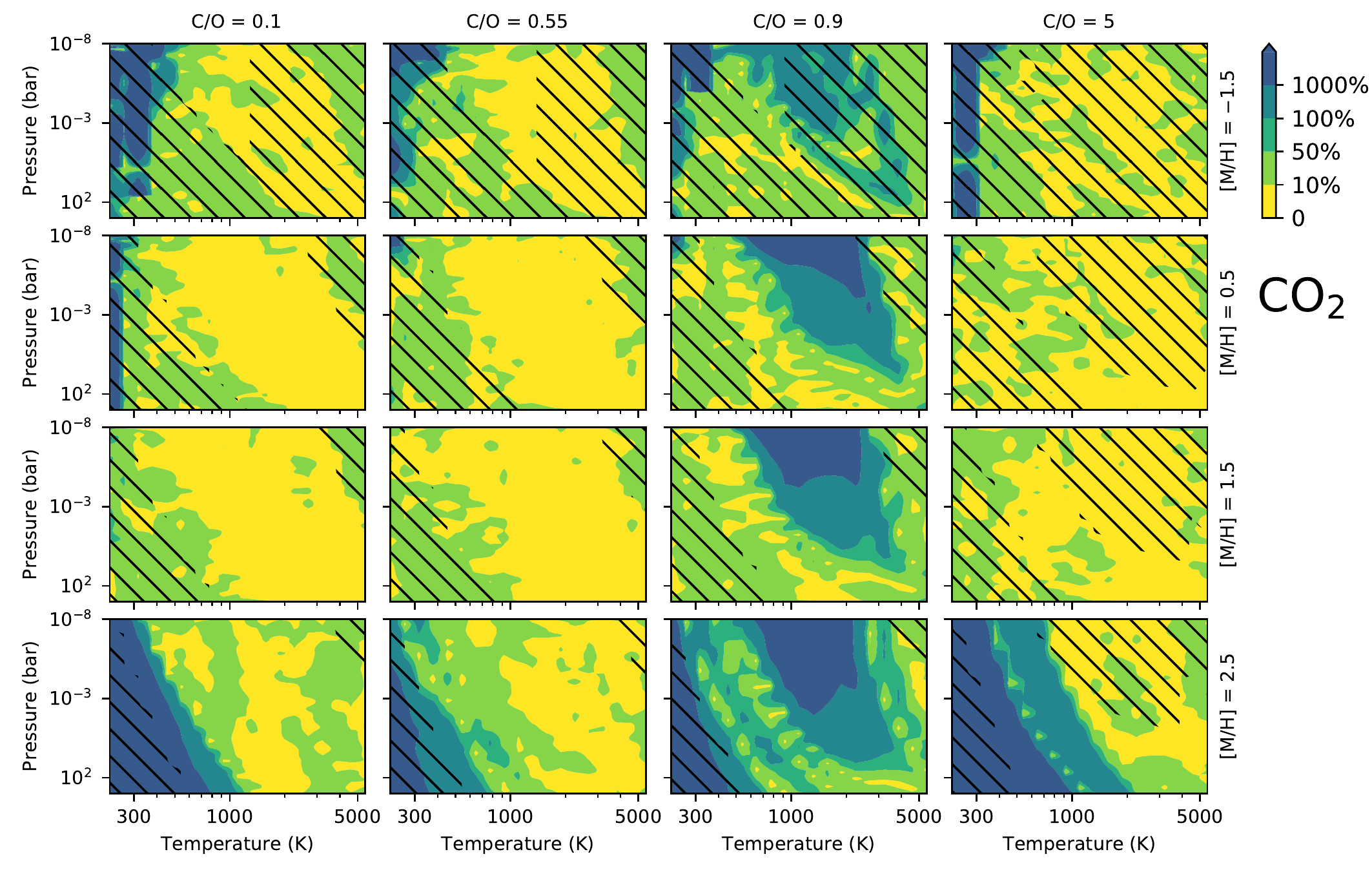}\\
\includegraphics[width=0.49\textwidth, clip]{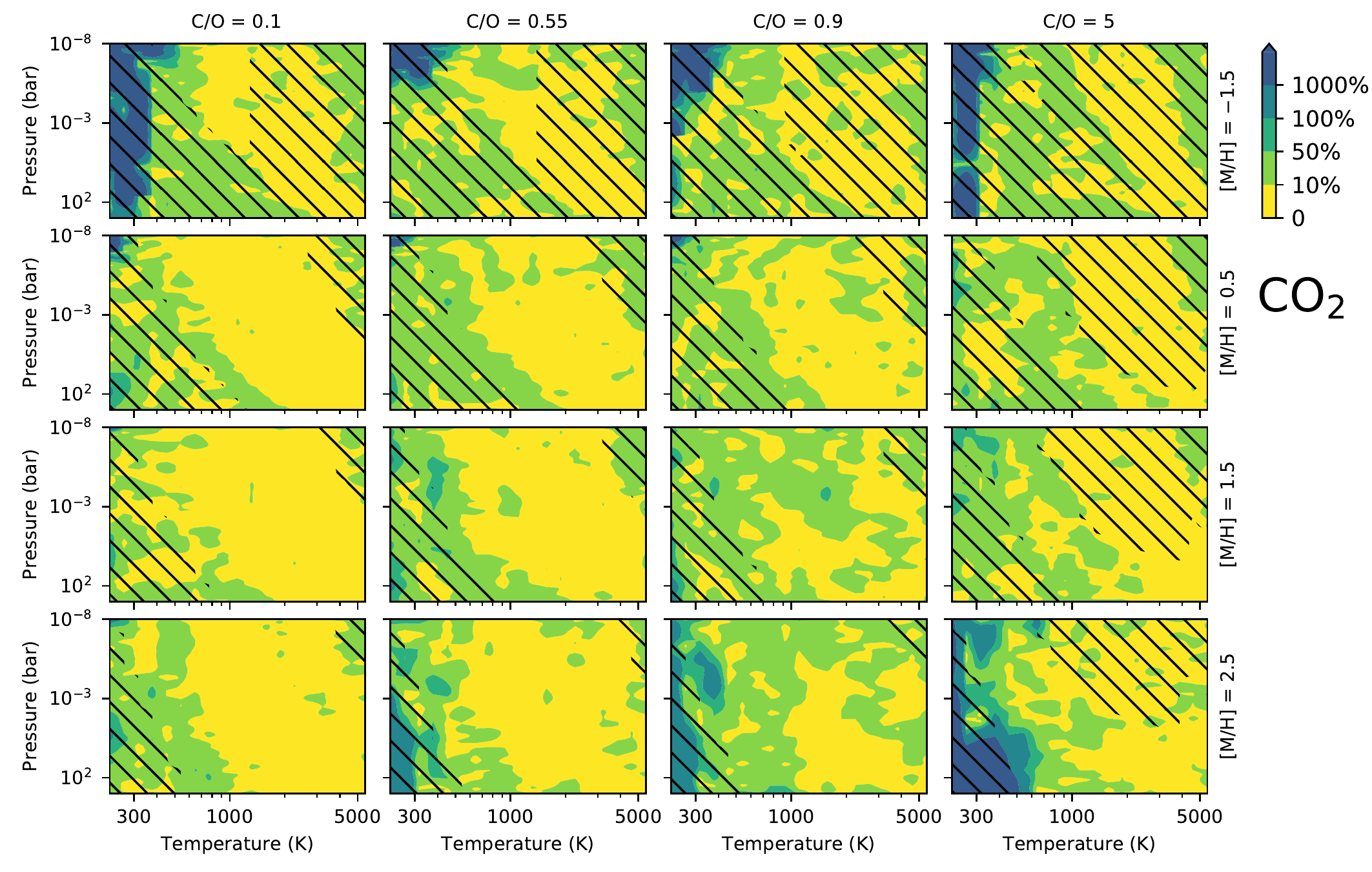}\hfill
\includegraphics[width=0.49\textwidth, clip]{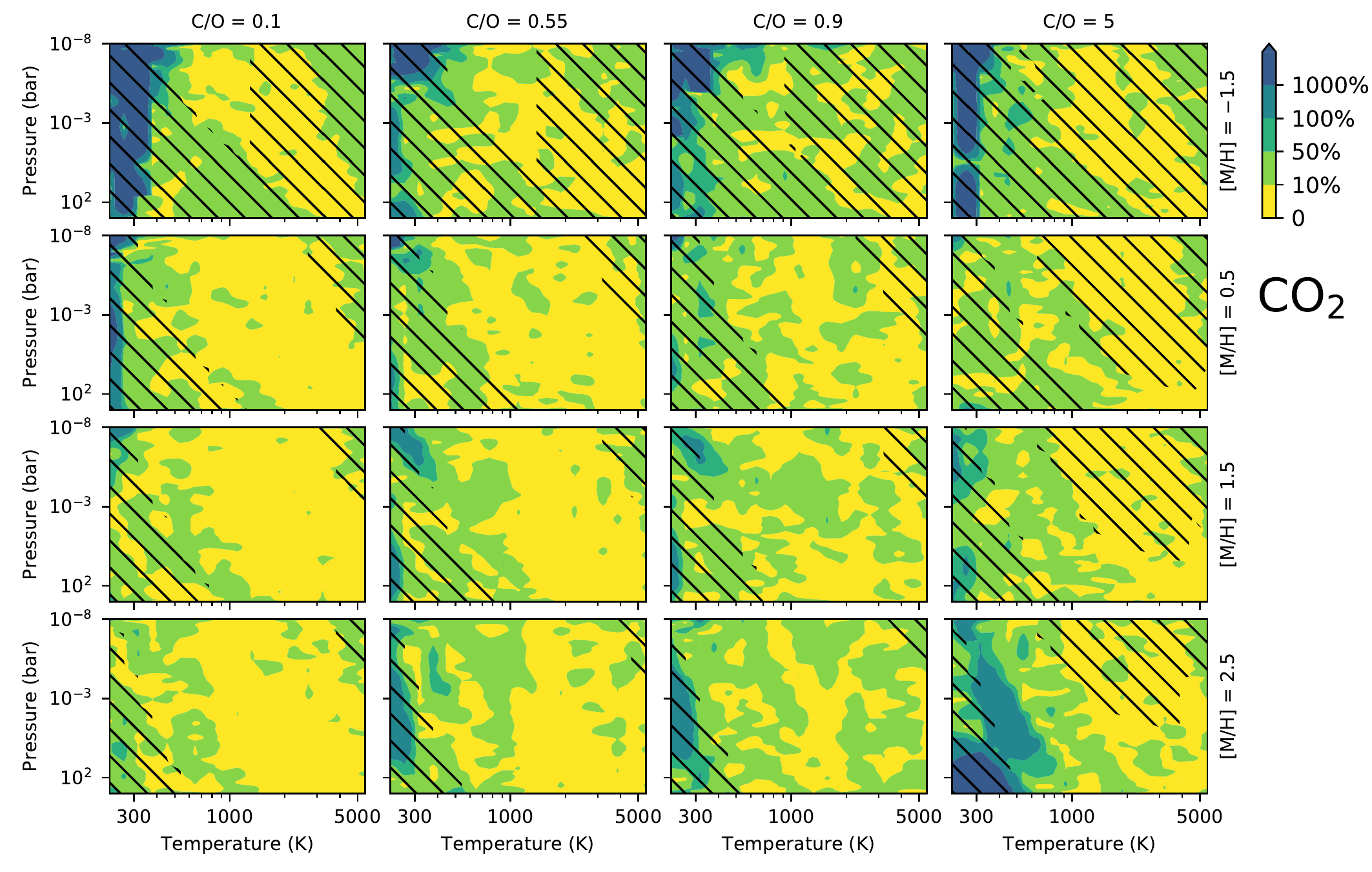}\\
\caption{As in Figure \ref{fig:H2O}, but for CO$_2$.
}
\label{fig:CO2}
\end{figure*}

\begin{figure*}[htb]
\centering
\includegraphics[width=0.49\textwidth, clip]{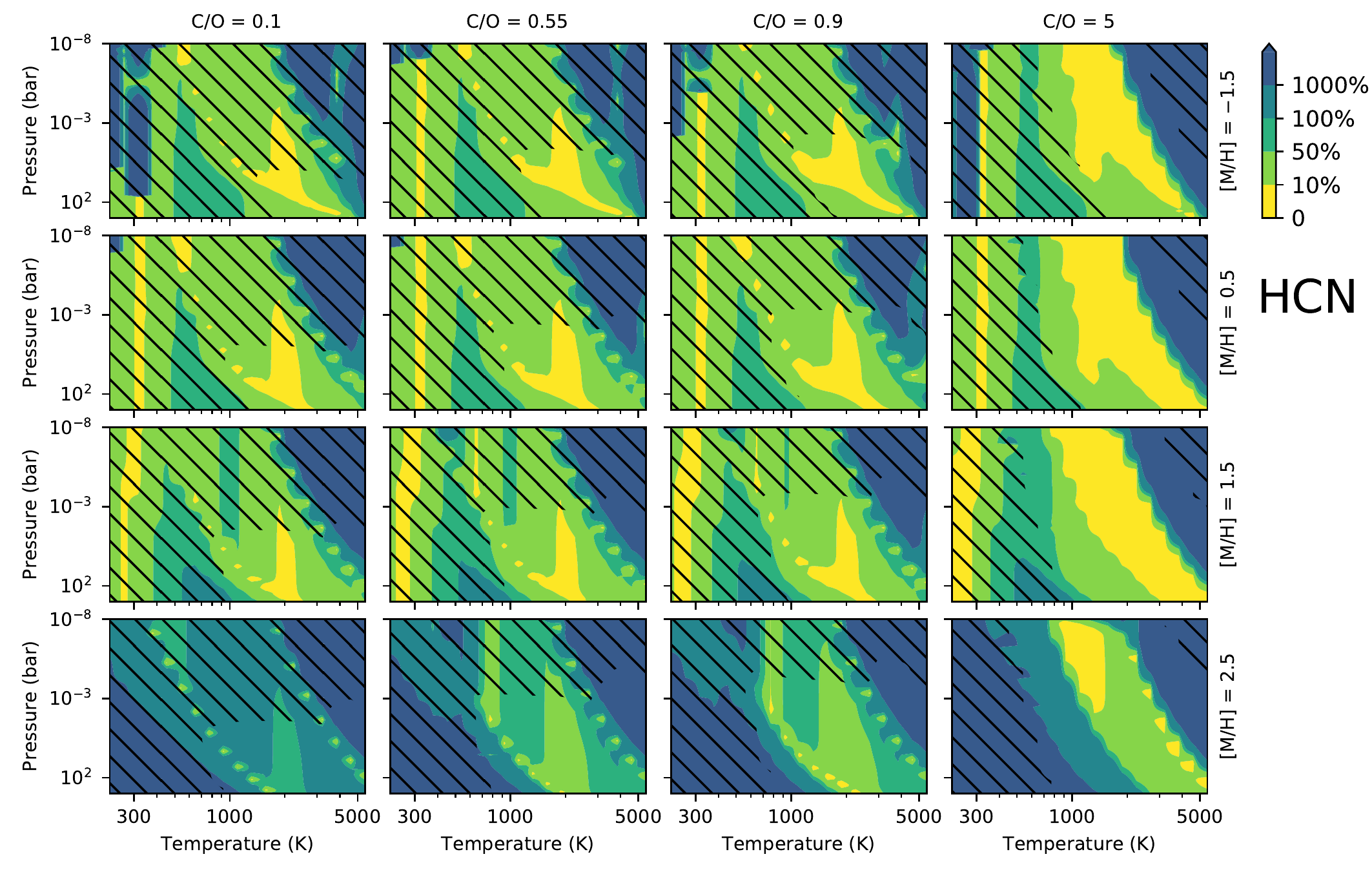}\hfill
\includegraphics[width=0.49\textwidth, clip]{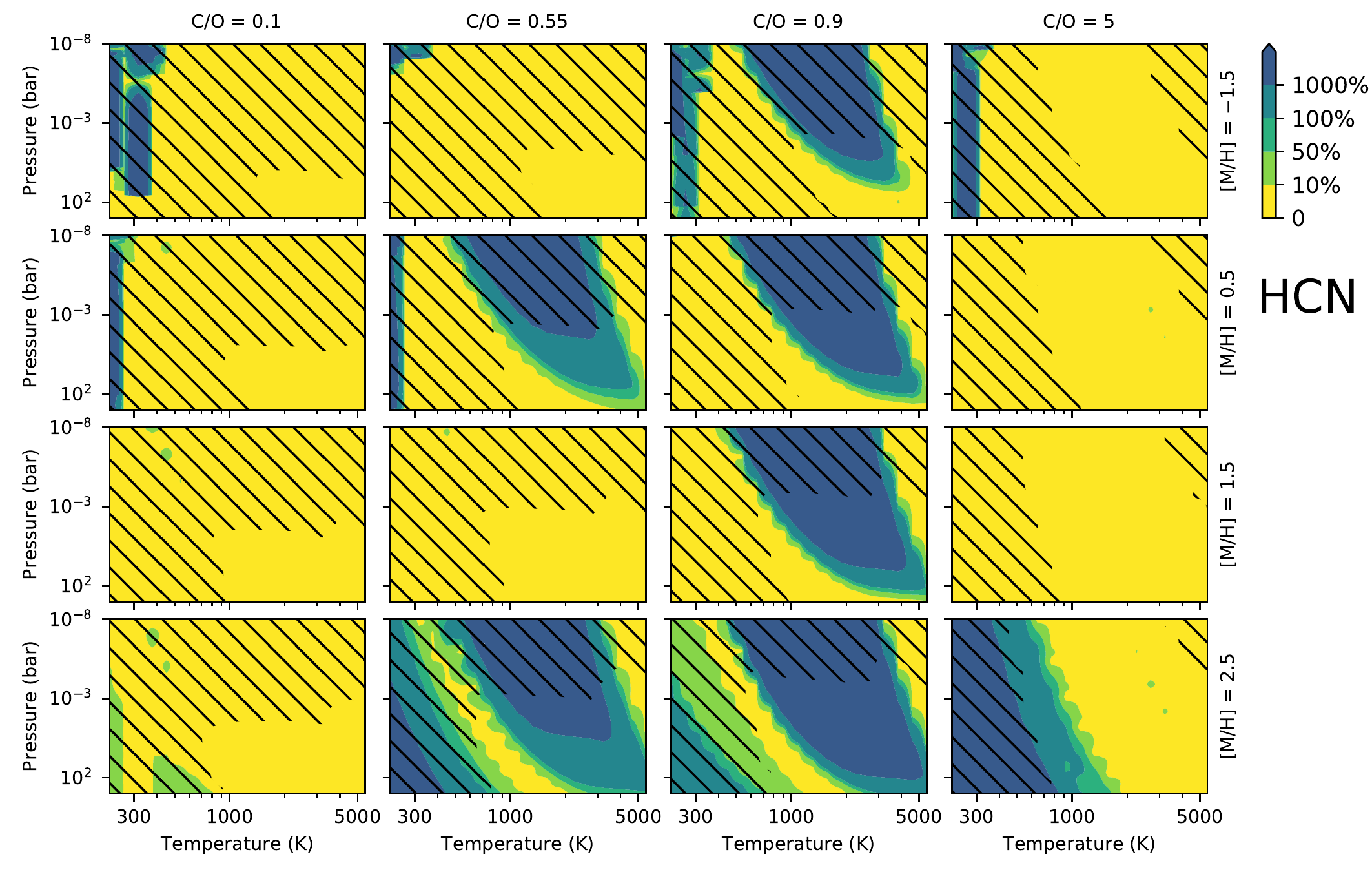}\\
\includegraphics[width=0.49\textwidth, clip]{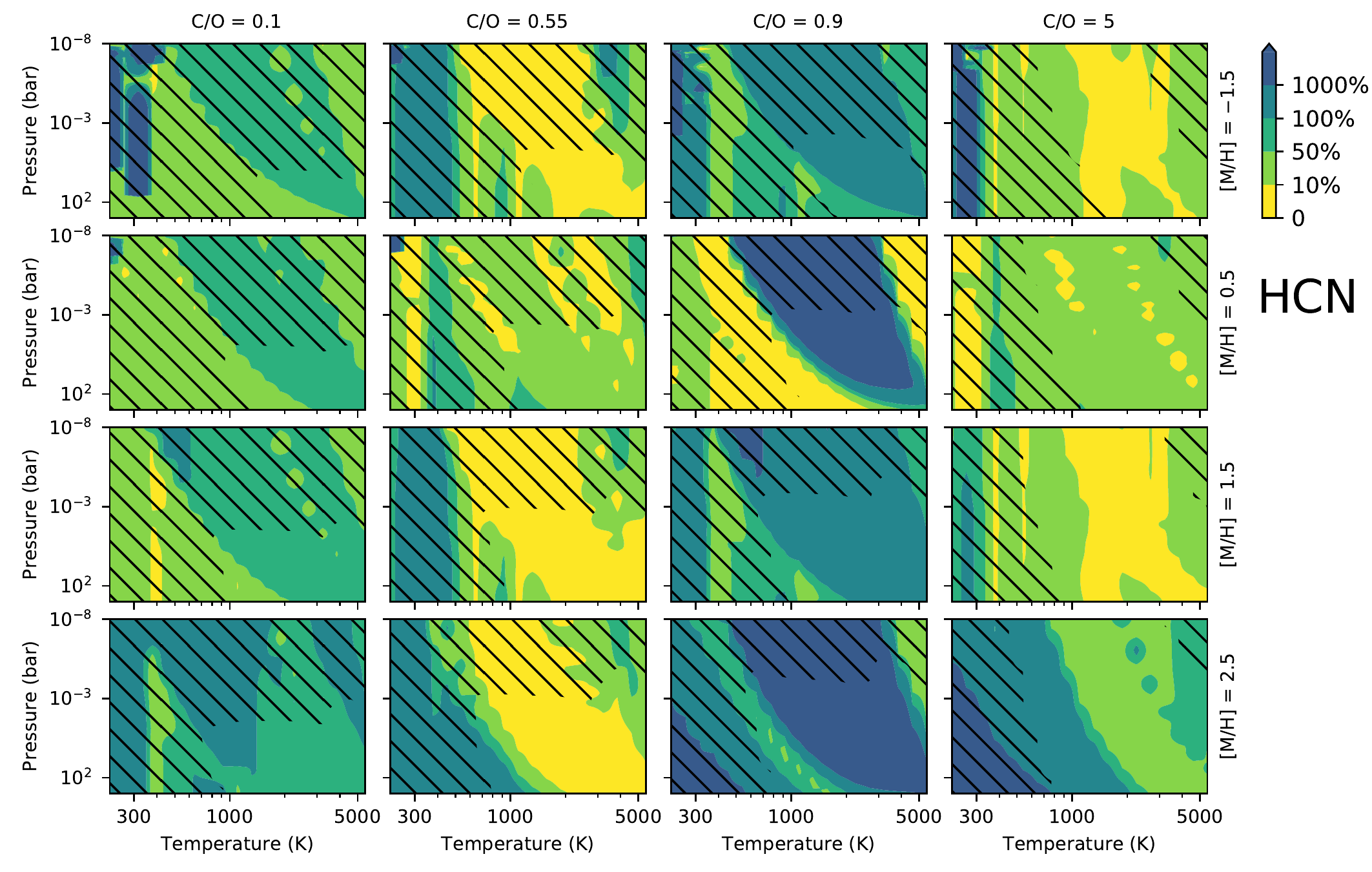}\hfill
\includegraphics[width=0.49\textwidth, clip]{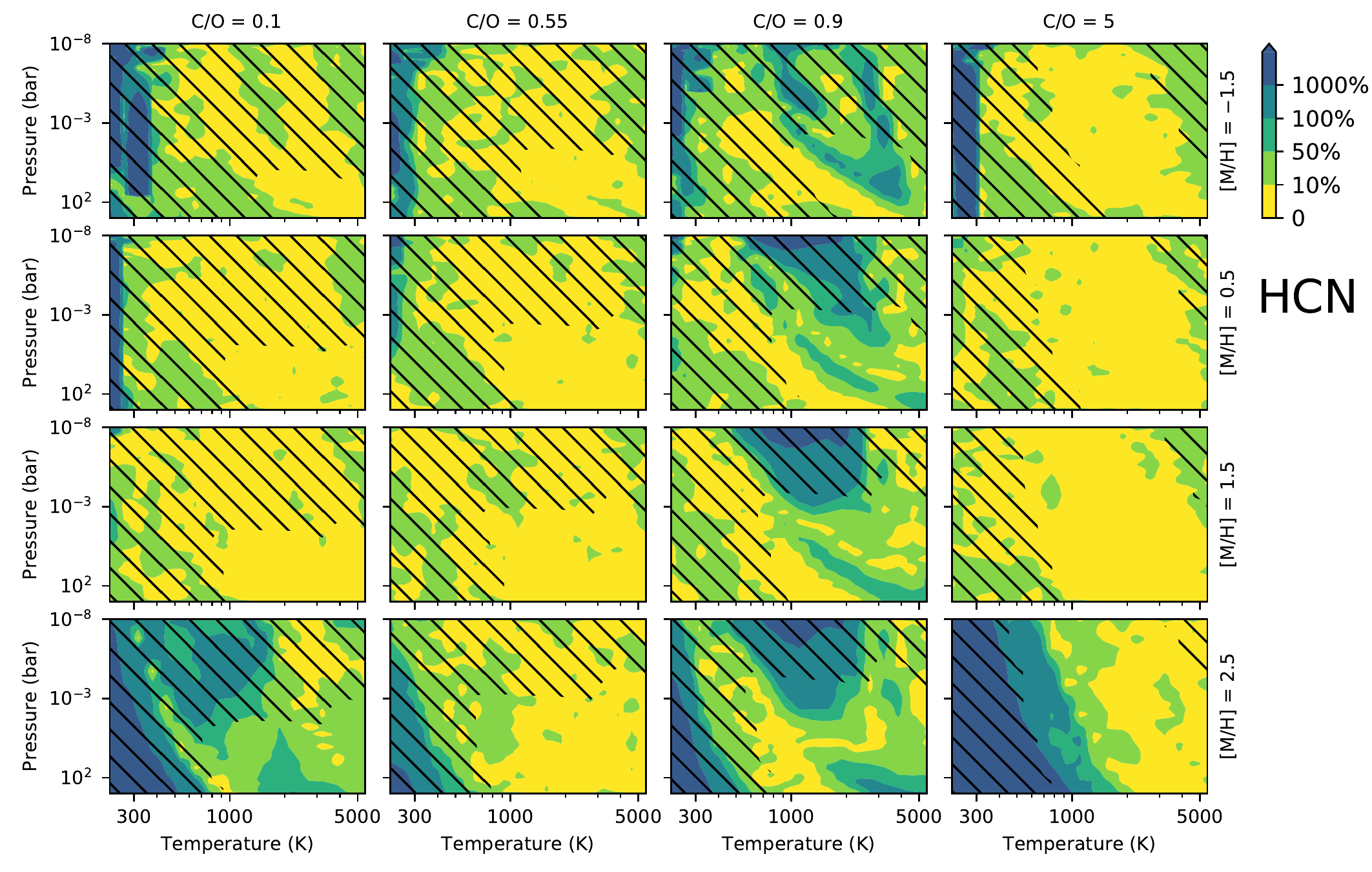}\\
\includegraphics[width=0.49\textwidth, clip]{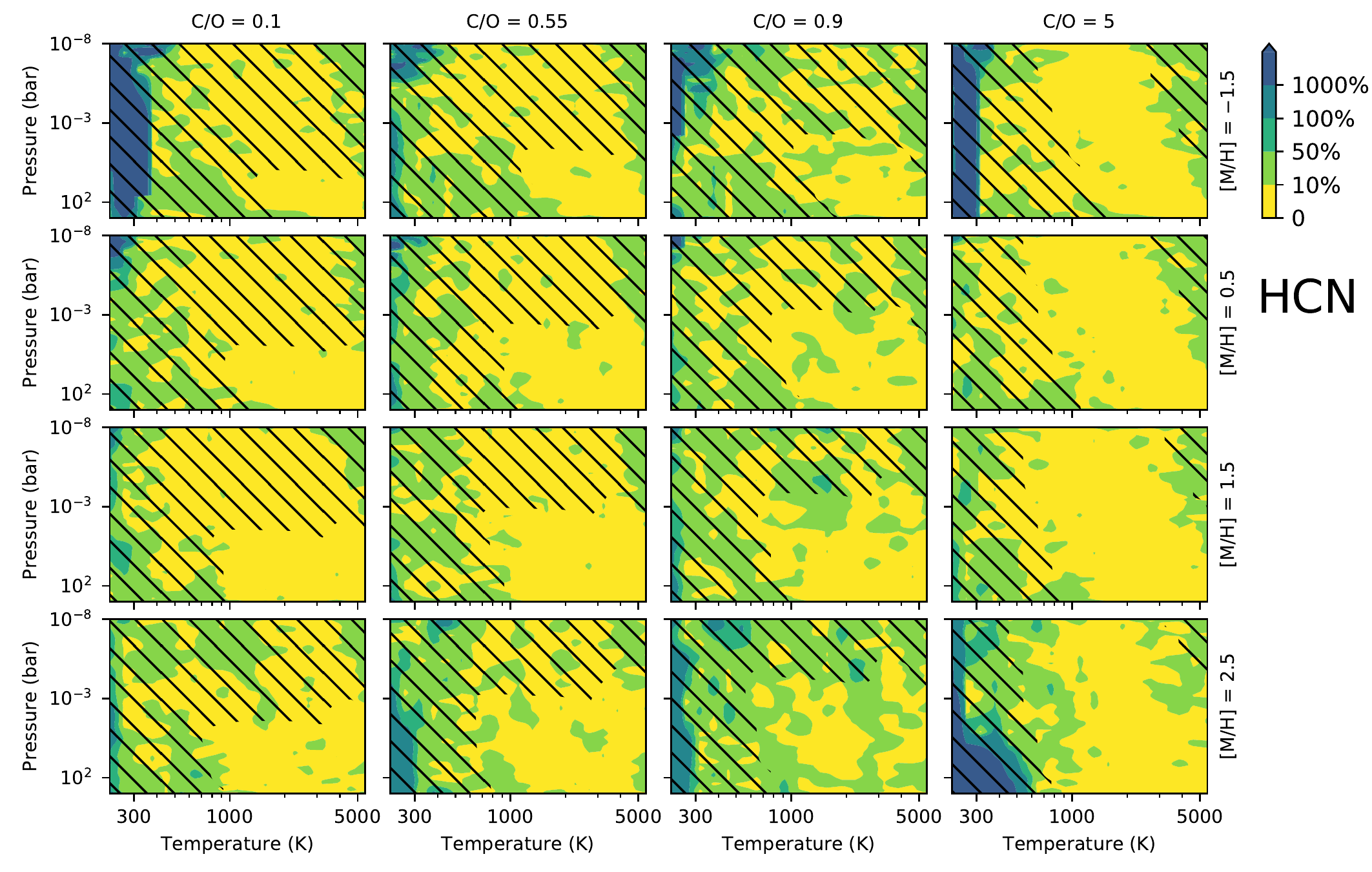}\hfill
\includegraphics[width=0.49\textwidth, clip]{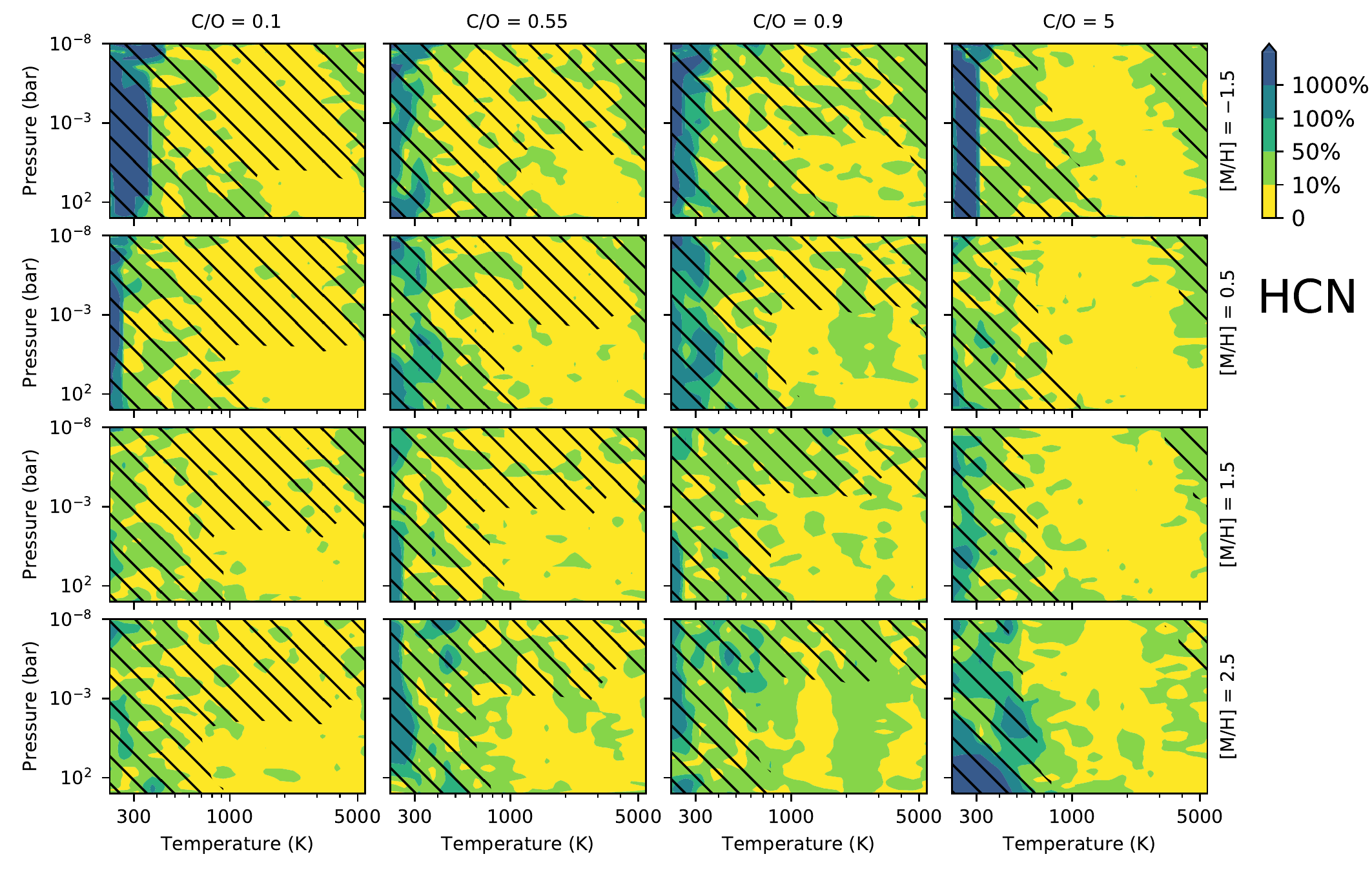}\\
\caption{As in Figure \ref{fig:H2O}, but for HCN.
}
\label{fig:HCN}
\end{figure*}

\begin{figure*}[htb]
\centering
\includegraphics[width=0.49\textwidth, clip]{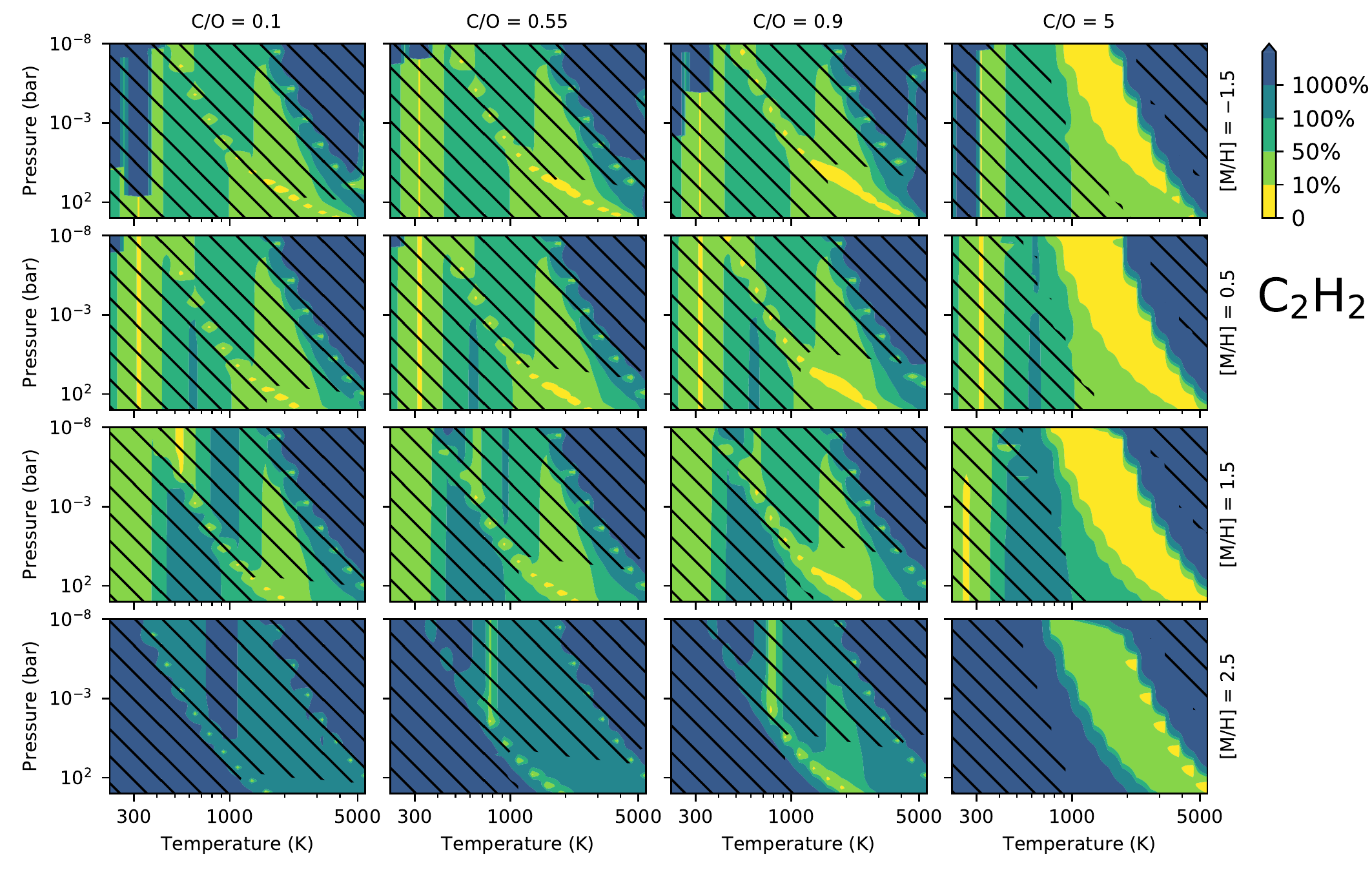}\hfill
\includegraphics[width=0.49\textwidth, clip]{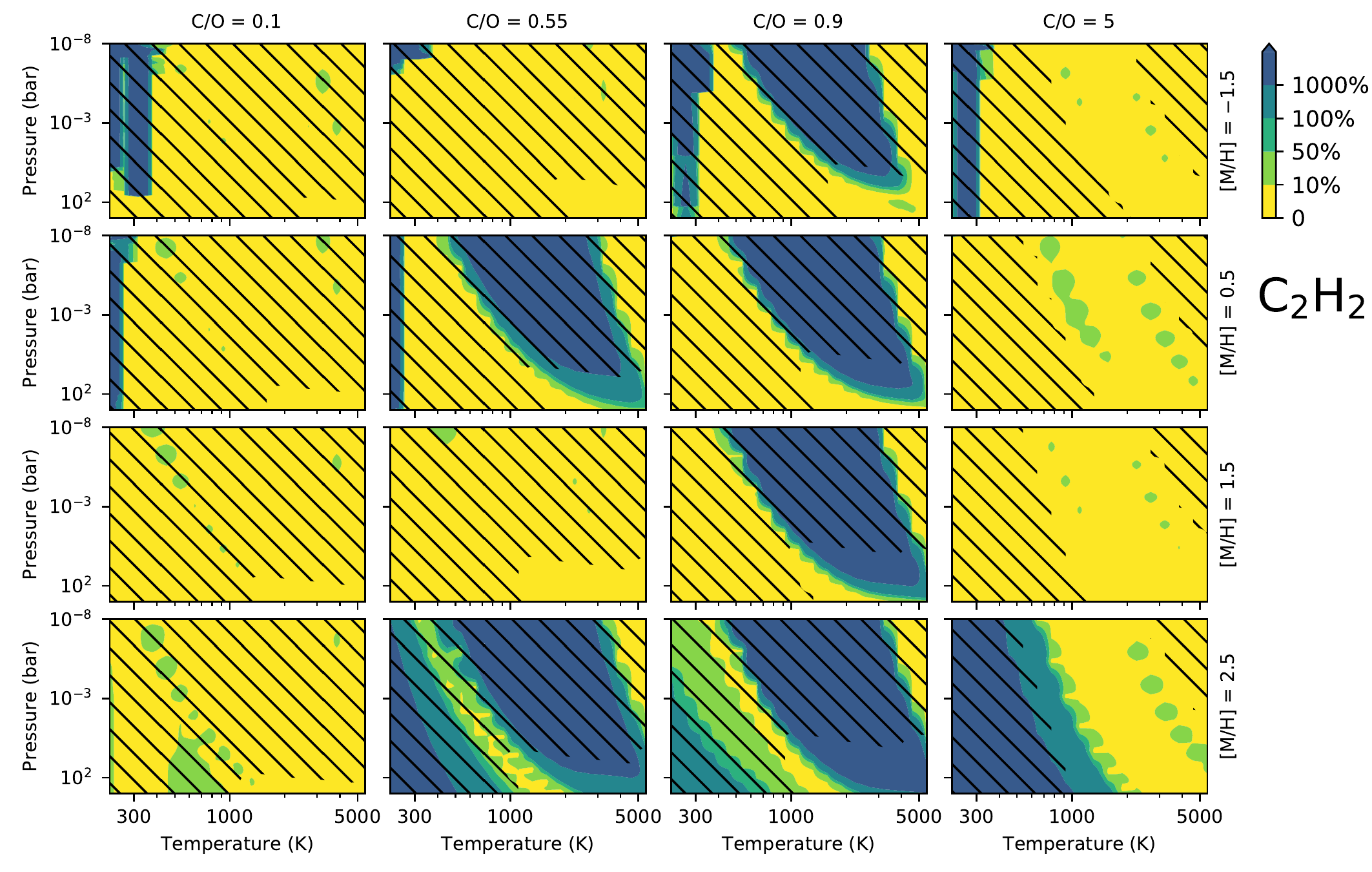}\\
\includegraphics[width=0.49\textwidth, clip]{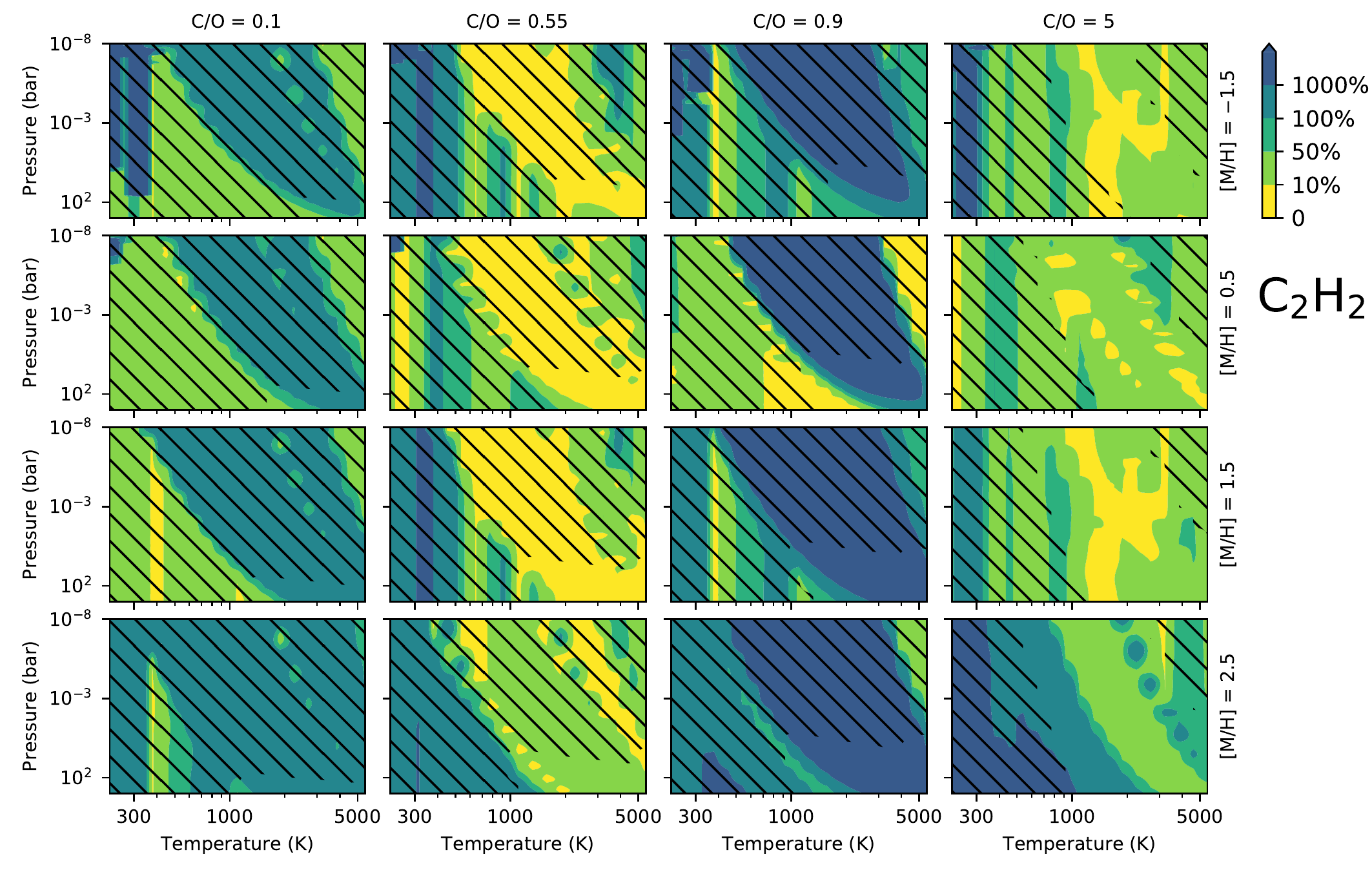}\hfill
\includegraphics[width=0.49\textwidth, clip]{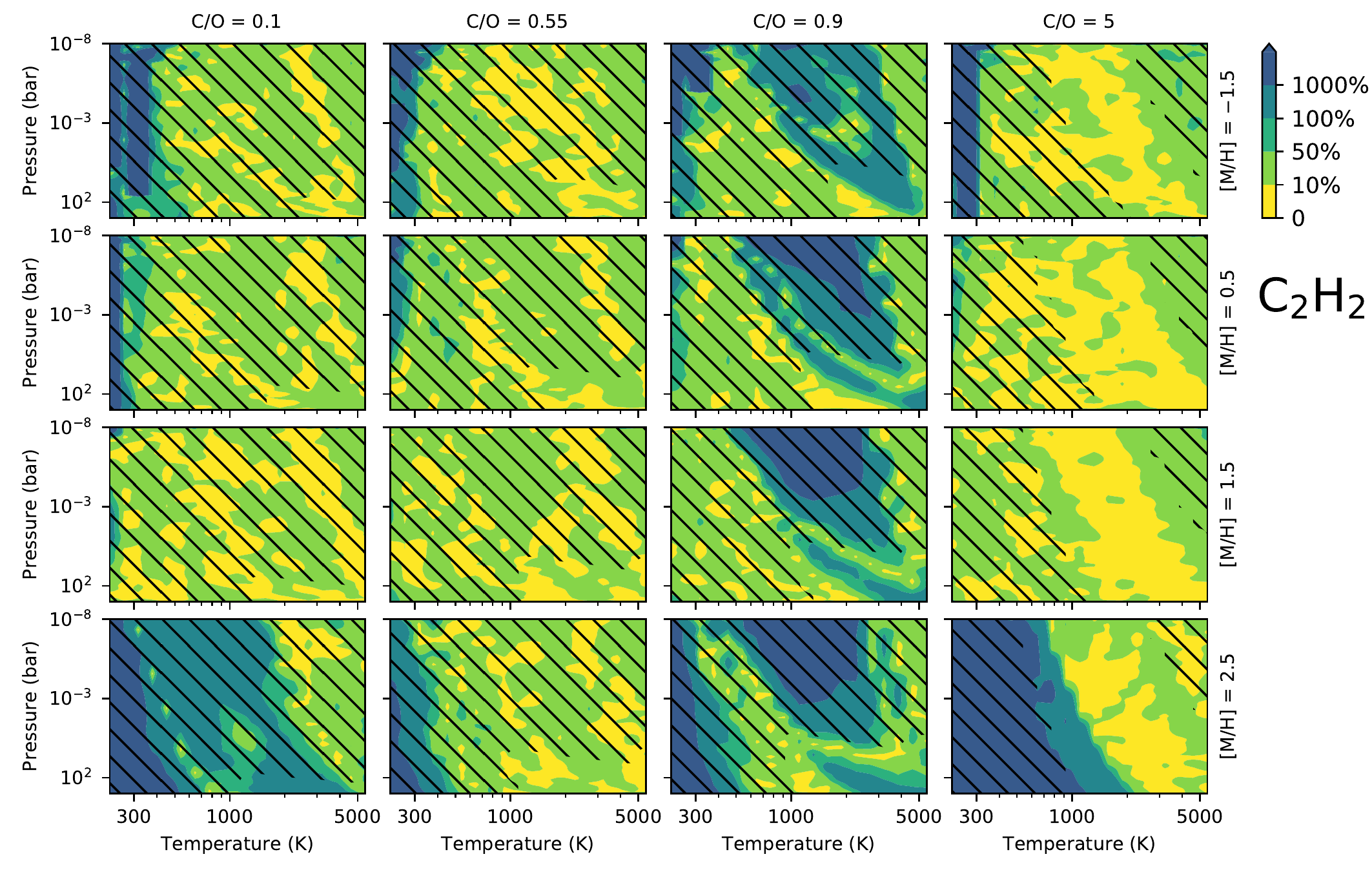}\\
\includegraphics[width=0.49\textwidth, clip]{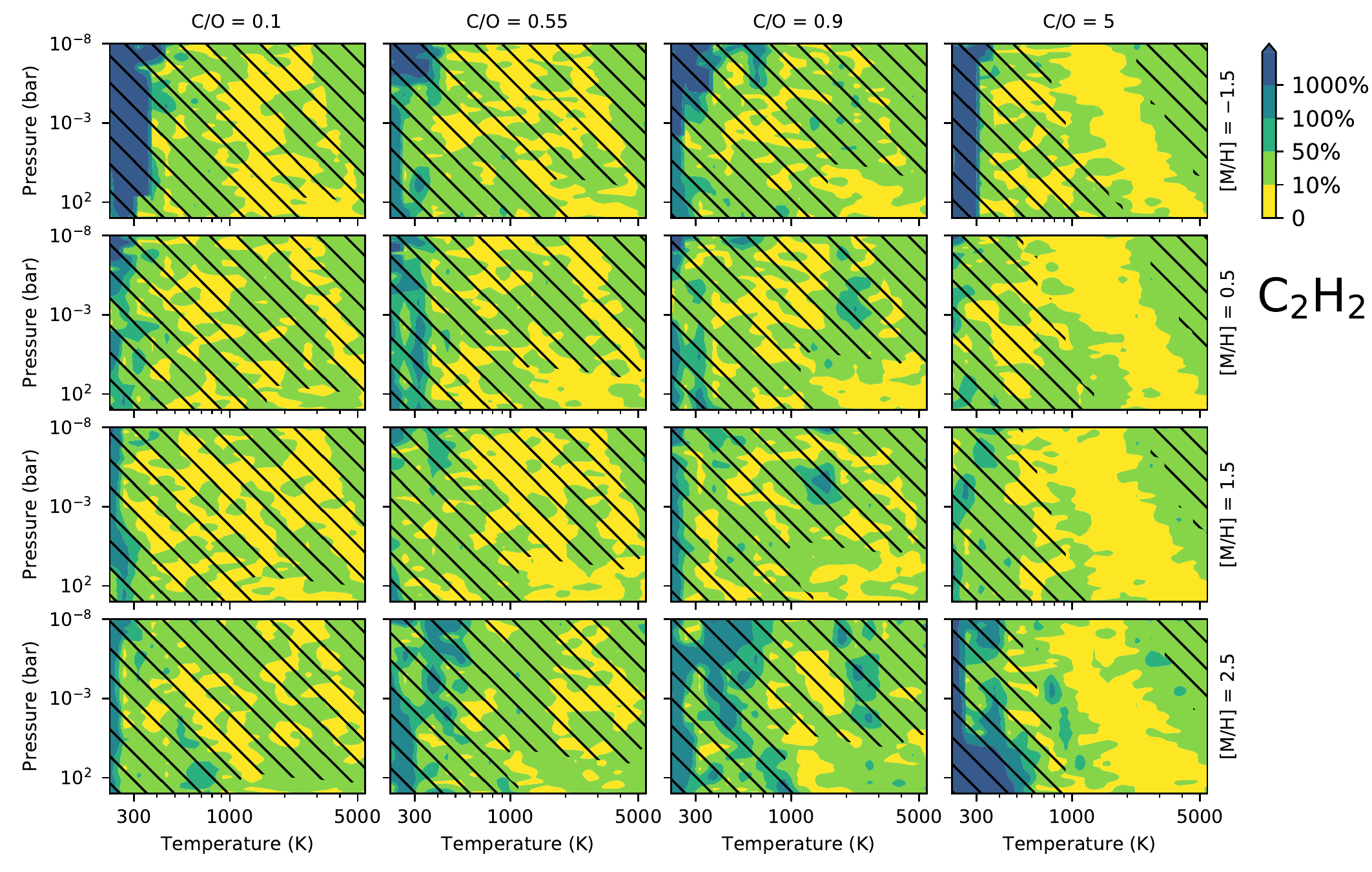}\hfill
\includegraphics[width=0.49\textwidth, clip]{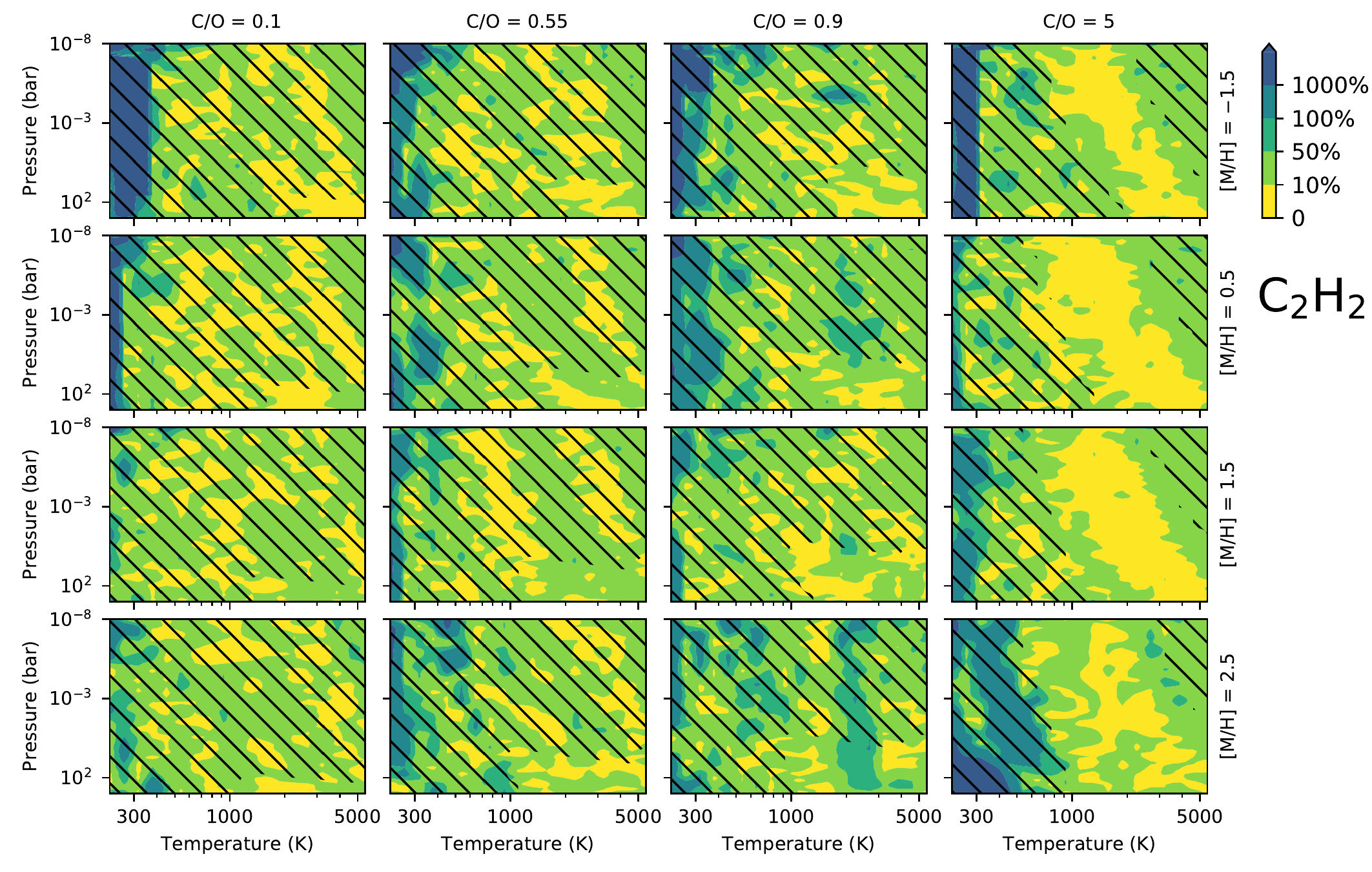}\\
\caption{As in Figure \ref{fig:H2O}, but for C$_2$H$_2$.
}
\label{fig:C2H2}
\end{figure*}

\begin{figure*}[htb]
\centering
\includegraphics[width=0.49\textwidth, clip]{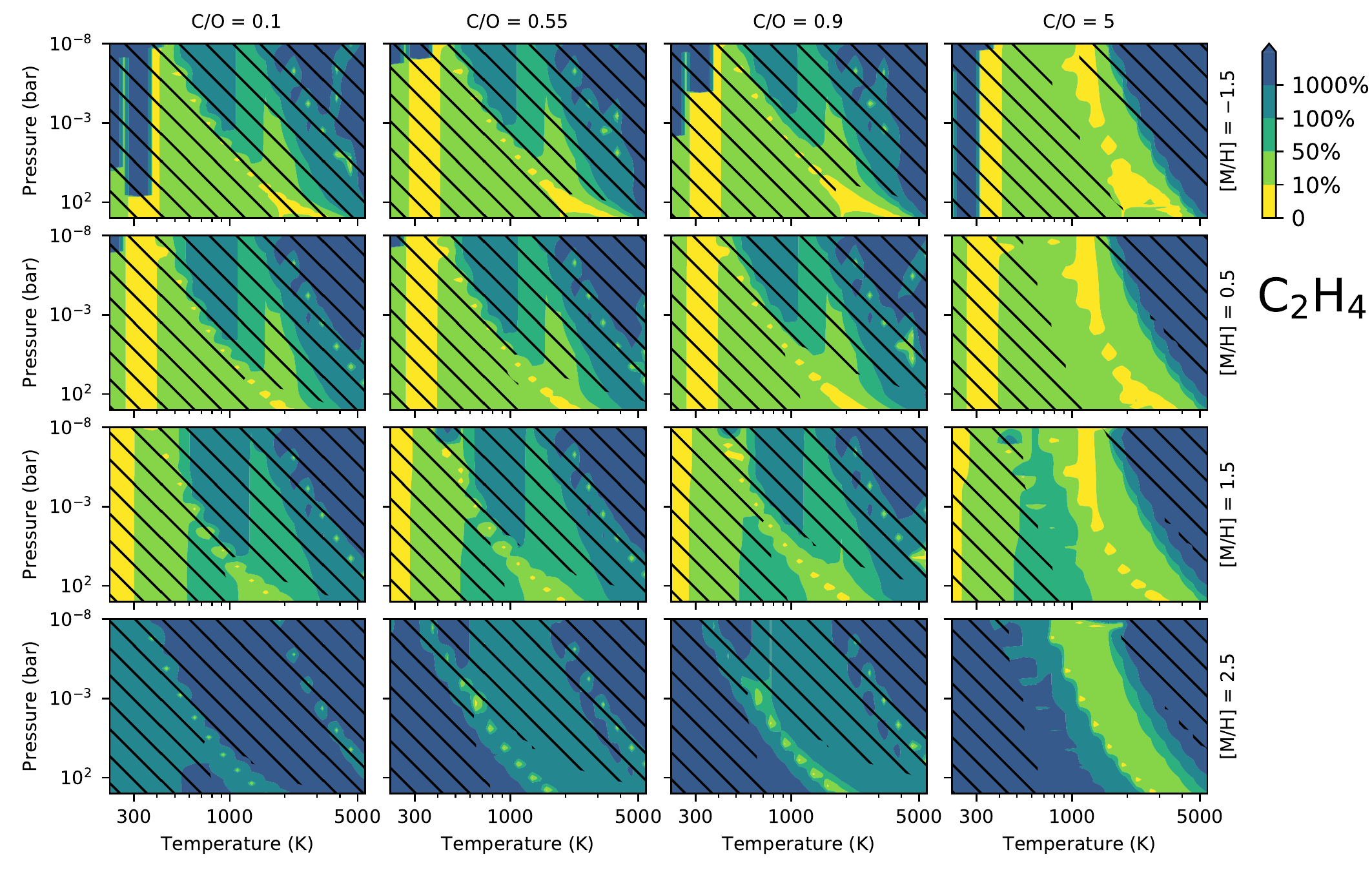}\hfill
\includegraphics[width=0.49\textwidth, clip]{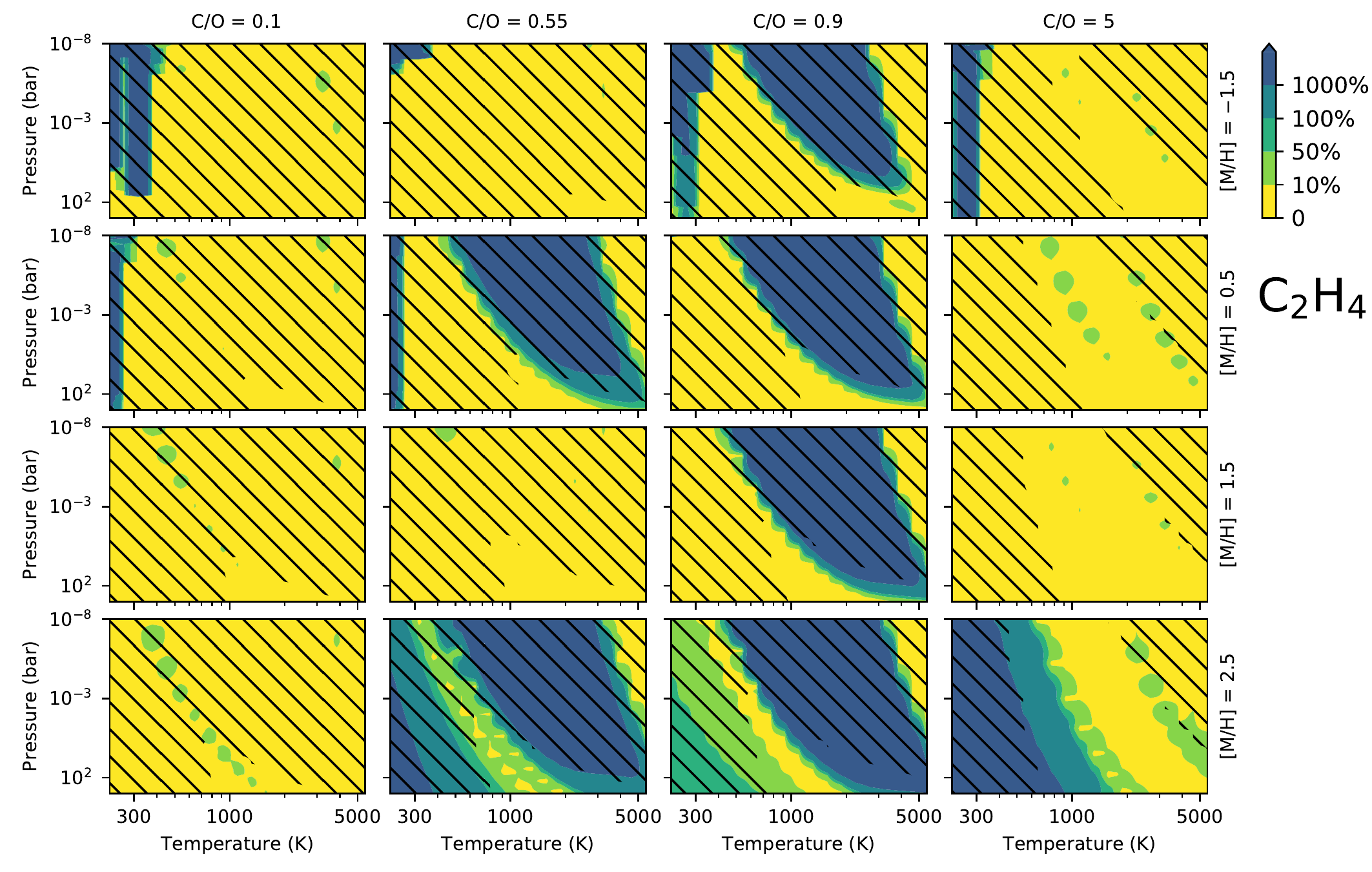}\\
\includegraphics[width=0.49\textwidth, clip]{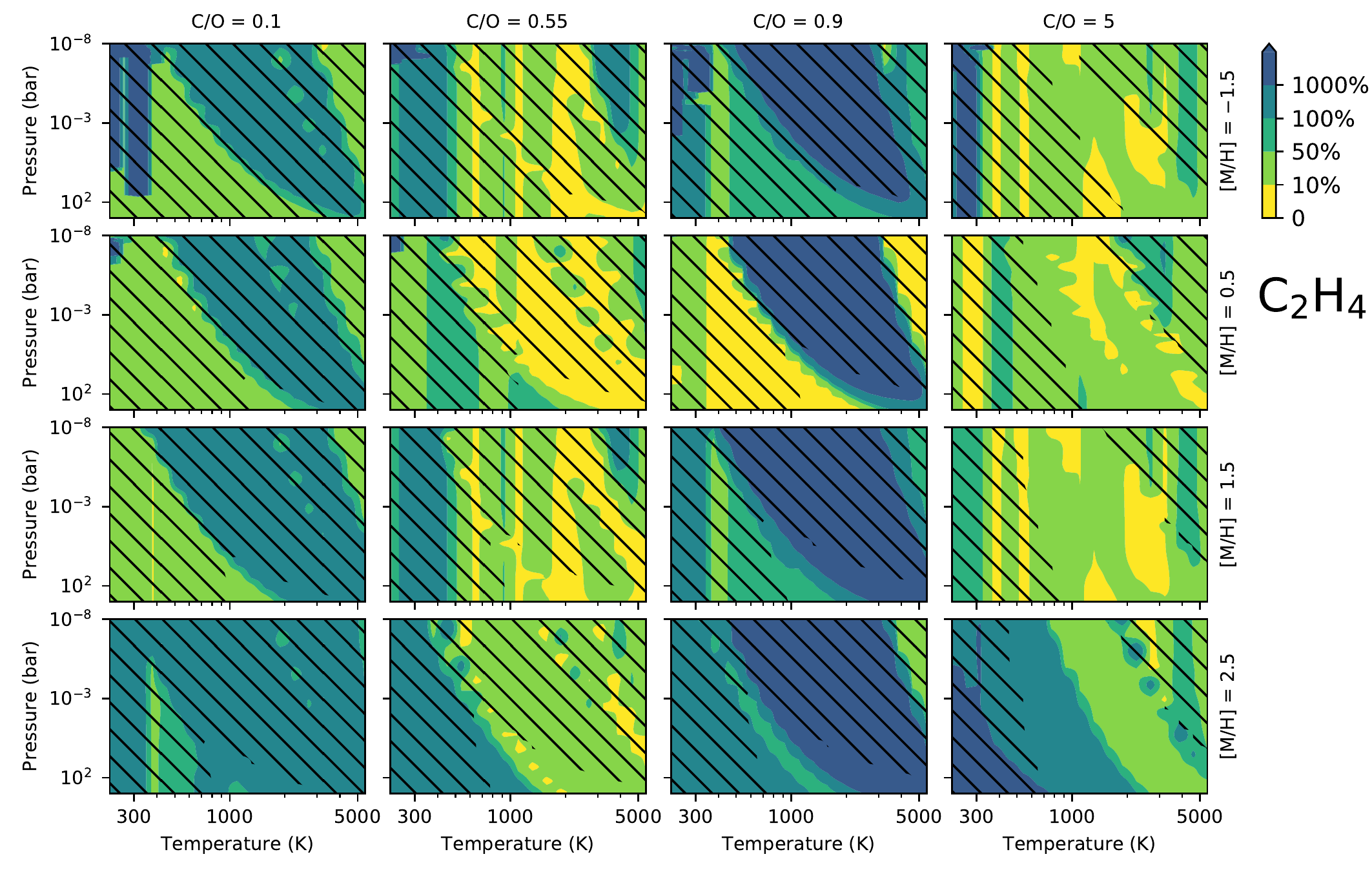}\hfill
\includegraphics[width=0.49\textwidth, clip]{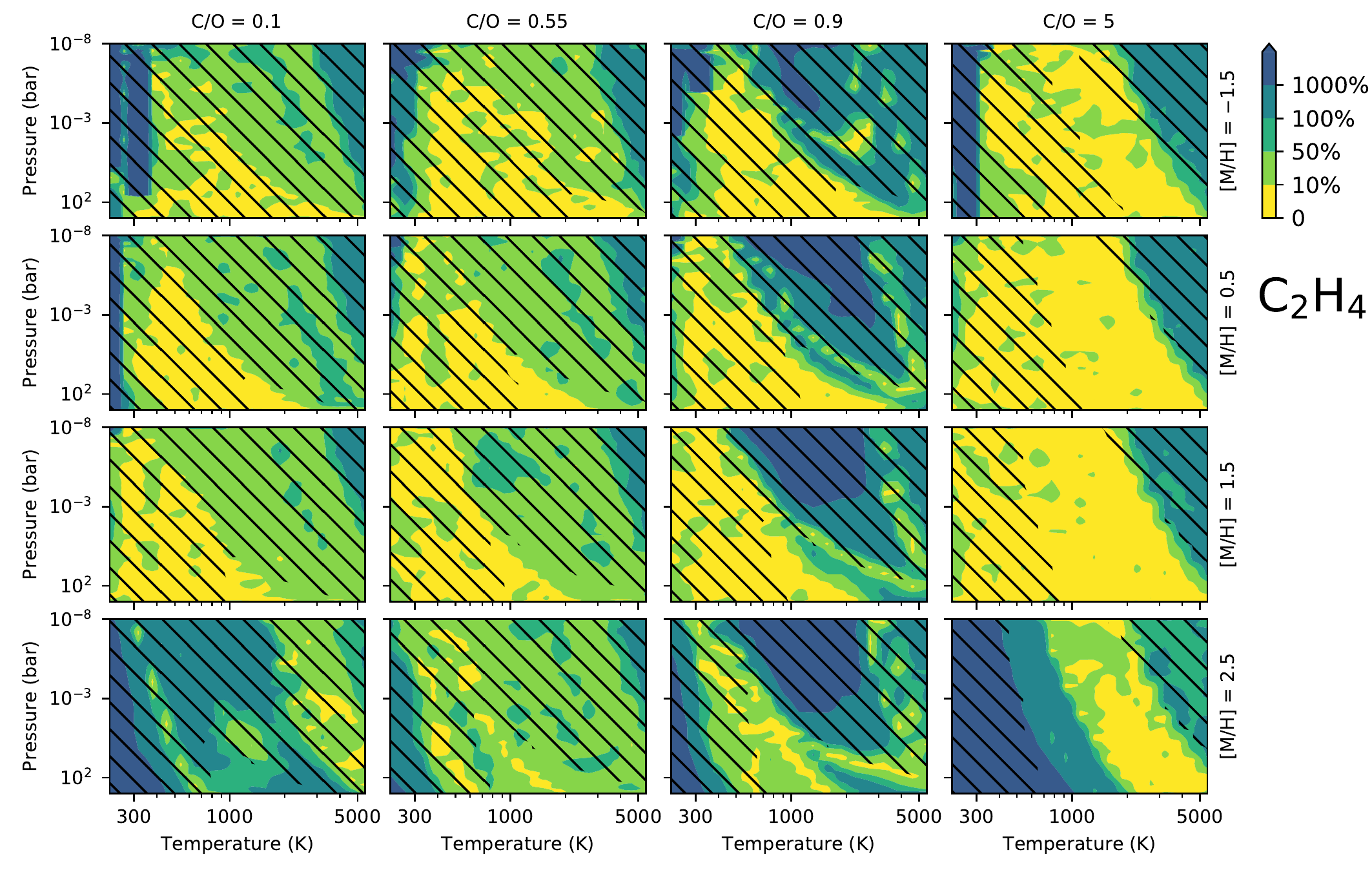}\\
\includegraphics[width=0.49\textwidth, clip]{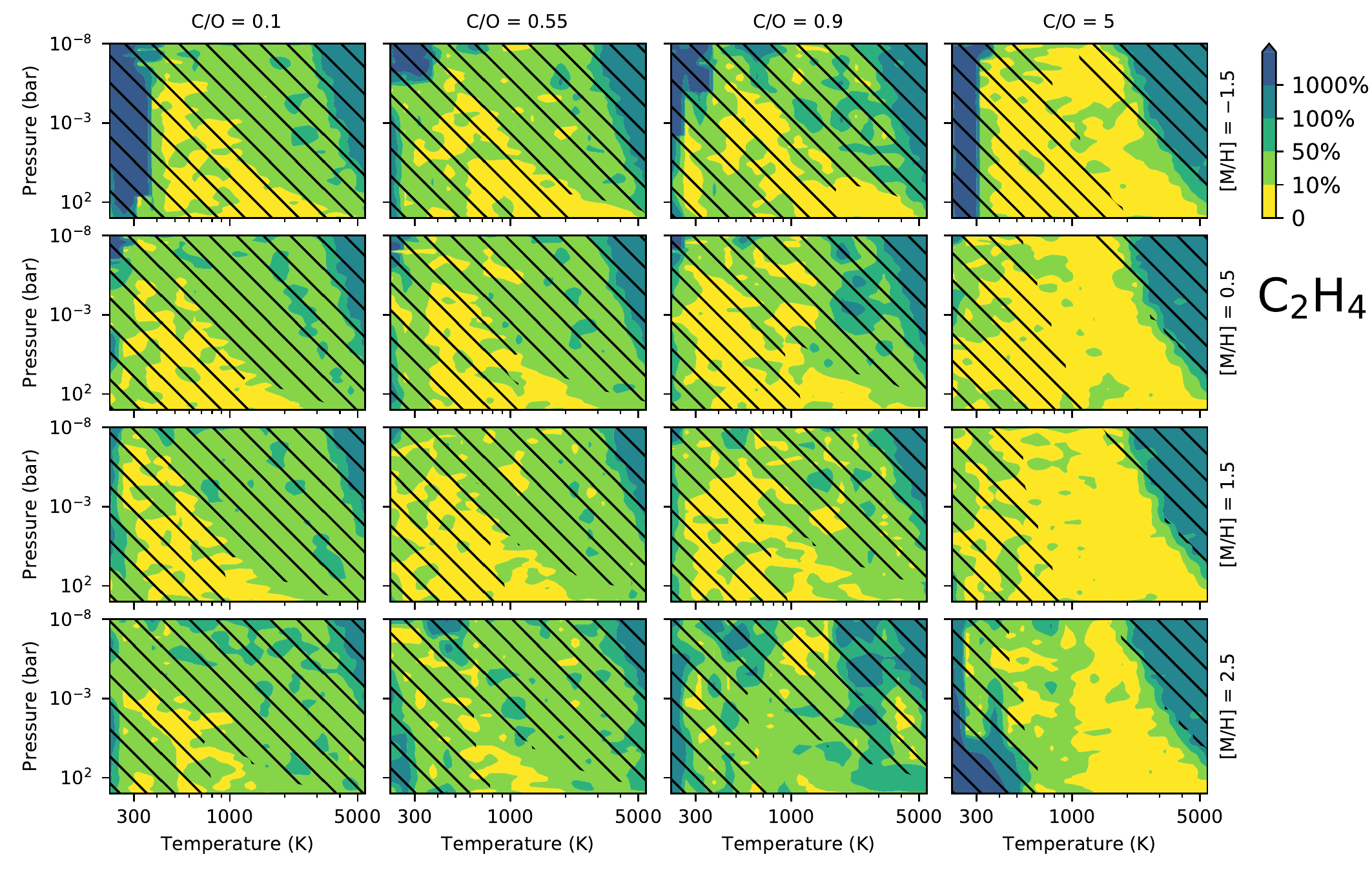}\hfill
\includegraphics[width=0.49\textwidth, clip]{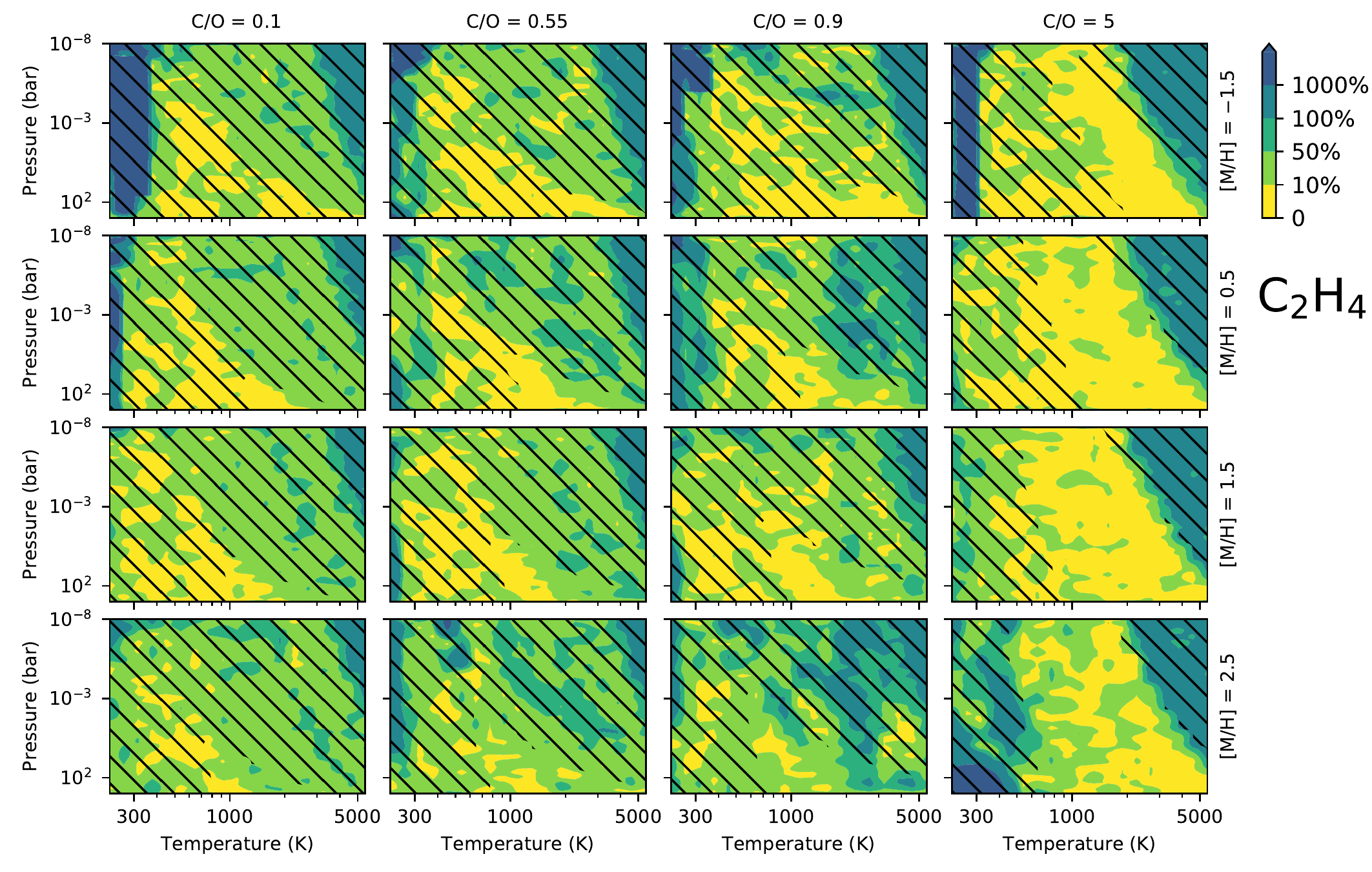}\\
\caption{As in Figure \ref{fig:H2O}, but for C$_2$H$_4$.
}
\label{fig:C2H4}
\end{figure*}

\begin{figure*}[htb]
\centering
\includegraphics[width=0.49\textwidth, clip]{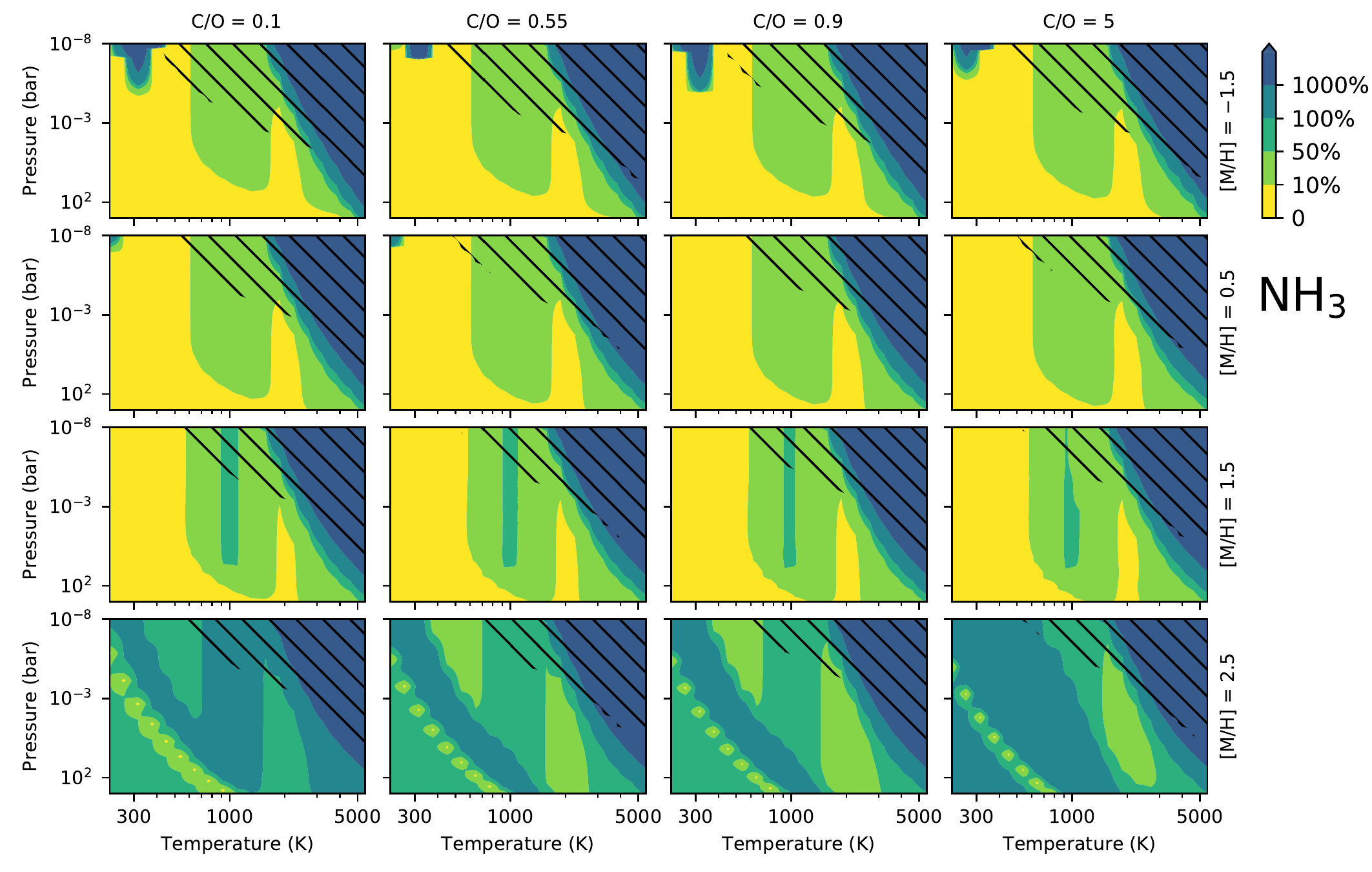}\hfill
\includegraphics[width=0.49\textwidth, clip]{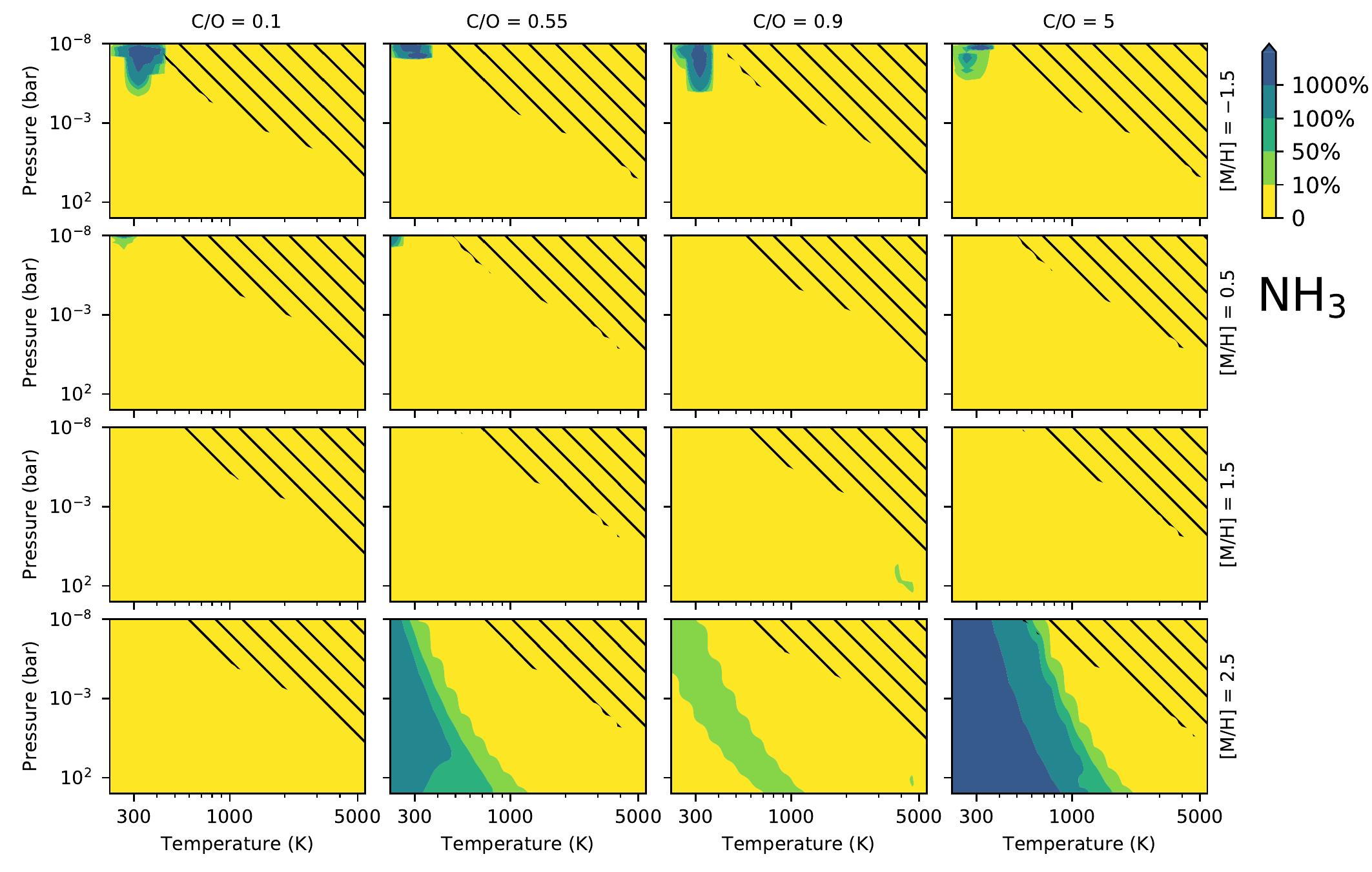}\\
\includegraphics[width=0.49\textwidth, clip]{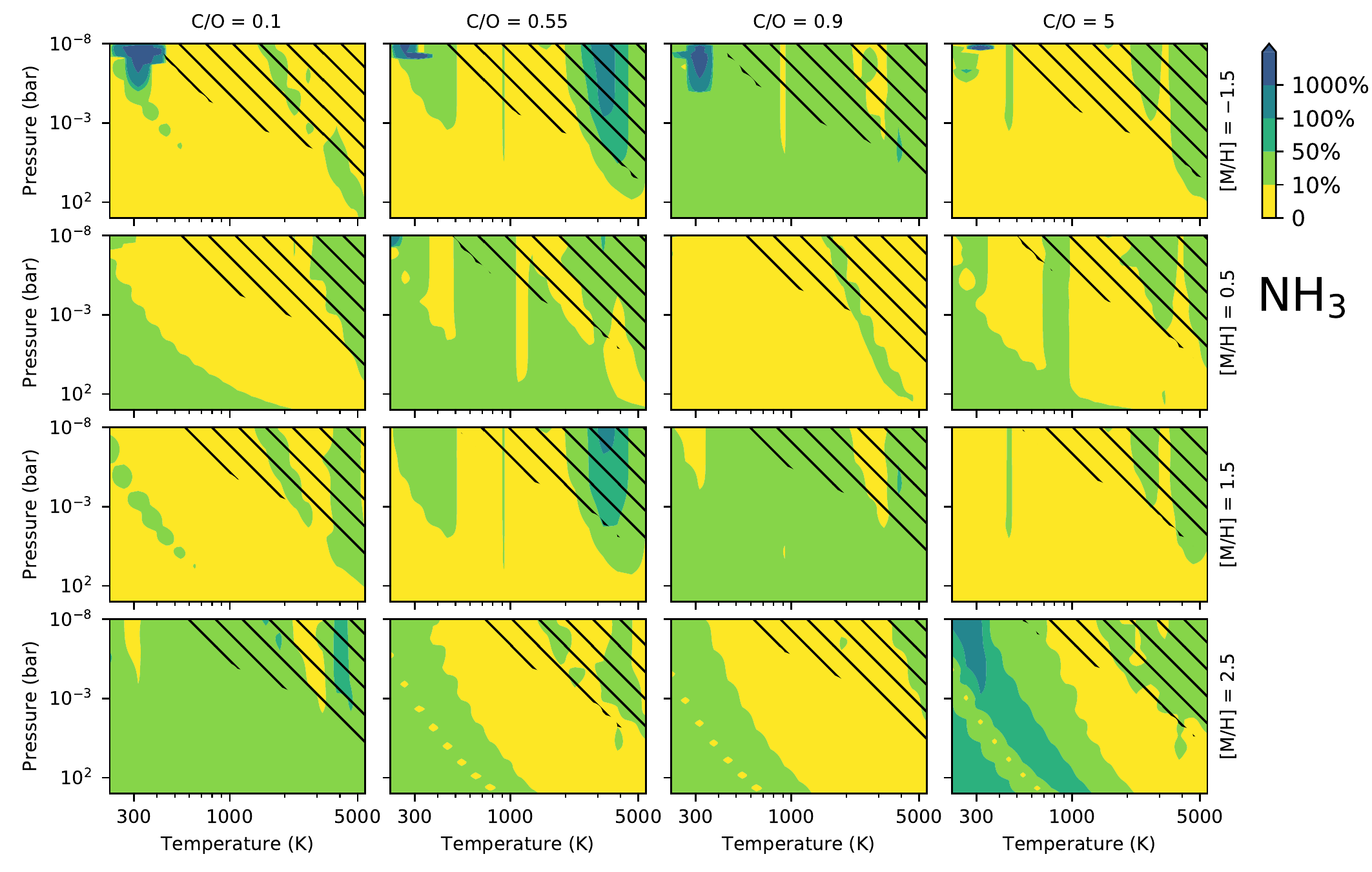}\hfill
\includegraphics[width=0.49\textwidth, clip]{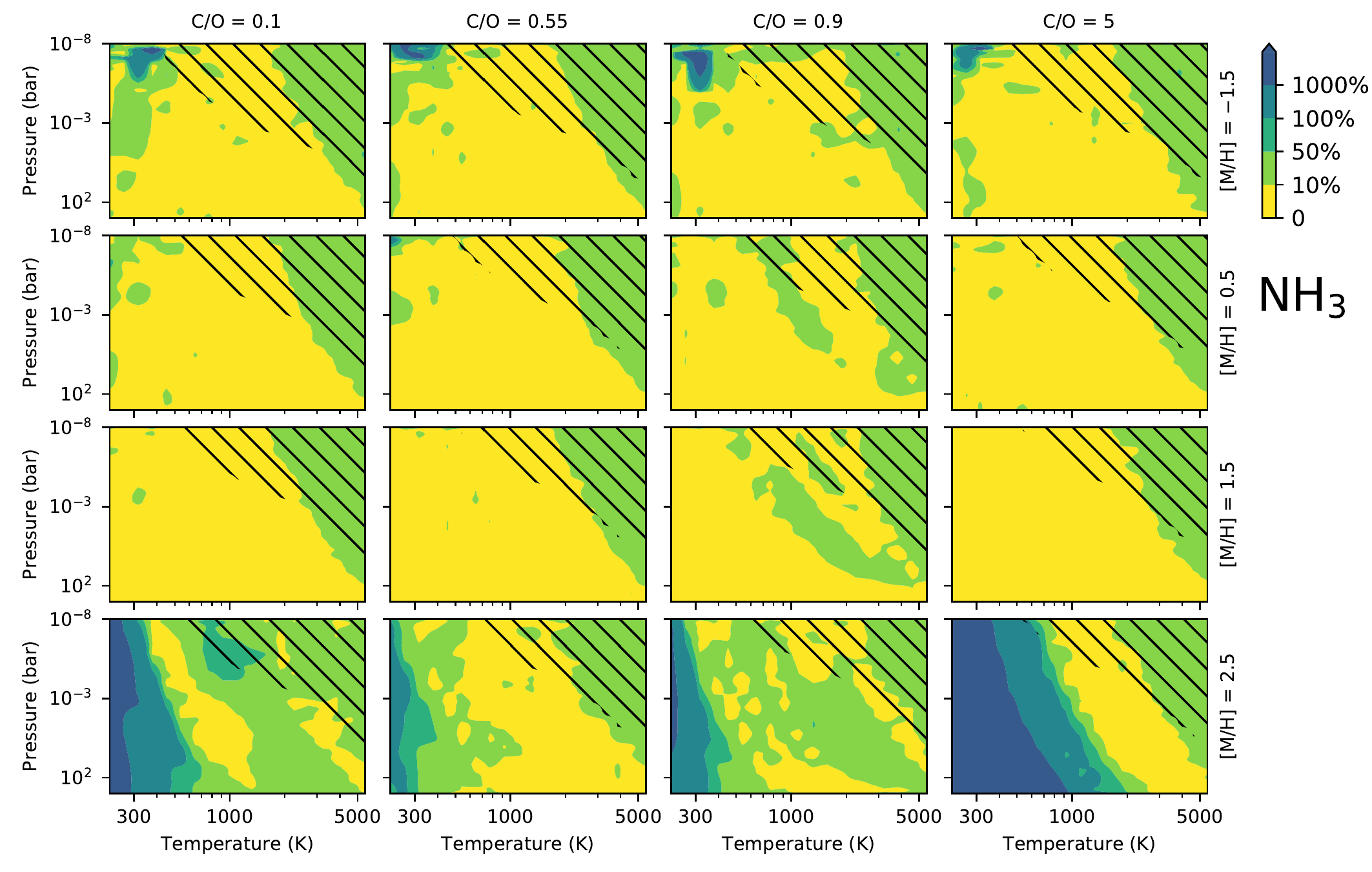}\\
\includegraphics[width=0.49\textwidth, clip]{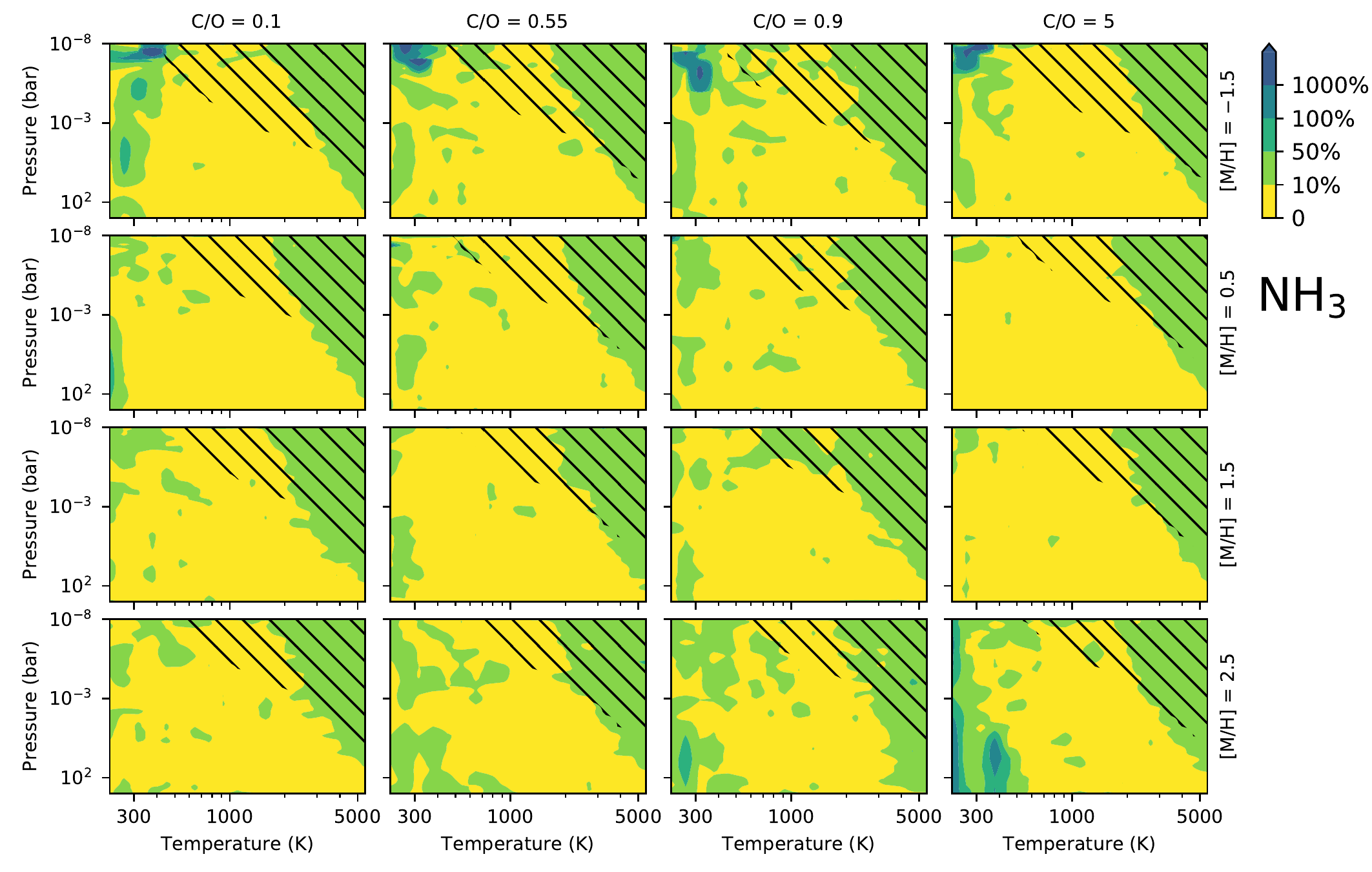}\hfill
\includegraphics[width=0.49\textwidth, clip]{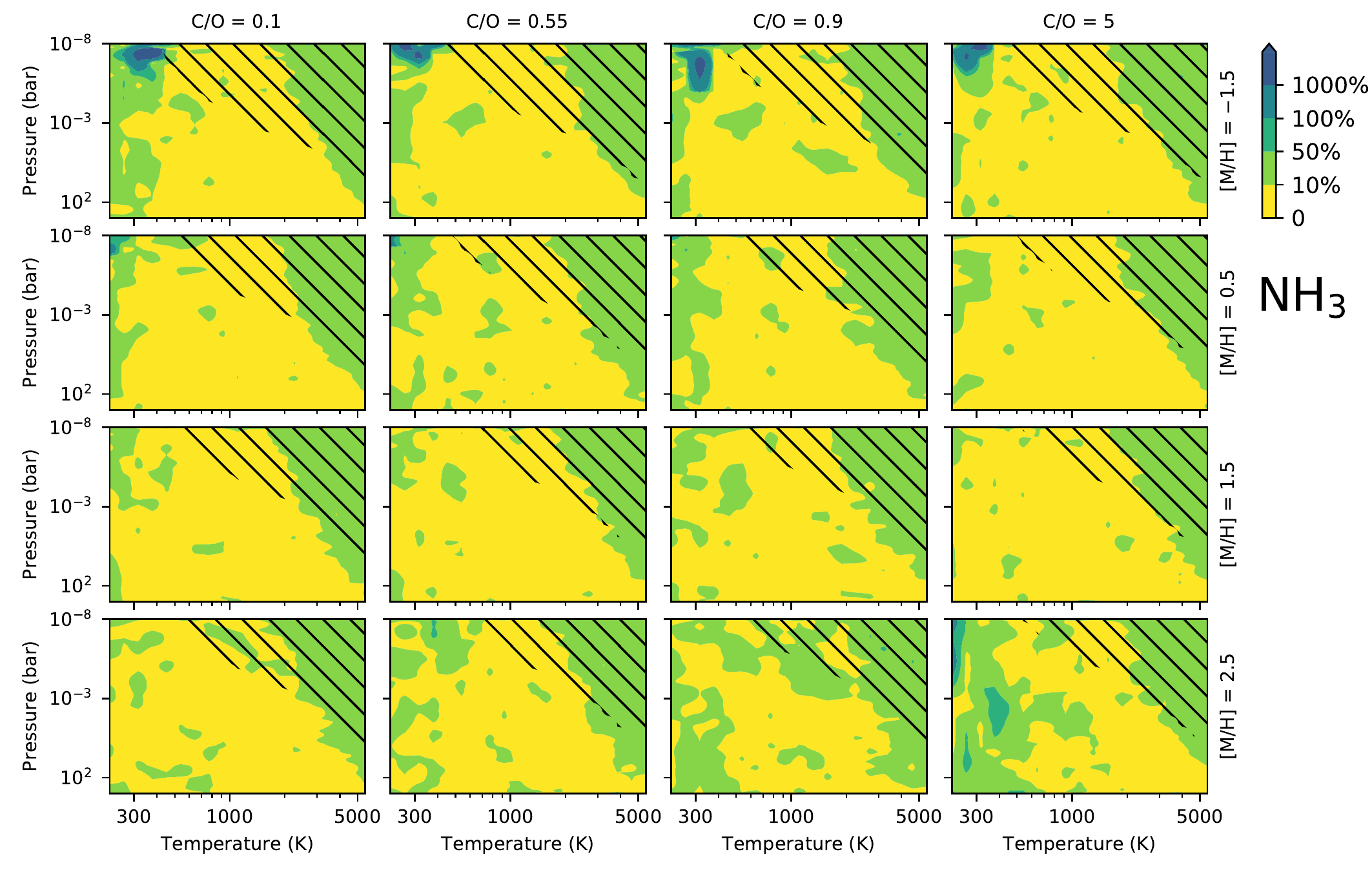}\\
\caption{As in Figure \ref{fig:H2O}, but for NH$_3$.
}
\label{fig:NH3}
\end{figure*}

\begin{figure*}[htb]
\centering
\includegraphics[width=0.49\textwidth, clip]{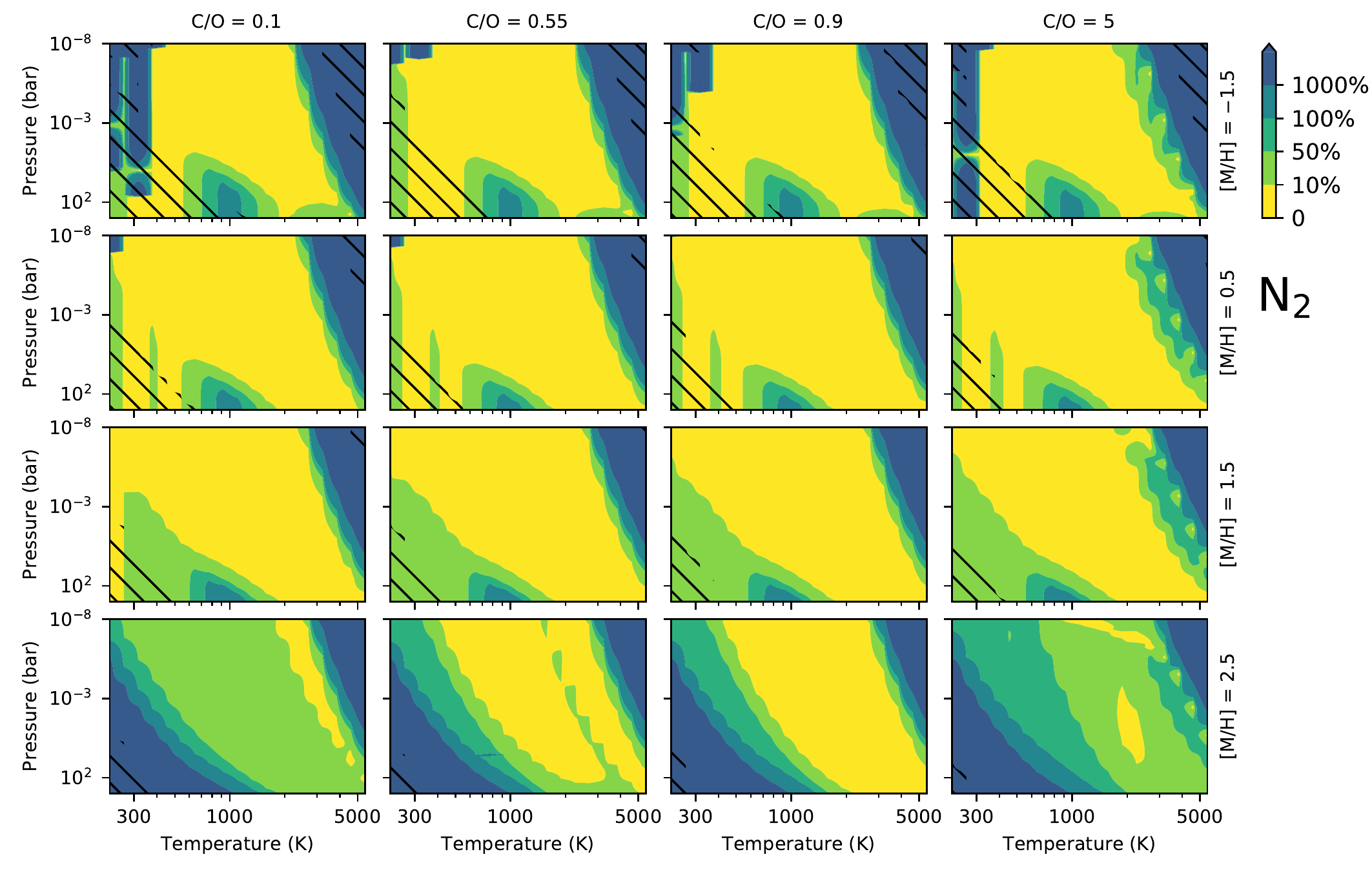}\hfill
\includegraphics[width=0.49\textwidth, clip]{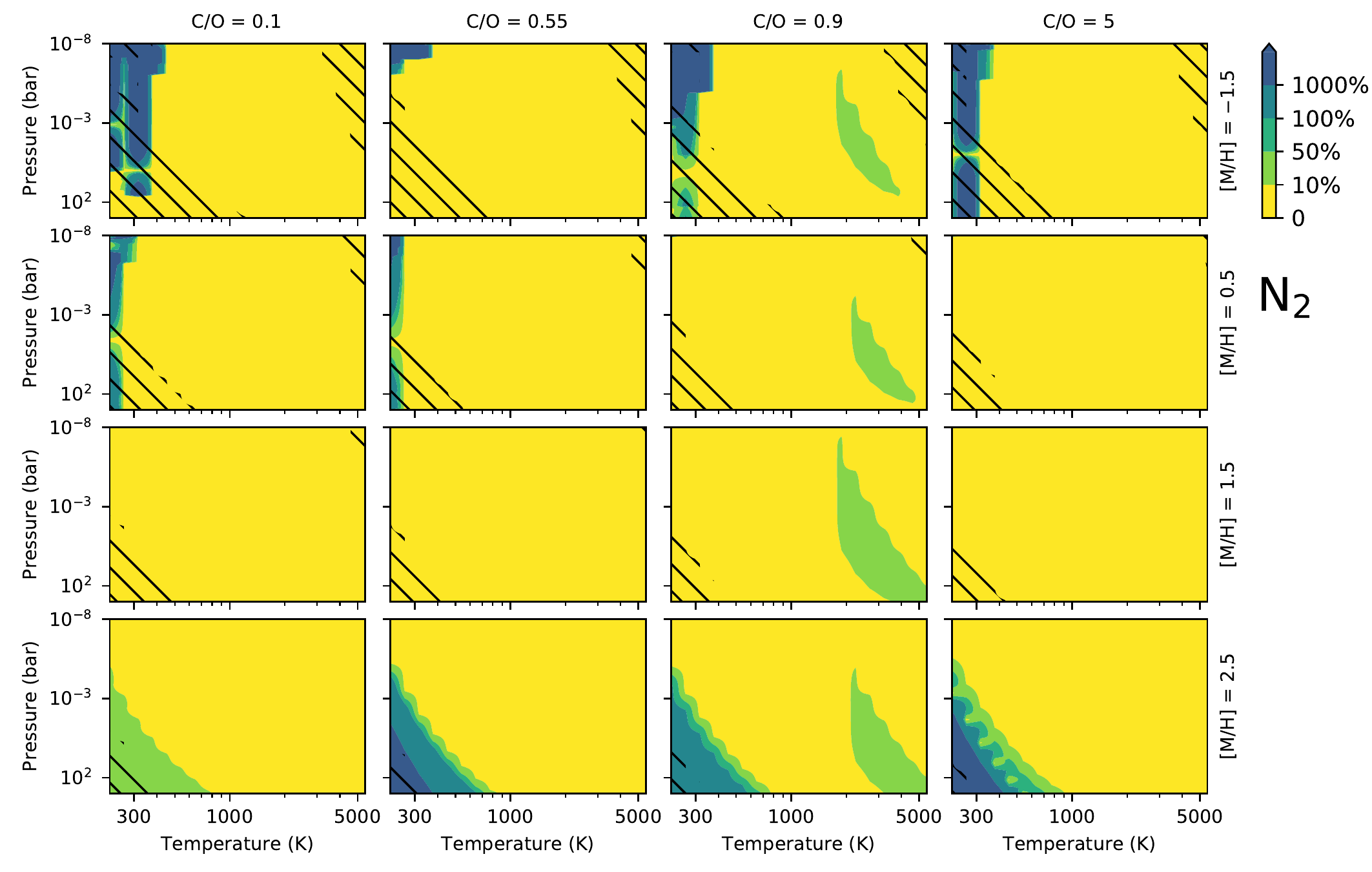}\\
\includegraphics[width=0.49\textwidth, clip]{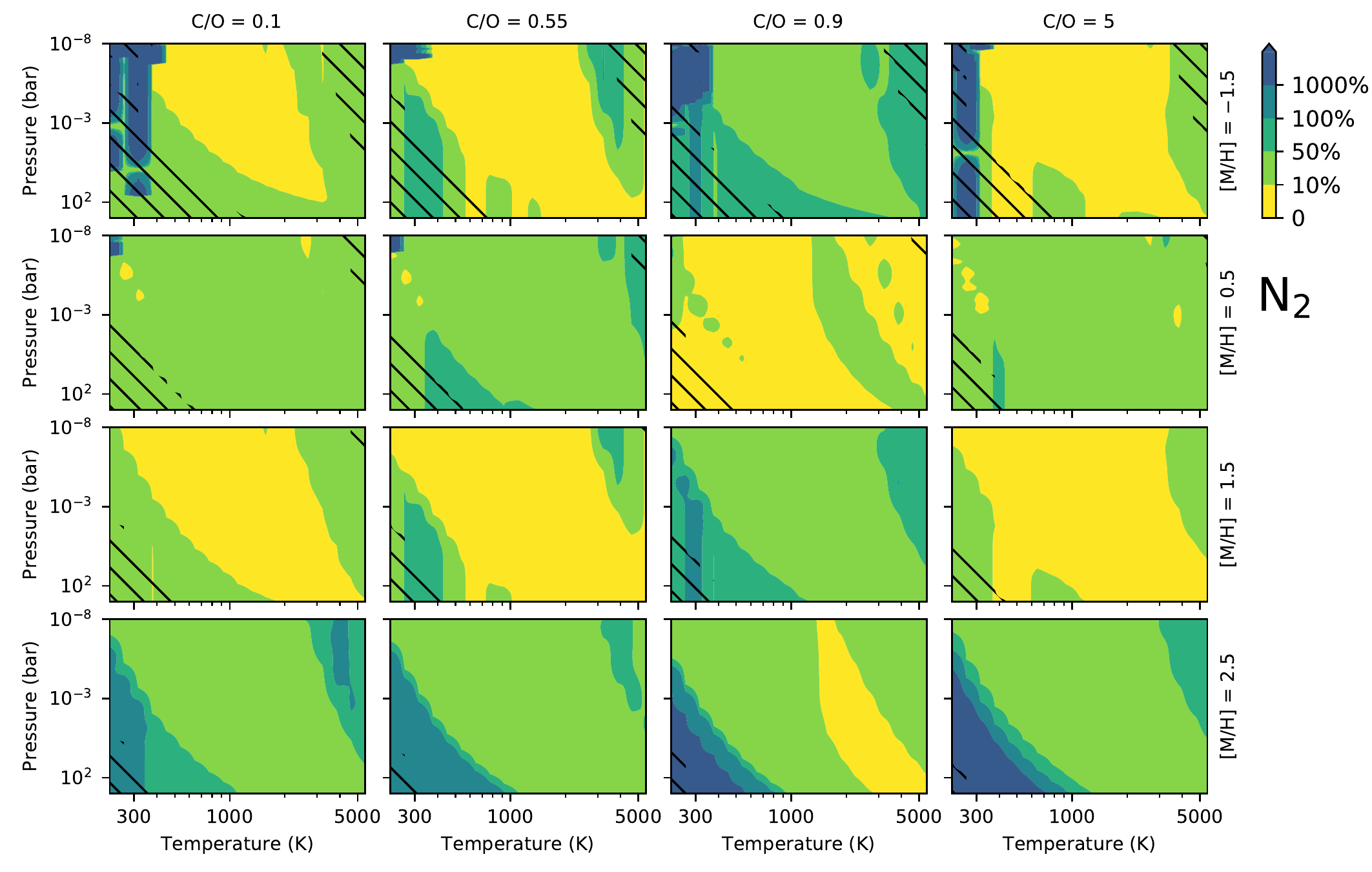}\hfill
\includegraphics[width=0.49\textwidth, clip]{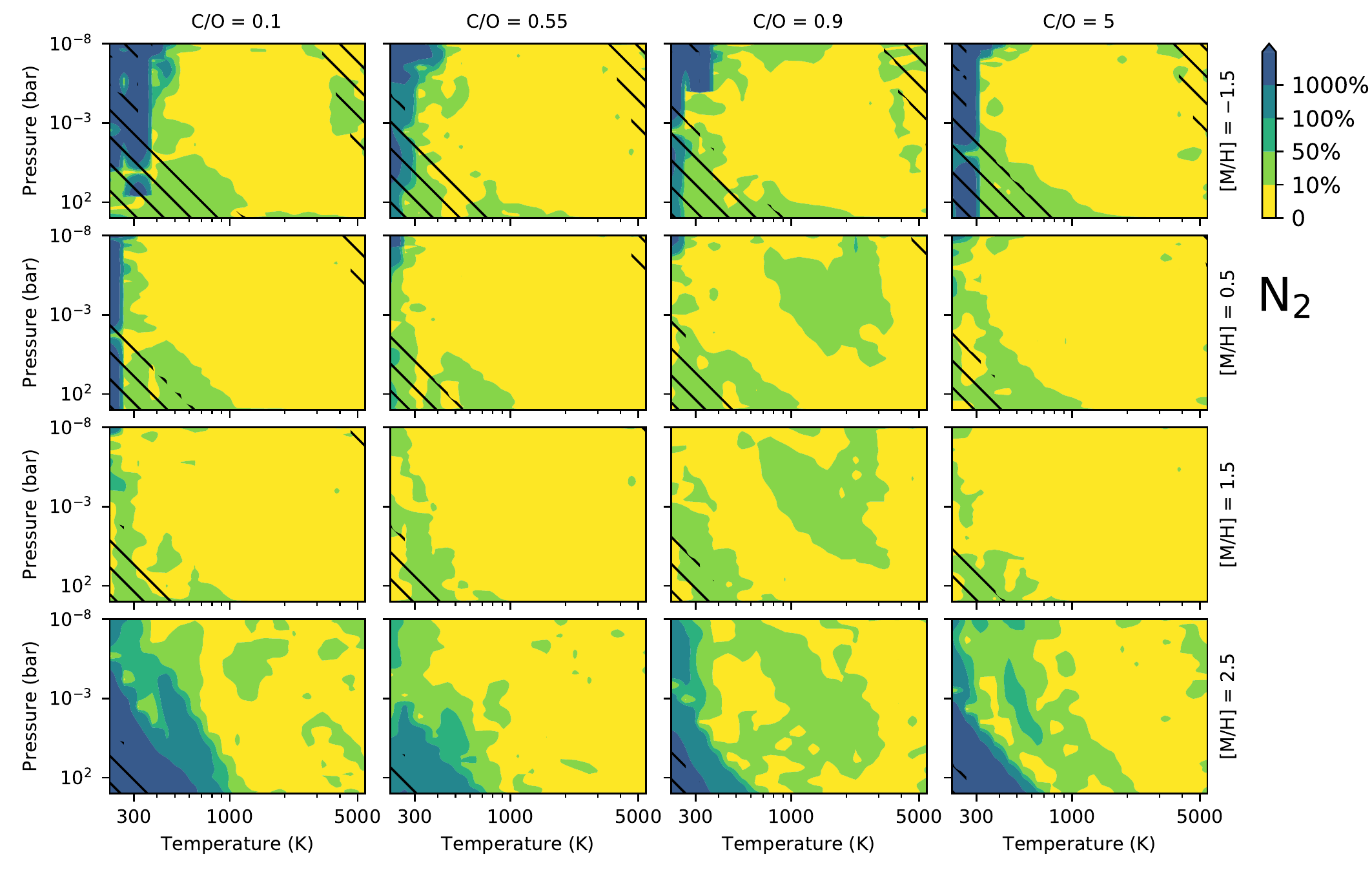}\\
\includegraphics[width=0.49\textwidth, clip]{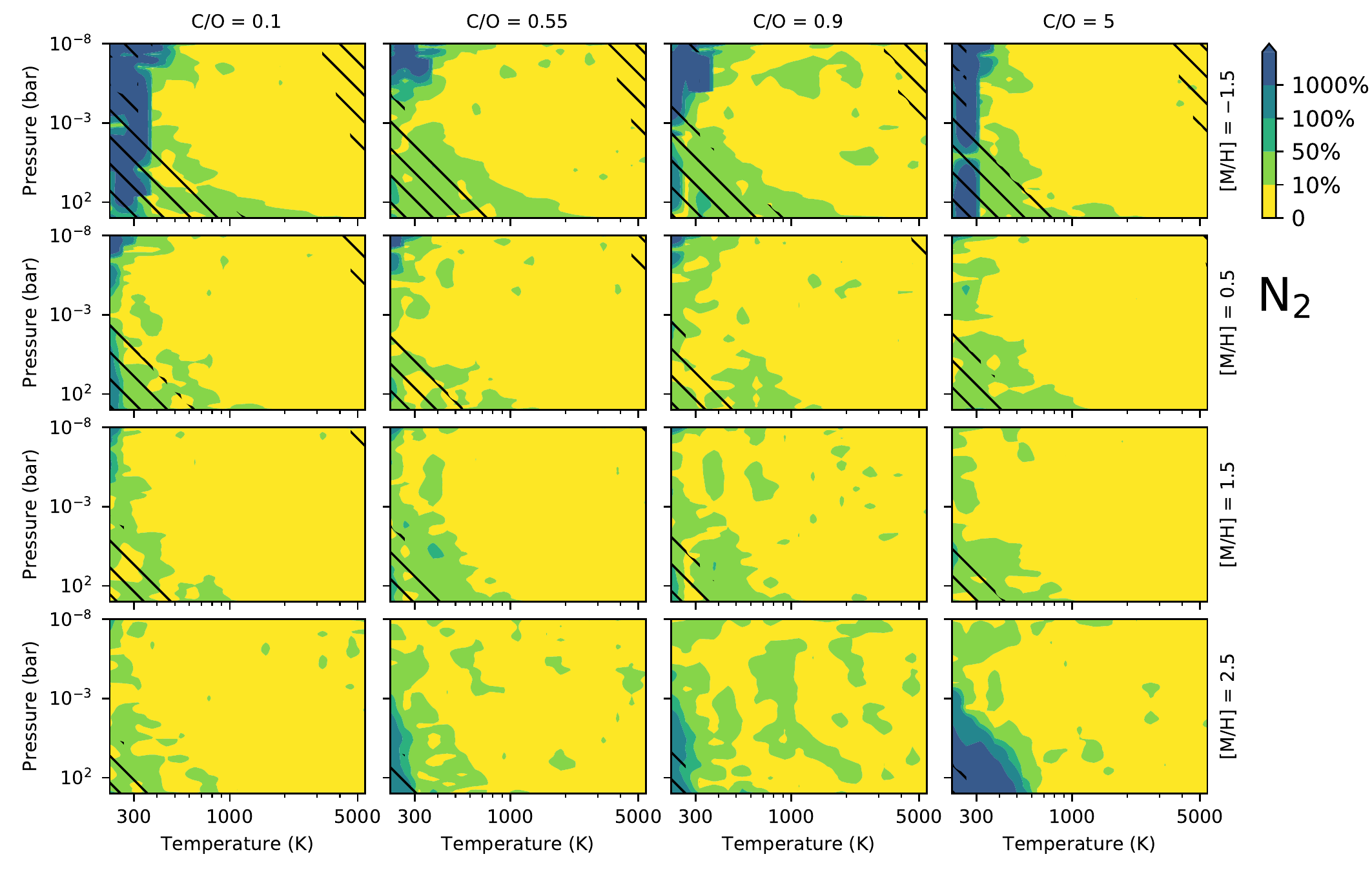}\hfill
\includegraphics[width=0.49\textwidth, clip]{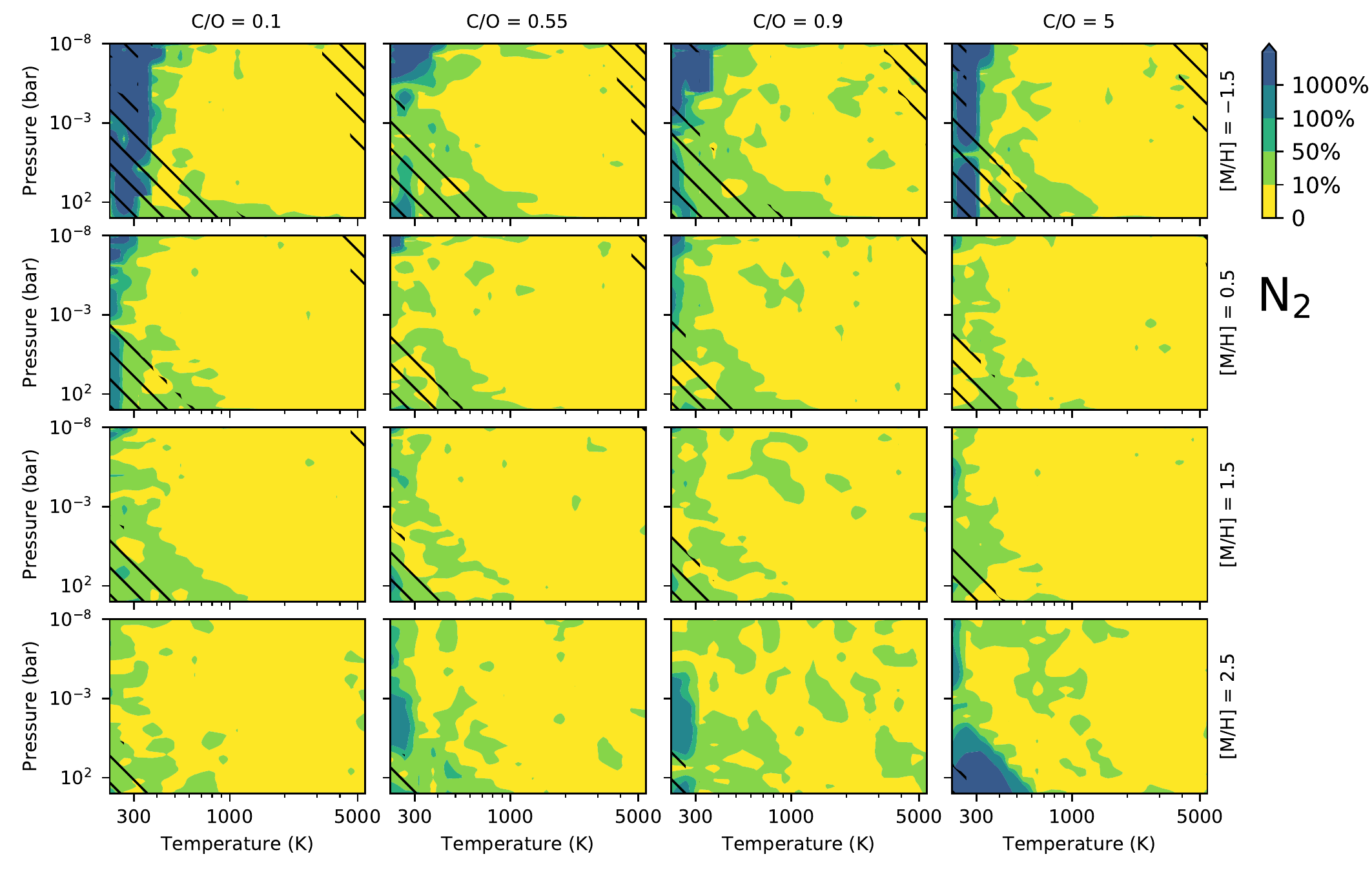}\\
\caption{As in Figure \ref{fig:H2O}, but for N$_2$.
}
\label{fig:N2}
\end{figure*}

\begin{figure*}[htb]
\centering
\includegraphics[width=0.49\textwidth, clip]{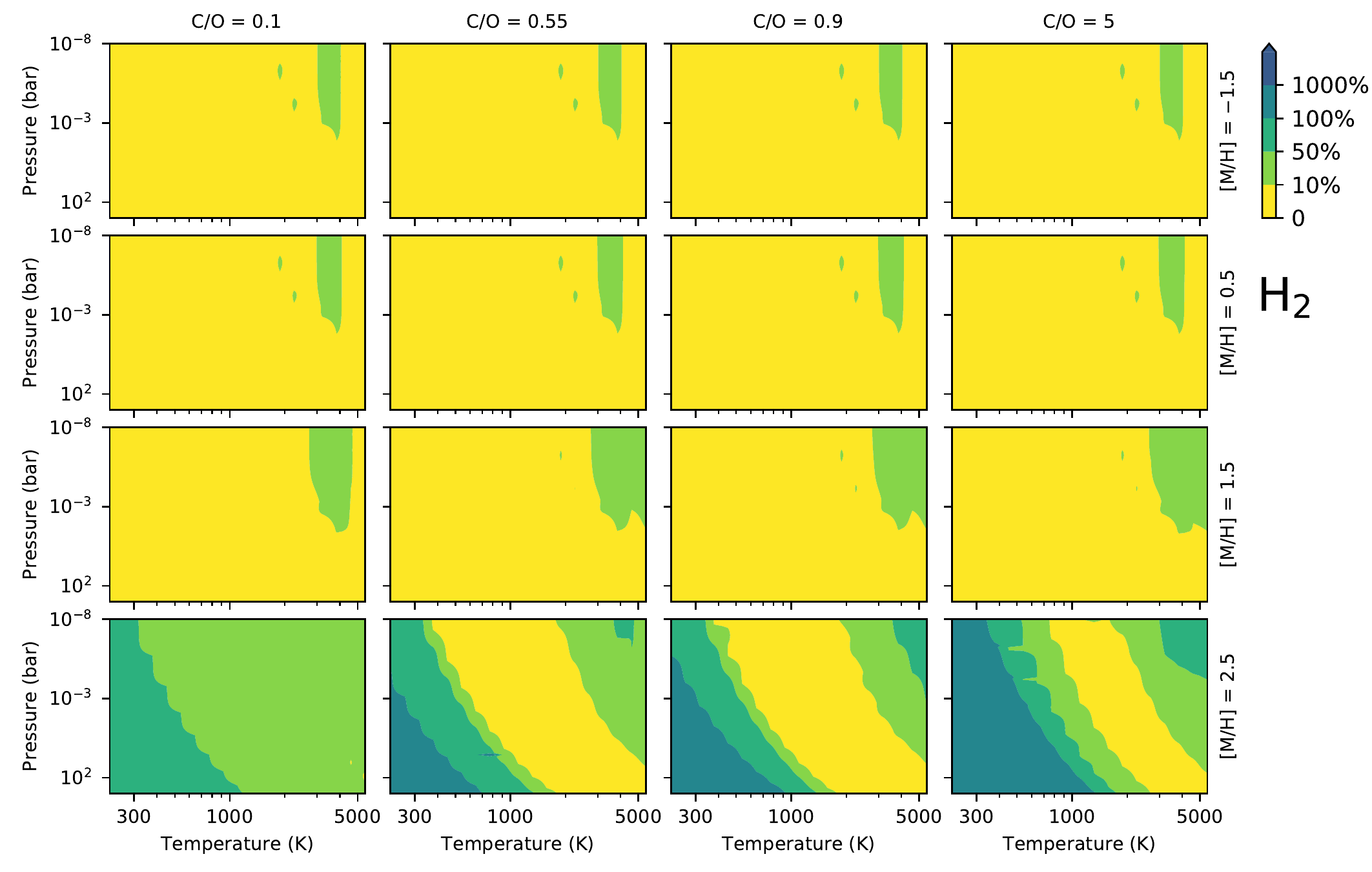}\hfill
\includegraphics[width=0.49\textwidth, clip]{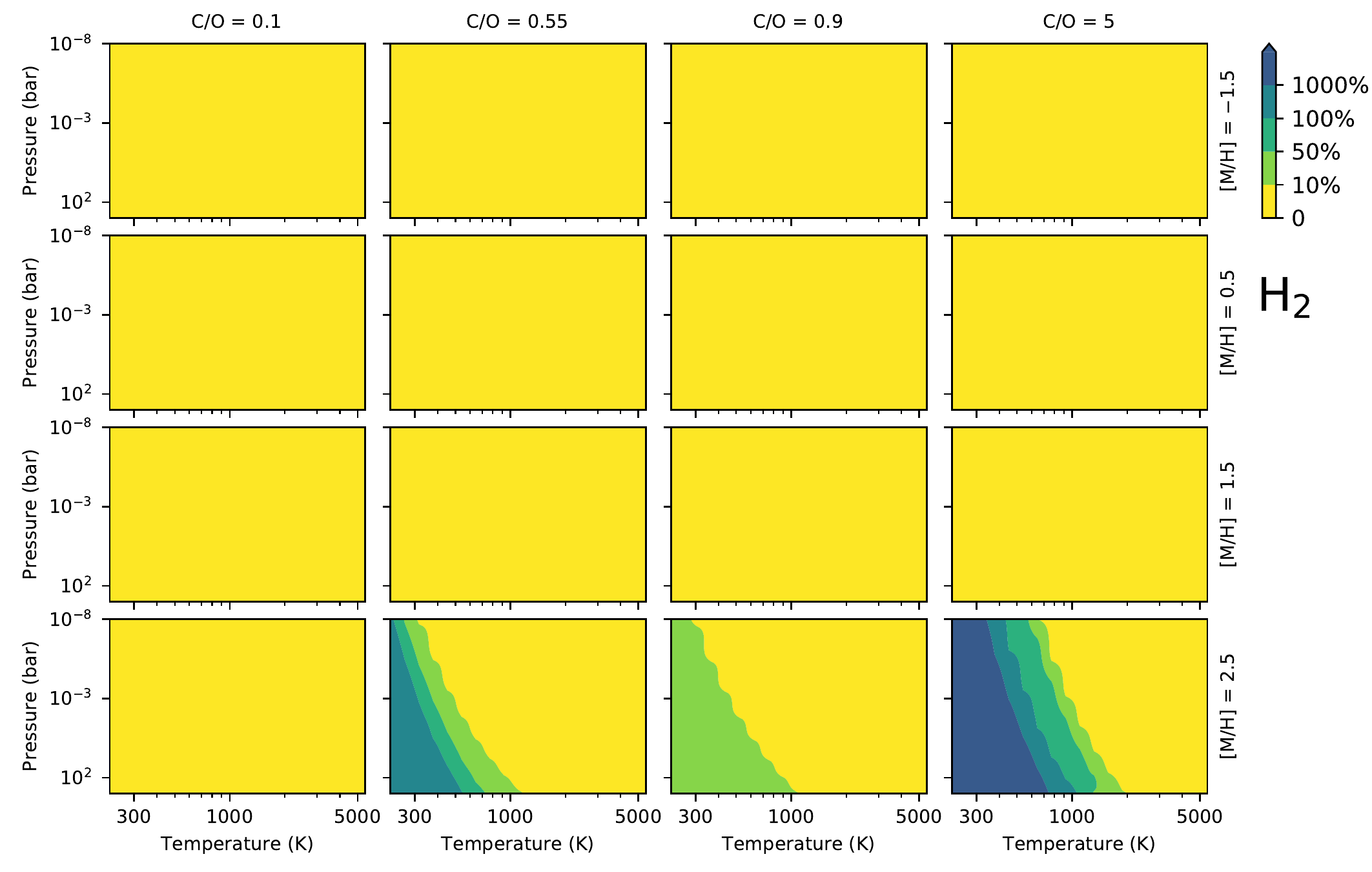}\\
\includegraphics[width=0.49\textwidth, clip]{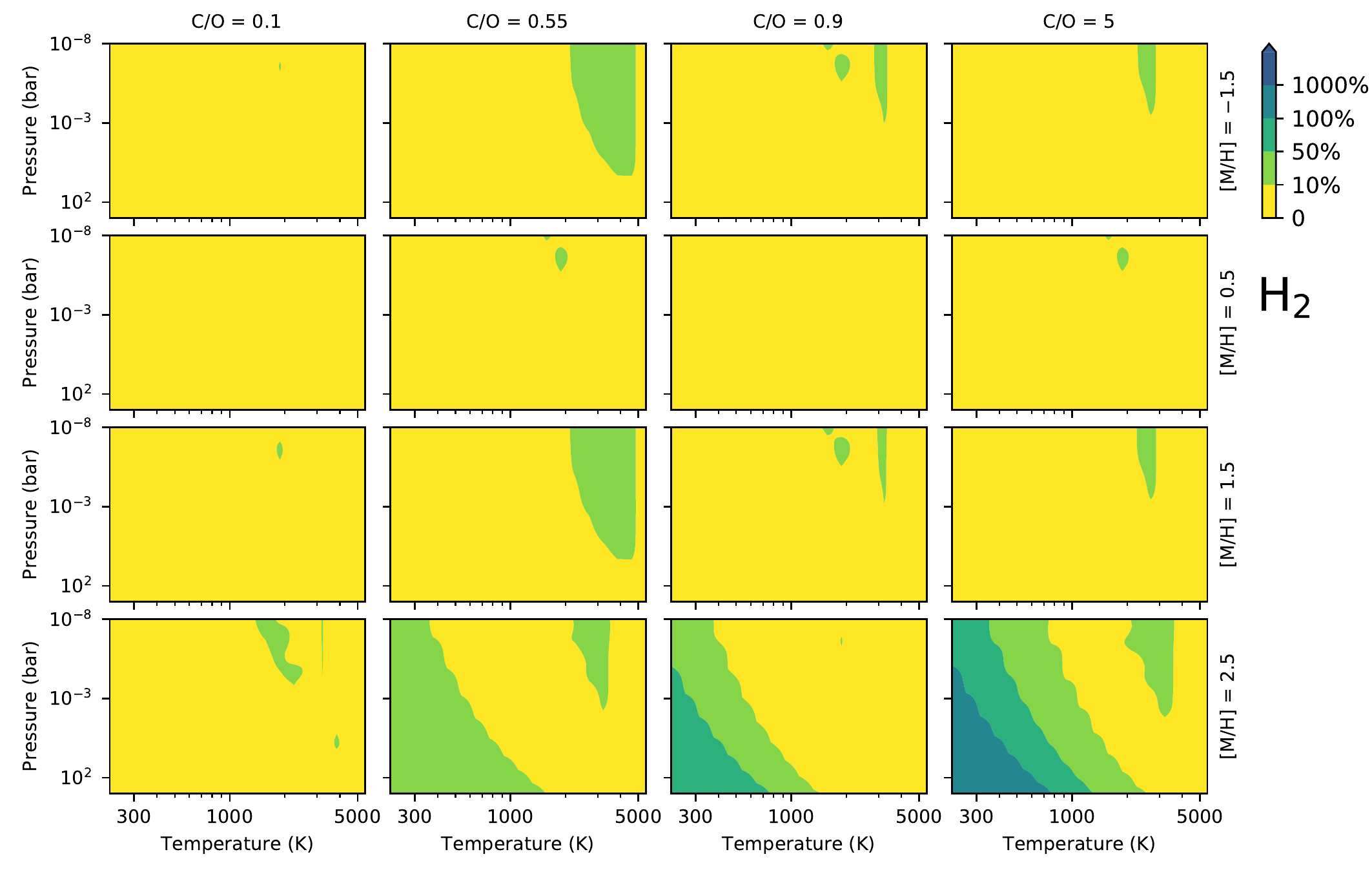}\hfill
\includegraphics[width=0.49\textwidth, clip]{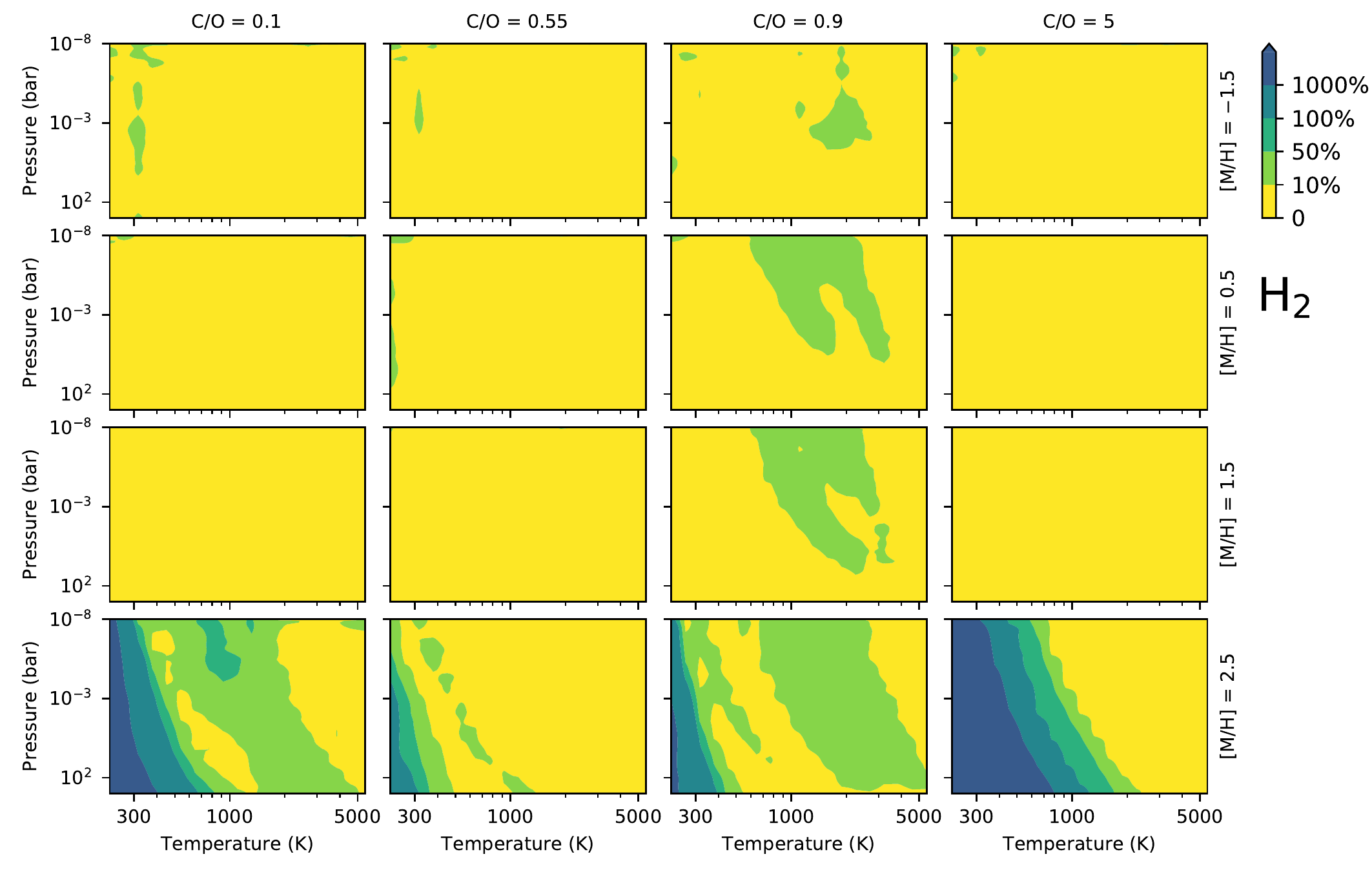}\\
\includegraphics[width=0.49\textwidth, clip]{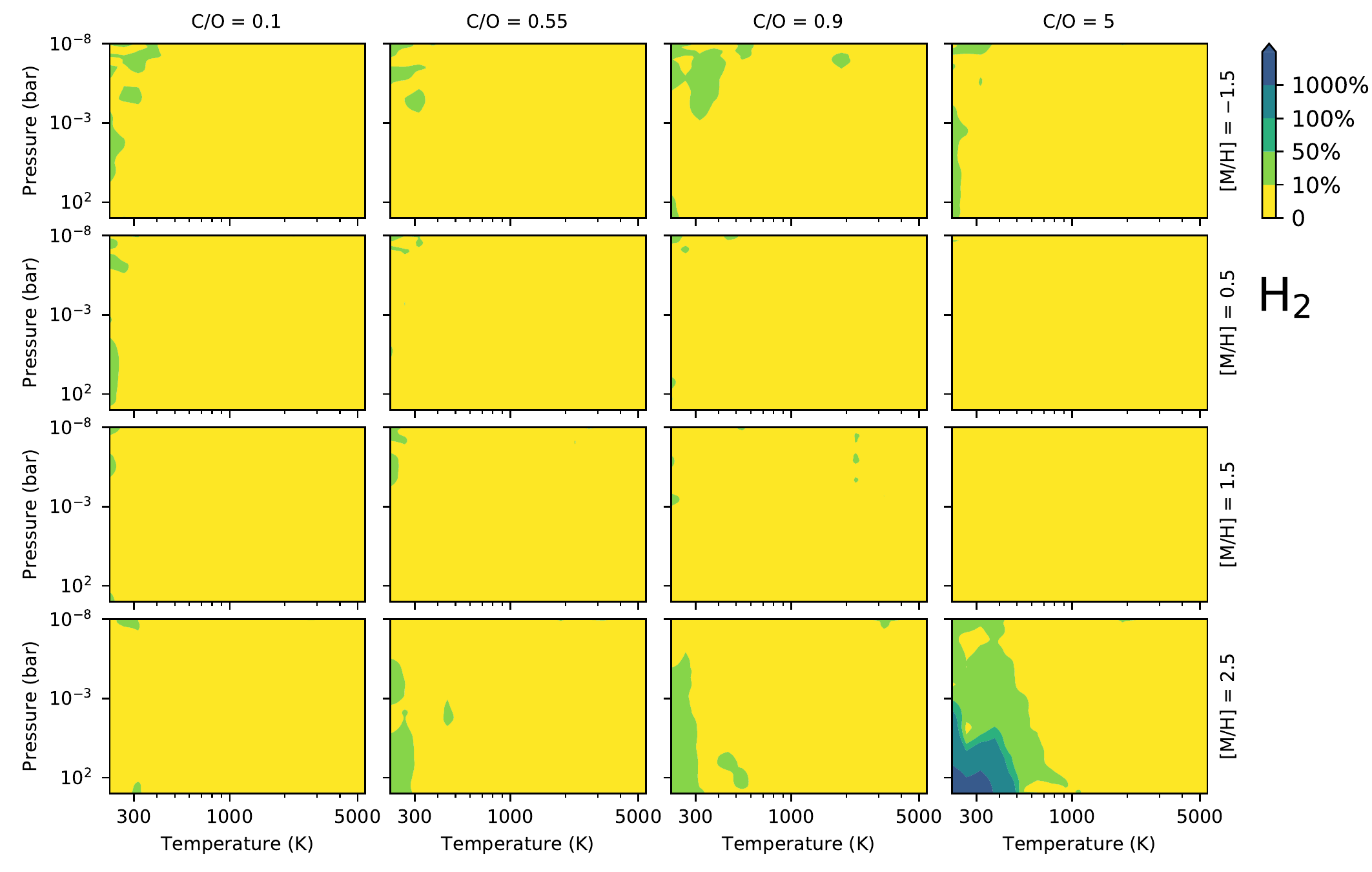}\hfill
\includegraphics[width=0.49\textwidth, clip]{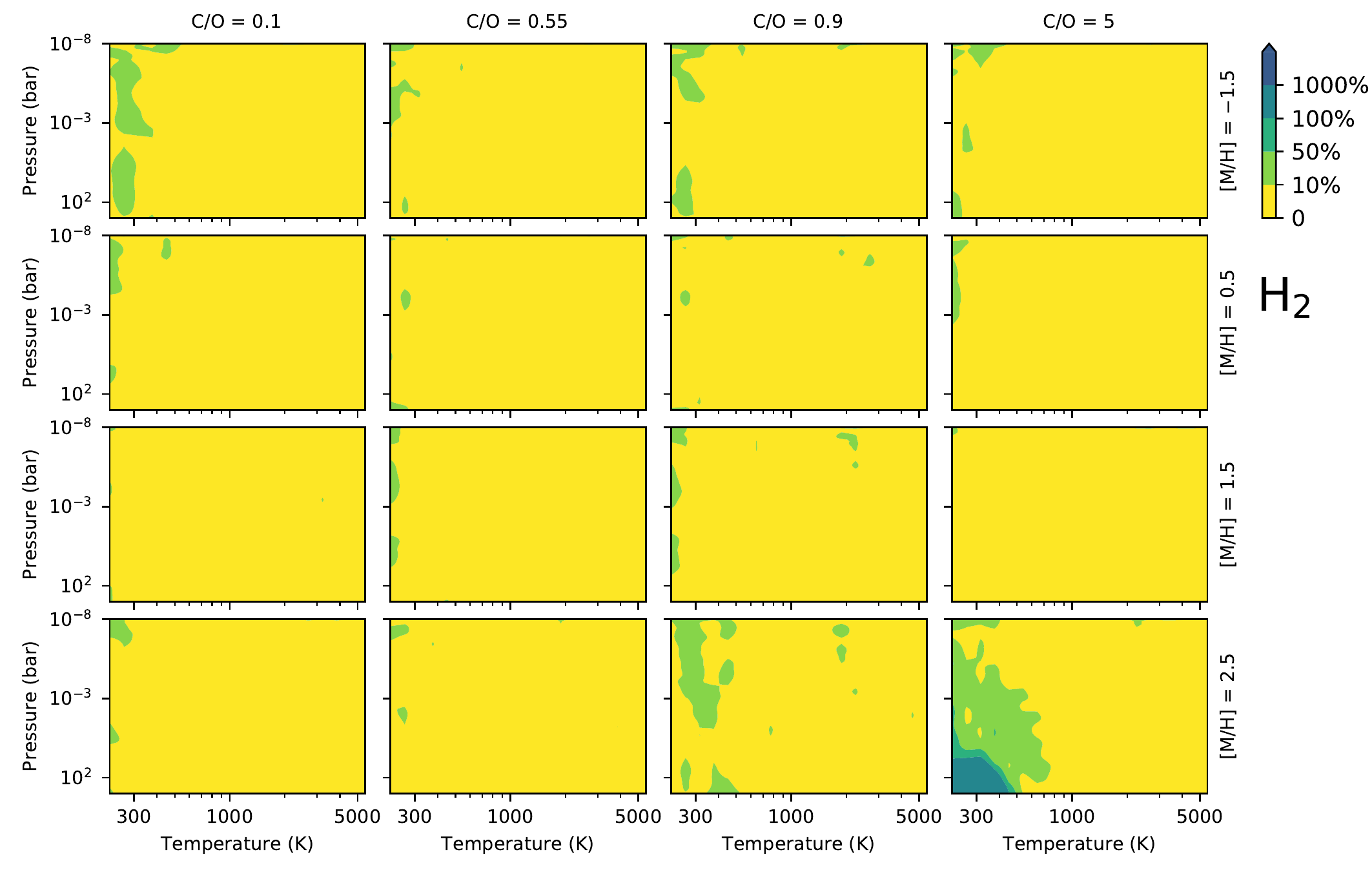}\\
\caption{PAs in Figure \ref{fig:H2O}, but for H$_2$.
}
\label{fig:H2}
\end{figure*}

\begin{figure*}[htb]
\centering
\includegraphics[width=0.49\textwidth, clip]{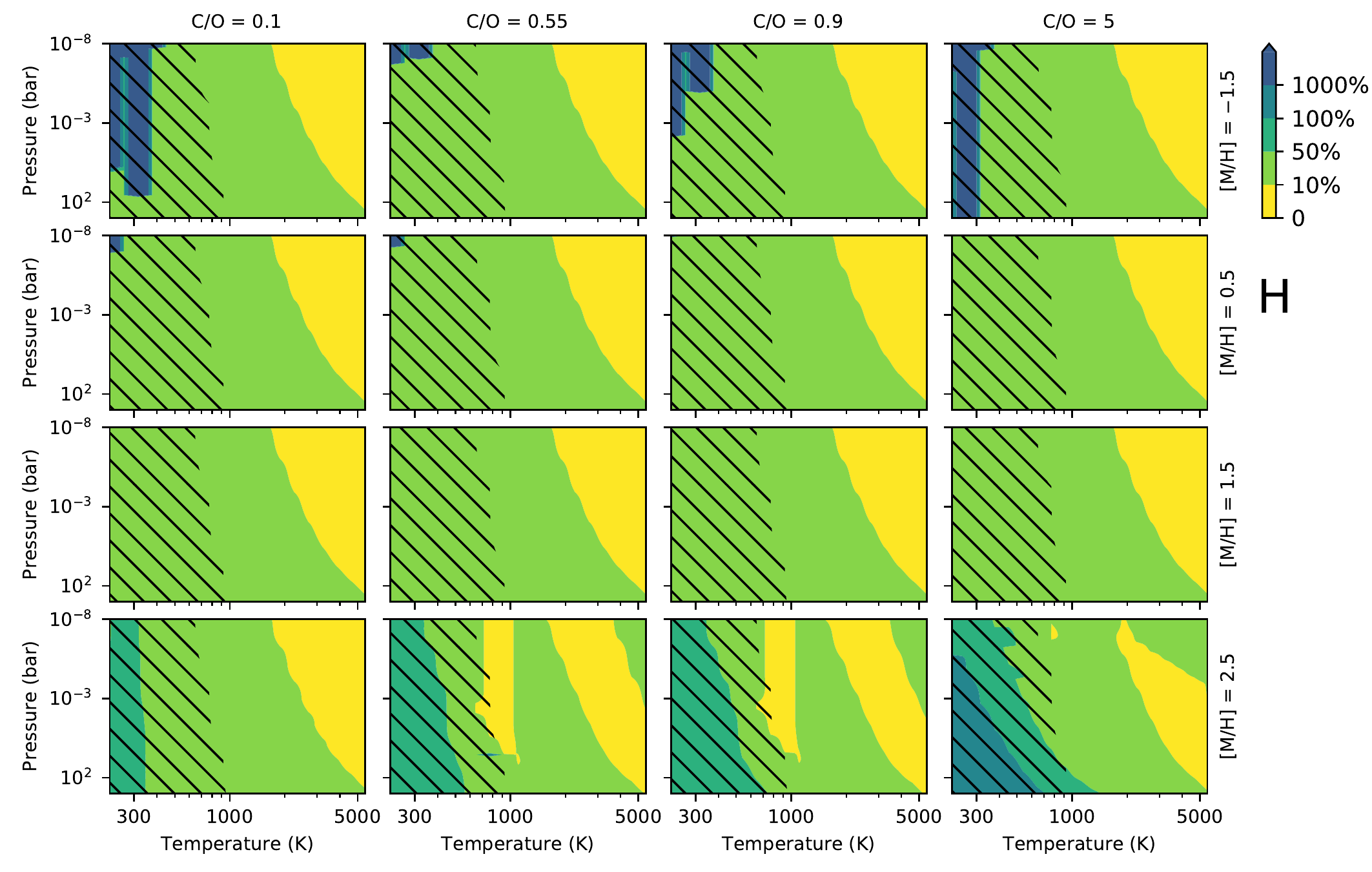}\hfill
\includegraphics[width=0.49\textwidth, clip]{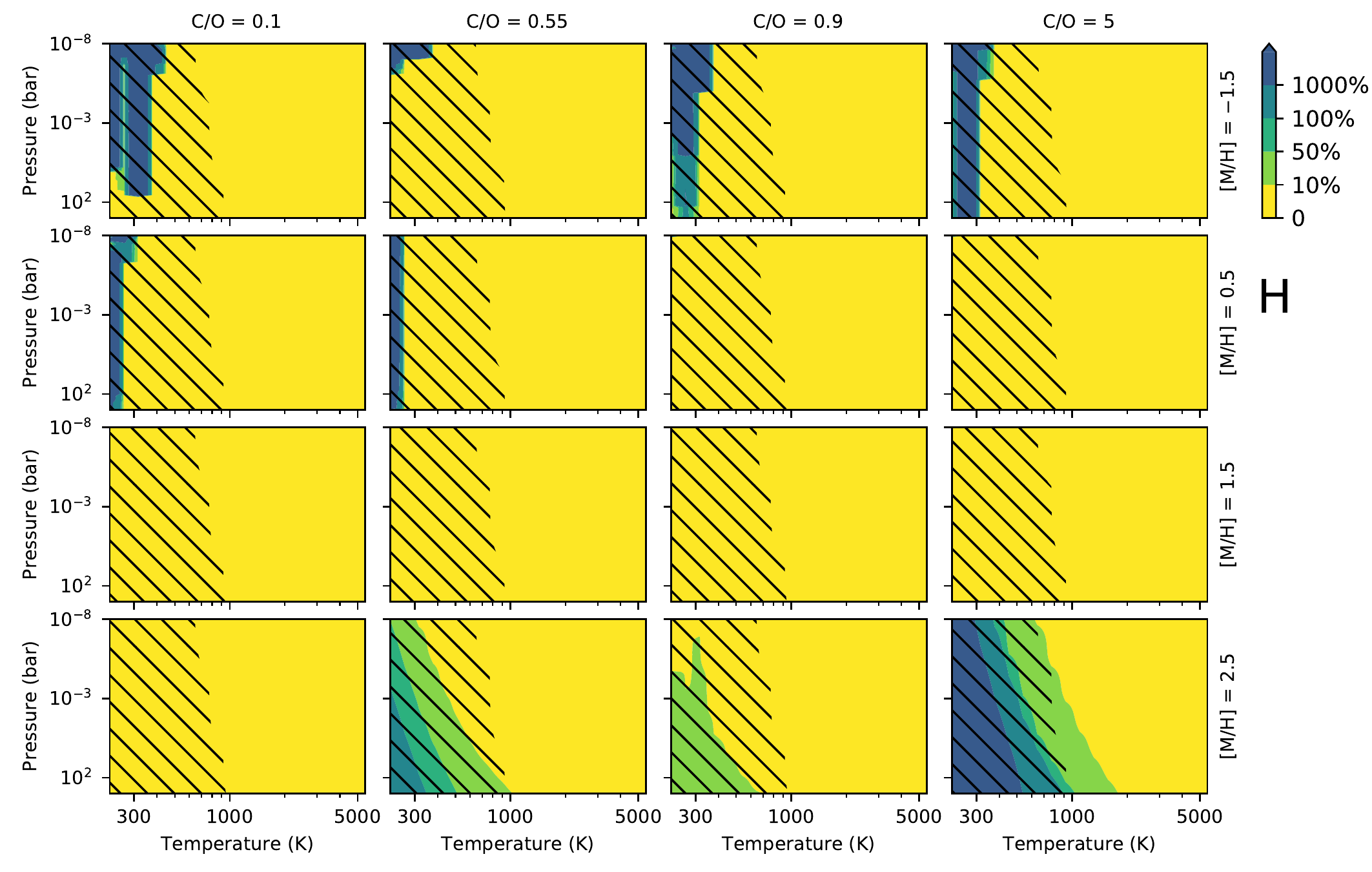}\\
\includegraphics[width=0.49\textwidth, clip]{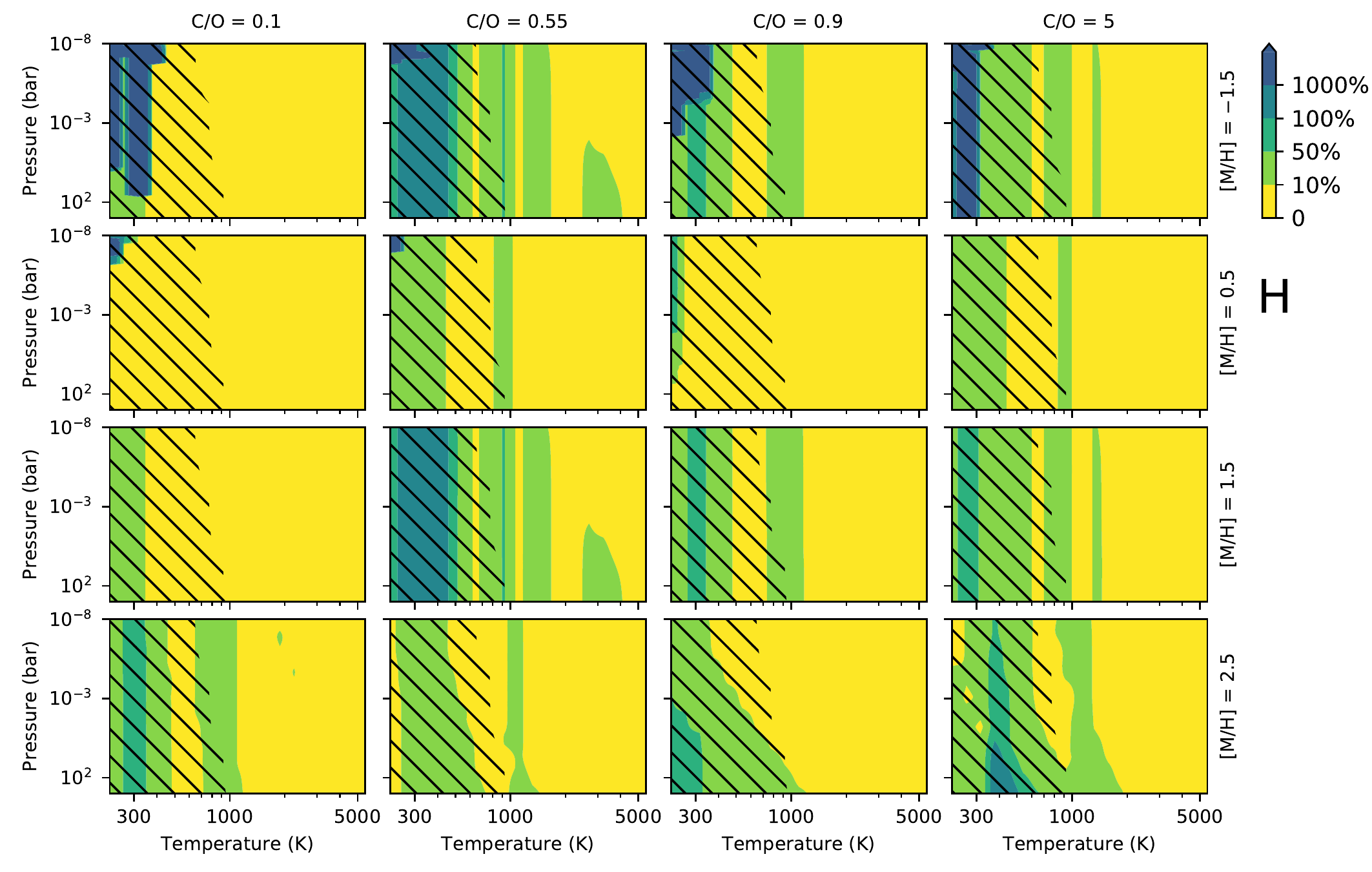}\hfill
\includegraphics[width=0.49\textwidth, clip]{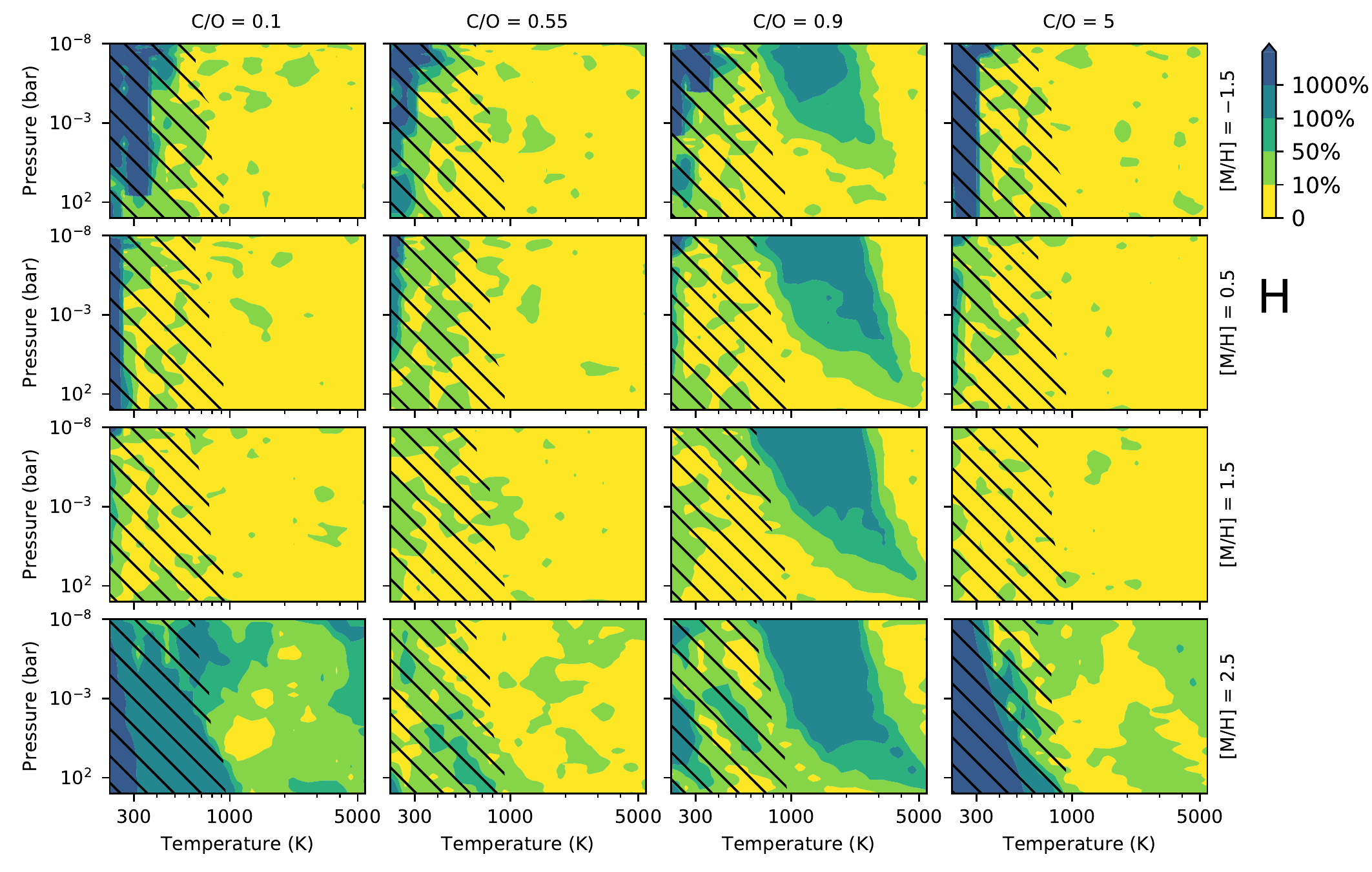}\\
\includegraphics[width=0.49\textwidth, clip]{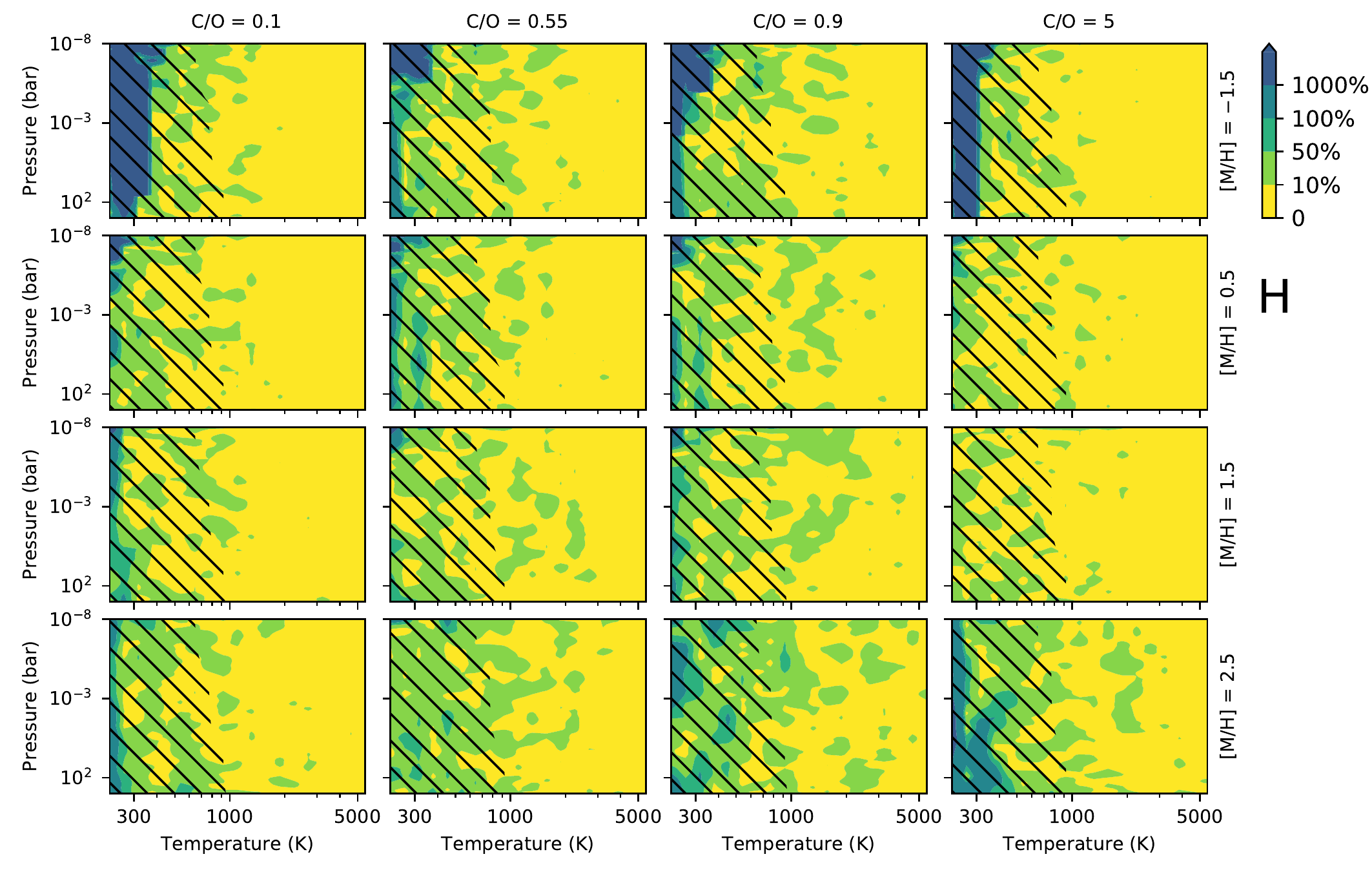}\hfill
\includegraphics[width=0.49\textwidth, clip]{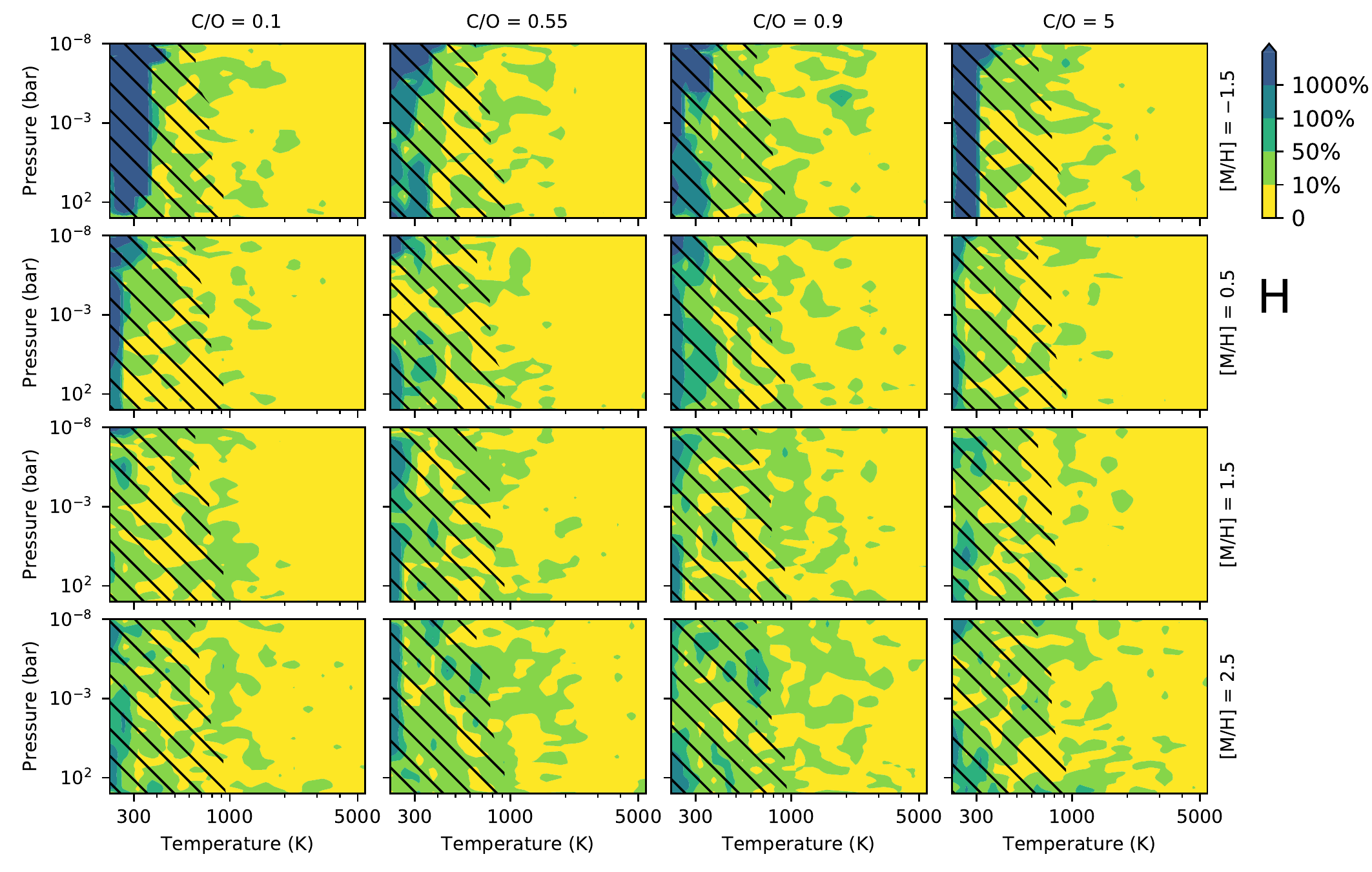}\\
\caption{As in Figure \ref{fig:H2O}, but for H.
}
\label{fig:H}
\end{figure*}

\begin{figure*}[htb]
\centering
\includegraphics[width=0.49\textwidth, clip]{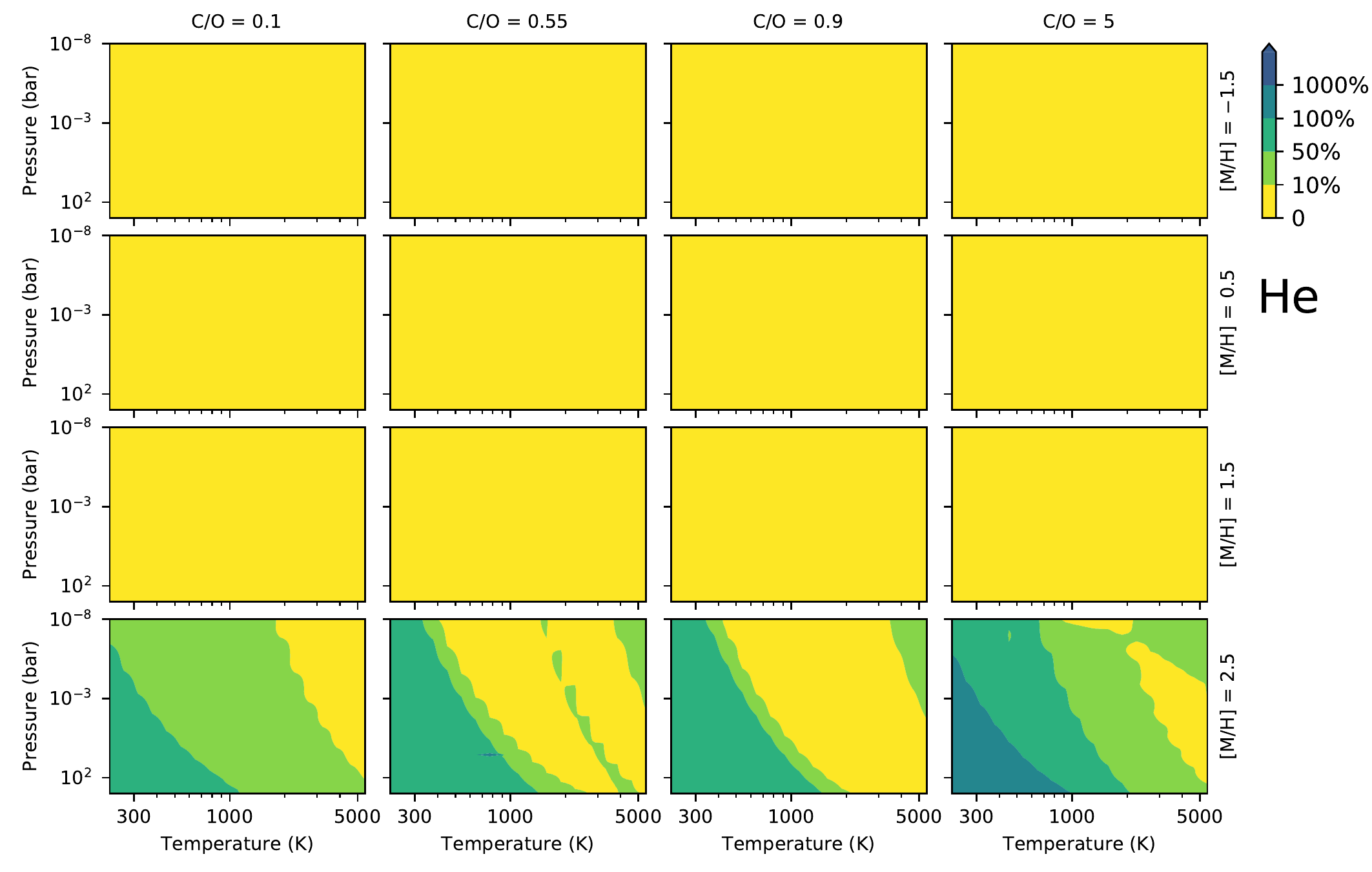}\hfill
\includegraphics[width=0.49\textwidth, clip]{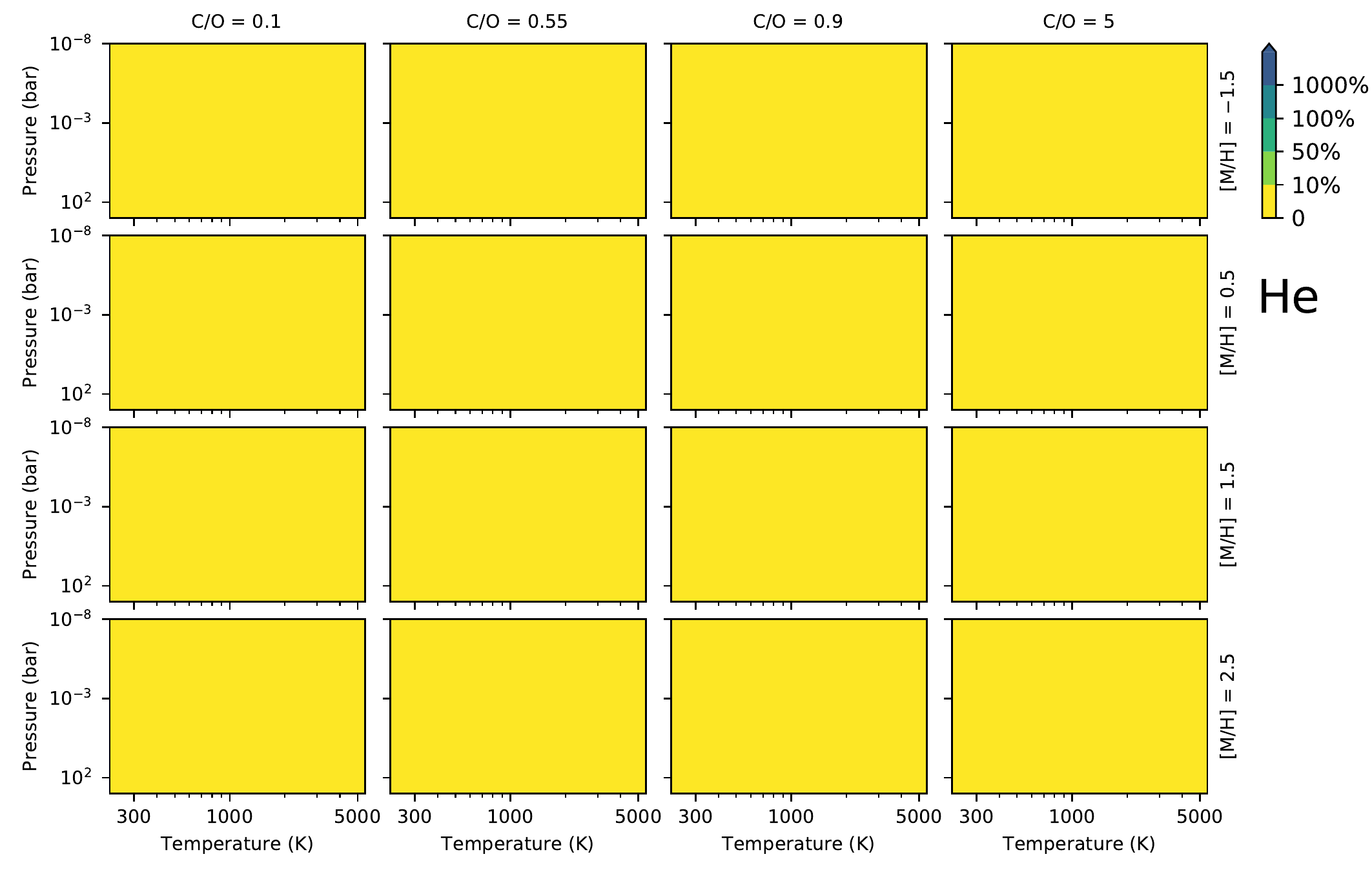}\\
\includegraphics[width=0.49\textwidth, clip]{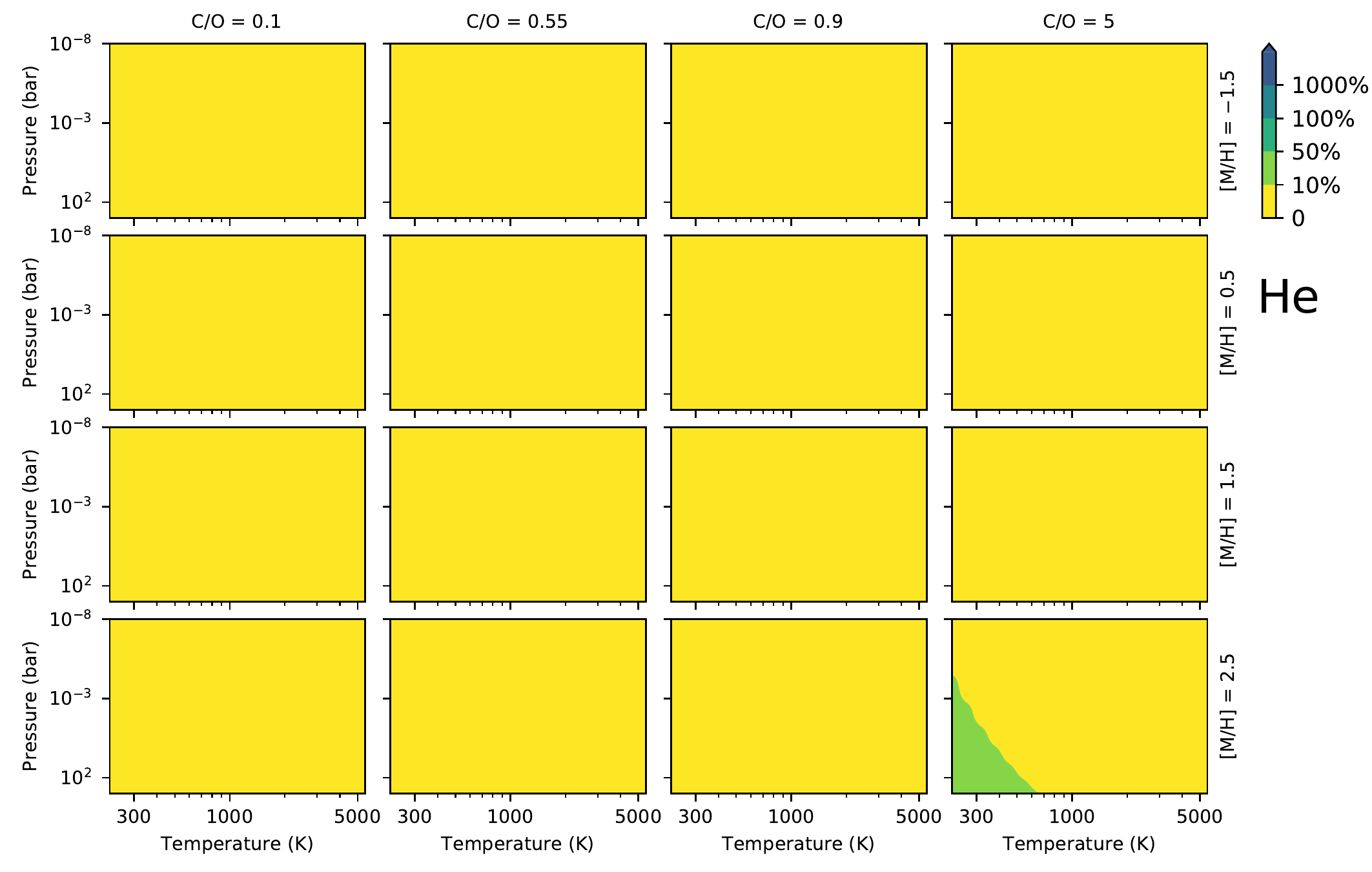}\hfill
\includegraphics[width=0.49\textwidth, clip]{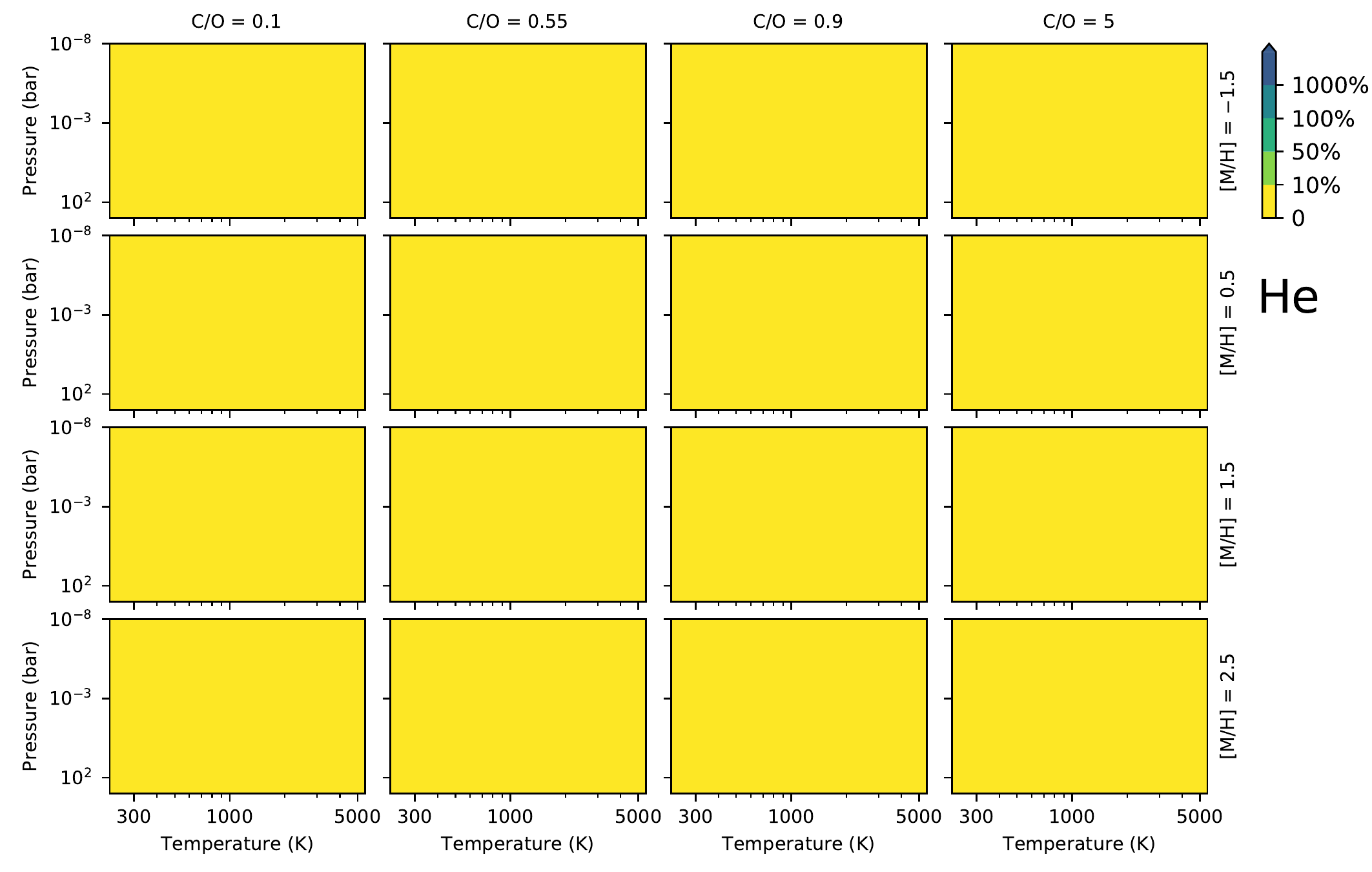}\\
\includegraphics[width=0.49\textwidth, clip]{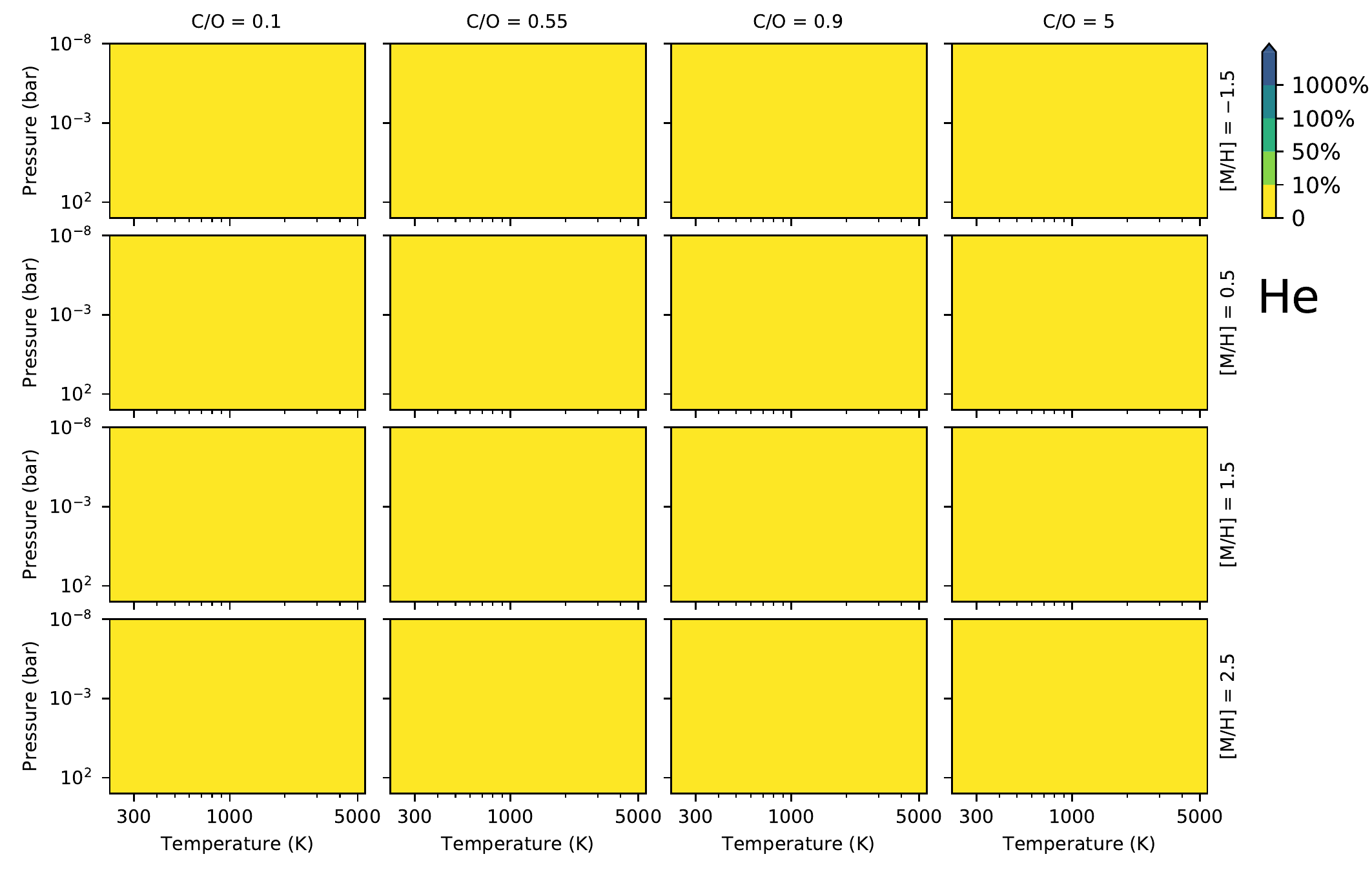}\hfill
\includegraphics[width=0.49\textwidth, clip]{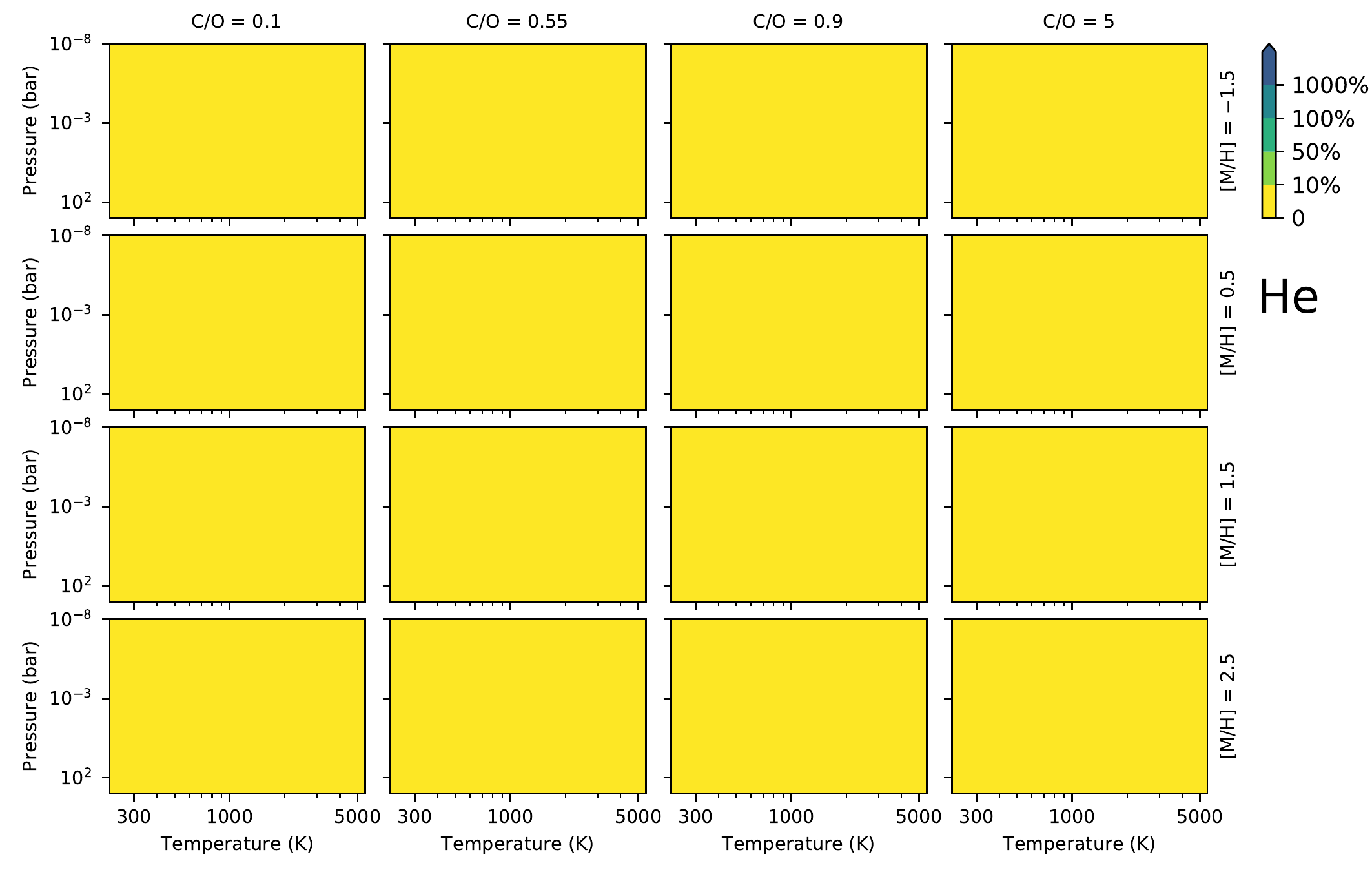}\\
\caption{As in Figure \ref{fig:H2O}, but for He.
}
\label{fig:He}
\end{figure*}

Table \ref{tbl:metrics} summarizes the median RMSE, median $R^2$, median ADE, and speedup factor relative to TEA for the considered models over the grid of test cases. 
Note that for this setup, TEA required roughly $11.7 \pm 1$ seconds to estimate equilibrium at a given temperature, C/H, N/H, and O/H over the 100 pressures considered.  
In general, we find that all of the considered models offer orders of magnitude reductions in computational cost compared to TEA, while still accurately approximating TEA over many regions of the phase space.  
Figure \ref{fig:example} provides two comparisons between the outputs of one NN model and the corresponding TEA case.
The left panel is representative of well modeled situations, showing close agreement between the NN and TEA for a temperature around 1000 K with solar metallicity and C/O; the right panel is representative of difficult situations for the NN to model, with a temperature around 2000 K with 2.5$\times$ solar metallicity and C/O $= 0.9$.
Figures \ref{fig:H2O} -- \ref{fig:He} visualize the model errors over the considered phase space, as in \citet{CubillosEtal2019apjRATE}, but with a different grid of metallicities.
Overall, we find that the NNs more accurately approximate TEA than the other models over the phase space considered.
Visually, the results of NN2, NN3, and NN4 look similar; consequently, we omit the figures from NN4, though they are available in the online compendium (see Section \ref{sec:conclusions6}).

Though the best linear-interpolation model achieves the lowest median absolute dex error, its RMSE and $R^2$ indicate it to be less accurate than the best IDW interpolation model as well as most of the NNs.
This is confirmed via the mean absolute dex errors: the best linear model's average error is about 0.38 dex, while the worst of the NN models features an average error of about 0.33 dex (NN2 and NN3 are even less, around 0.13 dex).
Thus, while the linear model performs well in many cases (as indicated by the low median dex error), the cases where it is inaccurate result in significant relative errors (as indicated by the greater average dex error).
This is confirmed by inspecting the error plots, as the model generally performs accurately but features significant errors in certain regions, e.g., for carbon- and oxygen-bearing species for C/O$ = 0.9$.
The interpolation error in non-linear regimes could be minimized by more finely sampling in that region, though this may not be feasible for more computationally expensive chemistry models than considered here.

Based on the calculated metrics, the best-performing linear and IDW interpolation models are comparable.
While linear interpolation is generally faster than IDW and achieves a lower median absolute dex error, IDW achieves a lower median RMSE and slightly higher median $R^2$.
Despite this, both models achieve similar average absolute dex errors of ${\sim}0.38$.
Table \ref{tbl:interp} details the performance of the other linear interpolation models considered in this investigation, while Table \ref{tbl:idw} details the performance of the other IDW interpolation models considered.
For simple linear interpolation, we find that increasing the grid density leads to improved model performance, as expected.
For IDW, we generally find that the value of the exponent is anti-correlated with the absolute dex error.  
However, at large values for the exponent, overflow can occur when the tested point is near one of the grid nodes.  
We also find that, for a given exponent, the average absolute dex error is typically minimized when considering the 2 nearest neighbors along each axis.  
While considering only the nearest neighbor along each axis performs equivalently or only slightly worse than the $n=2$ case, considering 3 or more of the nearest neighbors along each axis performs significantly worse than in the $n=2$ case.
This is consistent with intuition: at $n=1$, the interpolated result assumes the behavior is linear from the closest points, while at $n=2$ the interpolated result factors in the behavior of the closest points both above and below the target parameters.  
For $n>2$, the additional neighboring points are farther away and therefore less likely to follow a linear relation with respect to the point of interest, resulting in reduced accuracy.  

While RATE achieves a greater speedup factor than the NNs, it is less accurate than those NNs, as indicated by the other performance metrics.
Additionally, the speedup factors are computed for the NN using the central processing unit (CPU); on our machine, using an Nvidia Titan Xp graphics processing unit (GPU) resulted in a ${\sim}4\times$ speedup compared to using the AMD EPYC 7402P CPU, nearly matching RATE's speedup factor.  
Further, when computing the speedup factors, we only considered the computation time of the NN for a single atmospheric model (100 pressure layers), which utilizes only a fraction of the GPU's resources.  
For applications where multiple atmospheric models can be calculated in parallel (e.g., a Bayesian retrieval with $N$ parallel chains, a multidimensional retrieval model, or a global
circulation model), the NNs offer further speedup improvements, as their computational time scales less than linearly \citep{HimesEtal2022psjMARGEHOMER}.  

Among the NN models, we find that NN3 performs best, with NN2 and NN4 not far behind.  
NN1, which was trained on gridded data alone, performs worst among the NN models. 
Visually, NN1's error plots feature topography similar to the interpolation error plots.  
For example, in Figure \ref{fig:H2O}, at C/O $= 0.9$, the interpolation plots feature a region in the temperature--pressure space with significant error.  
However, its errors in these regimes are generally less than the interpolation approaches despite being trained on a data grid ${\sim}7$\% the size of the grid used for interpolation.
These similar topographies suggest that the model grid is undersampled in this region.  
As C/O approaches 1, oxygen-dominated chemistry gives way to carbon-dominated chemistry \citep{Madhusudhan2012apjCtoORatio, MosesEtal2013apjCtoORatio}.
Thus, it is likely that the model error in this regime is due to nonlinear changes in the chemical abundances, preventing the linear model from capturing it accurately.  
This is supported by the fact that the species lacking carbon and oxygen do not display this topography (Figures \ref{fig:NH3}, \ref{fig:N2}, \ref{fig:H2}, \ref{fig:H}, \ref{fig:He}), while all carbon- and oxygen-bearing species bear this topography at varying magnitudes (Figures \ref{fig:H2O}, \ref{fig:CO}, \ref{fig:CH4}, \ref{fig:CO2}, \ref{fig:HCN}, \ref{fig:C2H2}, \ref{fig:C2H4}).  
Similarly, NN1's poorer performance at M/H = 2.5 compared to the other NNs is likely also attributable to undersampling in that regime.  

The other NNs, which were trained on random data, do not feature this topography.  
NN3 was trained on a data set with ${\sim}7$\% of the number of cases in the gridded data set used by the interpolation methods.
Despite the same training data size as NN1 and training for fewer epochs, NN3 outperforms NN1 according to the performance metrics.
This suggests that NN surrogate models more efficiently learn a problem when the training data are randomly generated, rather than being generated on a fixed gridding. 
NN4 utilized both NN1's gridded data and NN3's random data---the most data out of any of the NN models.
NN3 and NN4 were trained for the same number of epochs, yet despite the additional data considered by NN4 (which led to a greater number of training iterations than NN3), NN4 performs worse than both NN2 and NN3. 
NN4's performance falls between NN1 and NN2, which, together with NN1's performance, suggests that the gridded data force the NN towards a solution that, while optimal for those data, does not properly generalize to approximate thermochemical equilibrium in all regimes.
Future studies should examine this in more detail to determine the optimal approach to generating data to train NN surrogate models.

\begin{figure*}[htb]
\centering
\includegraphics[width=0.49\textwidth, clip]{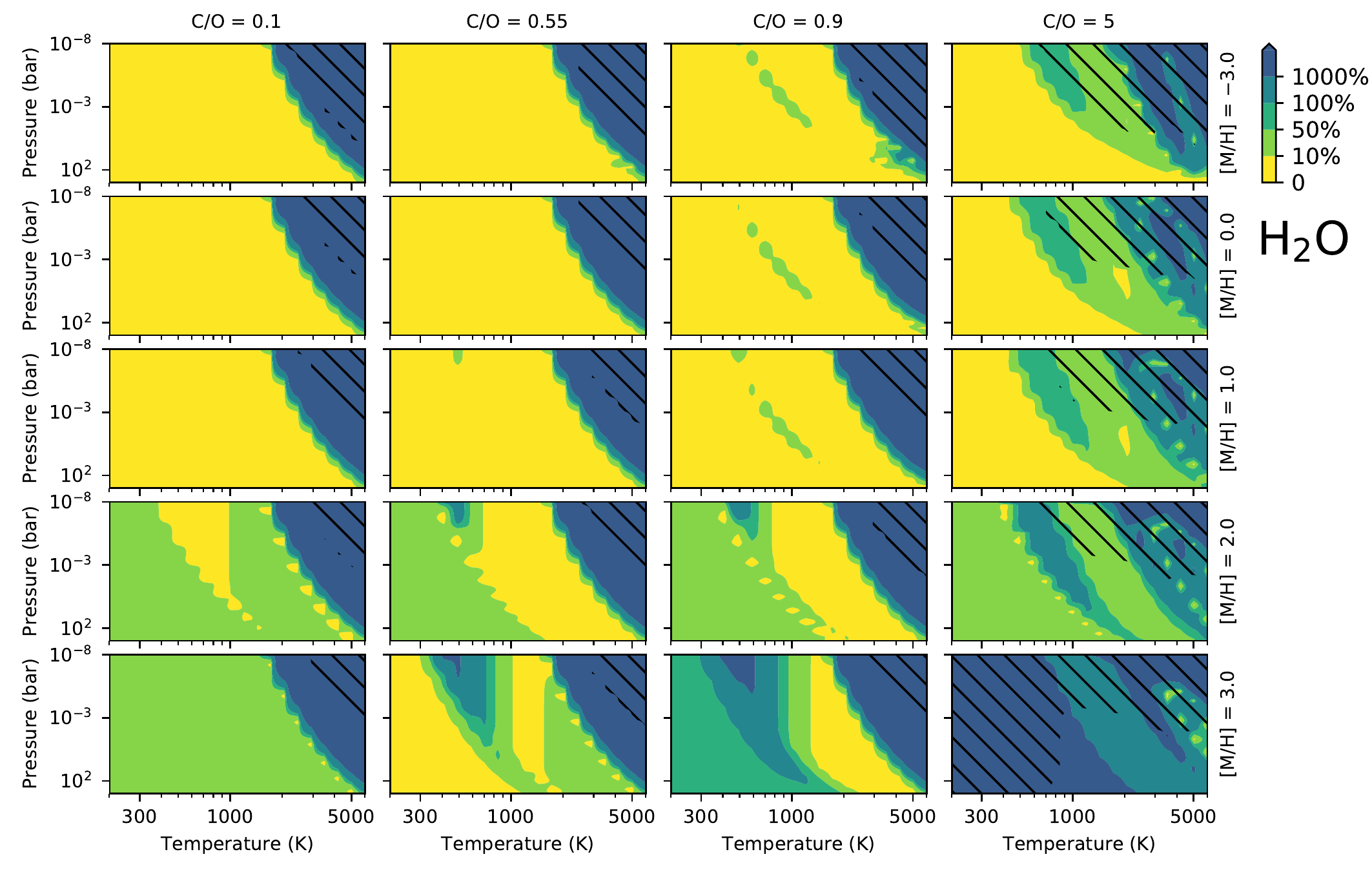}\hfill
\includegraphics[width=0.49\textwidth, clip]{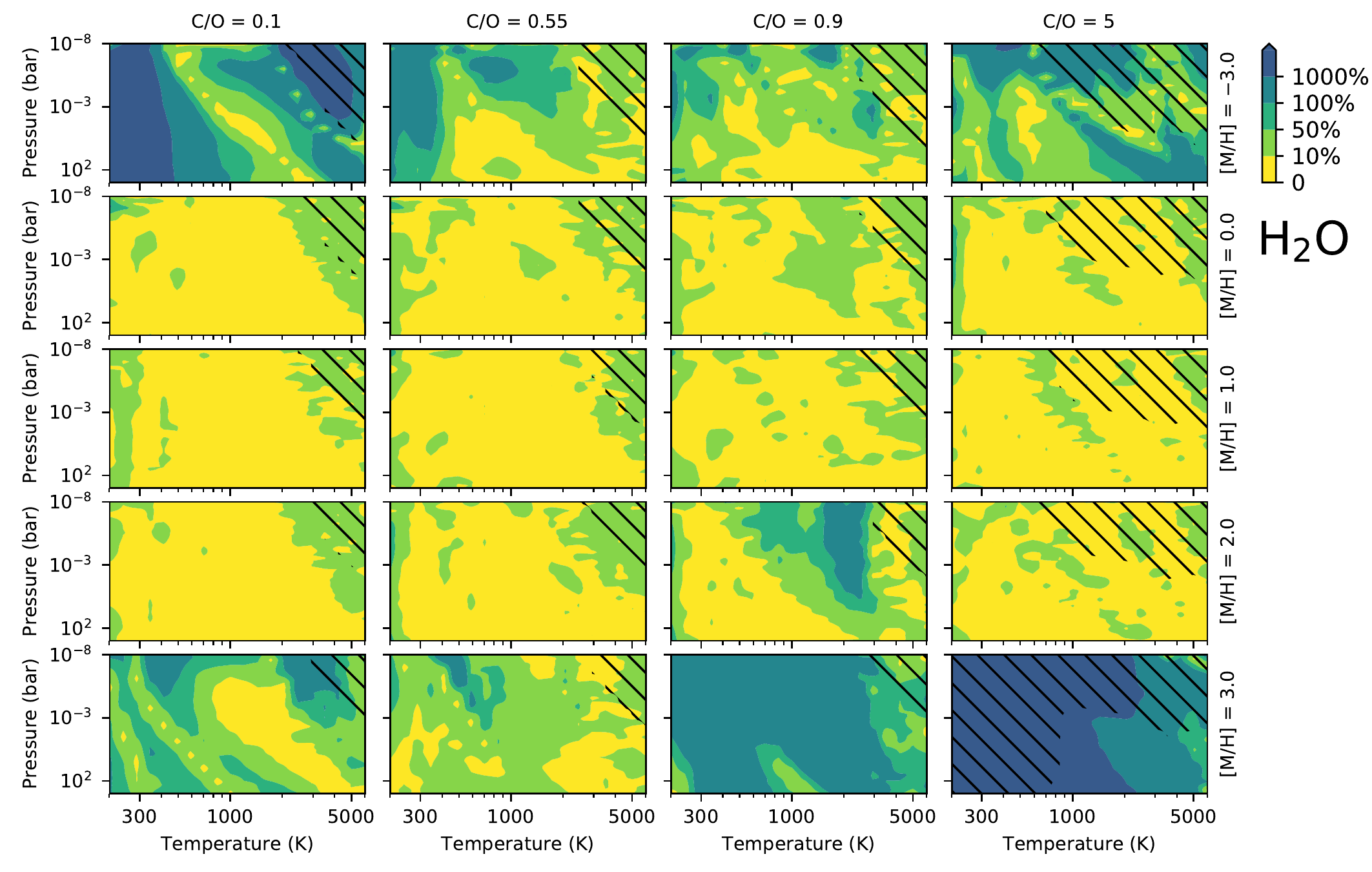}\\
\includegraphics[width=0.49\textwidth, clip]{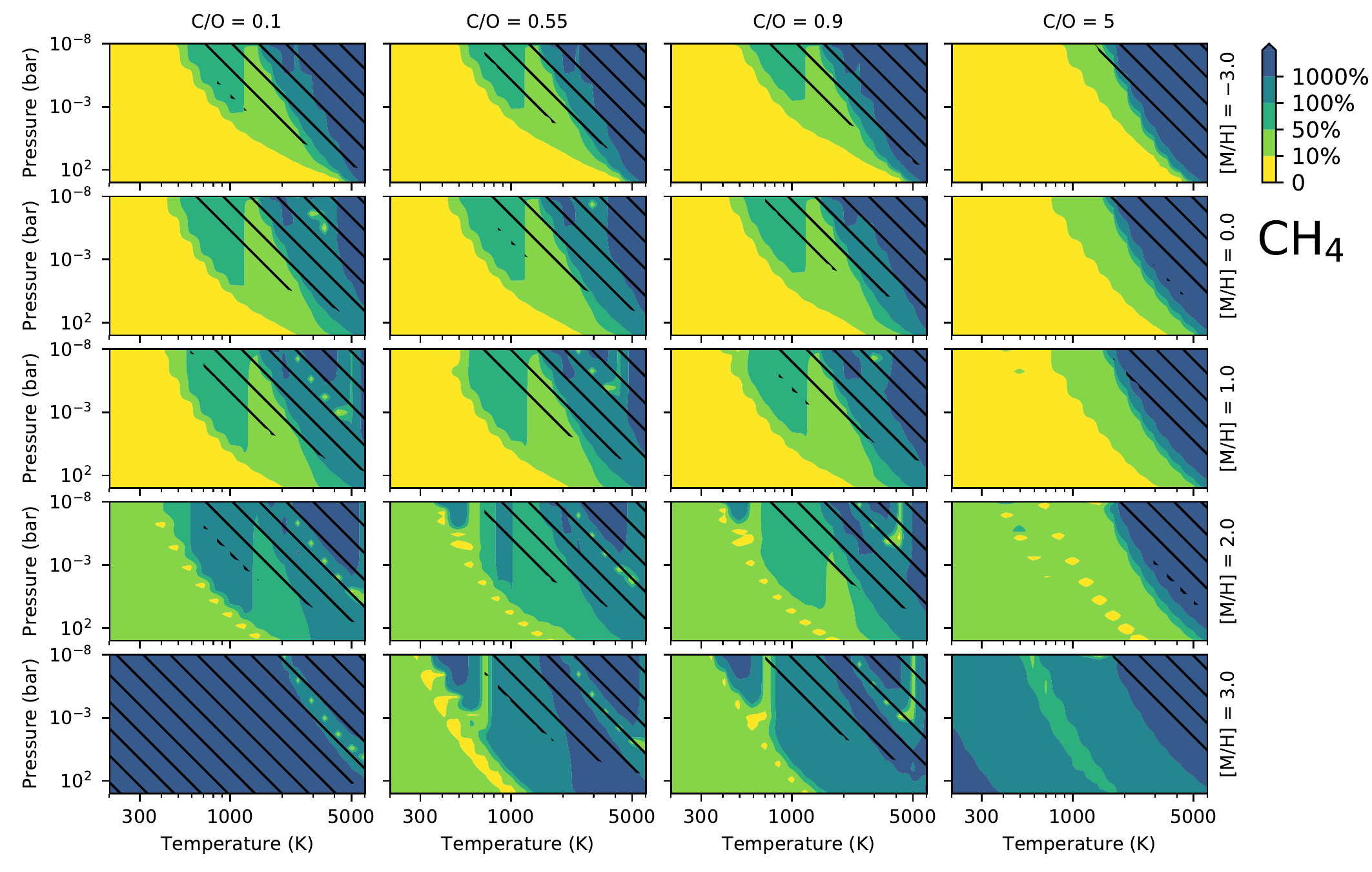}\hfill
\includegraphics[width=0.49\textwidth, clip]{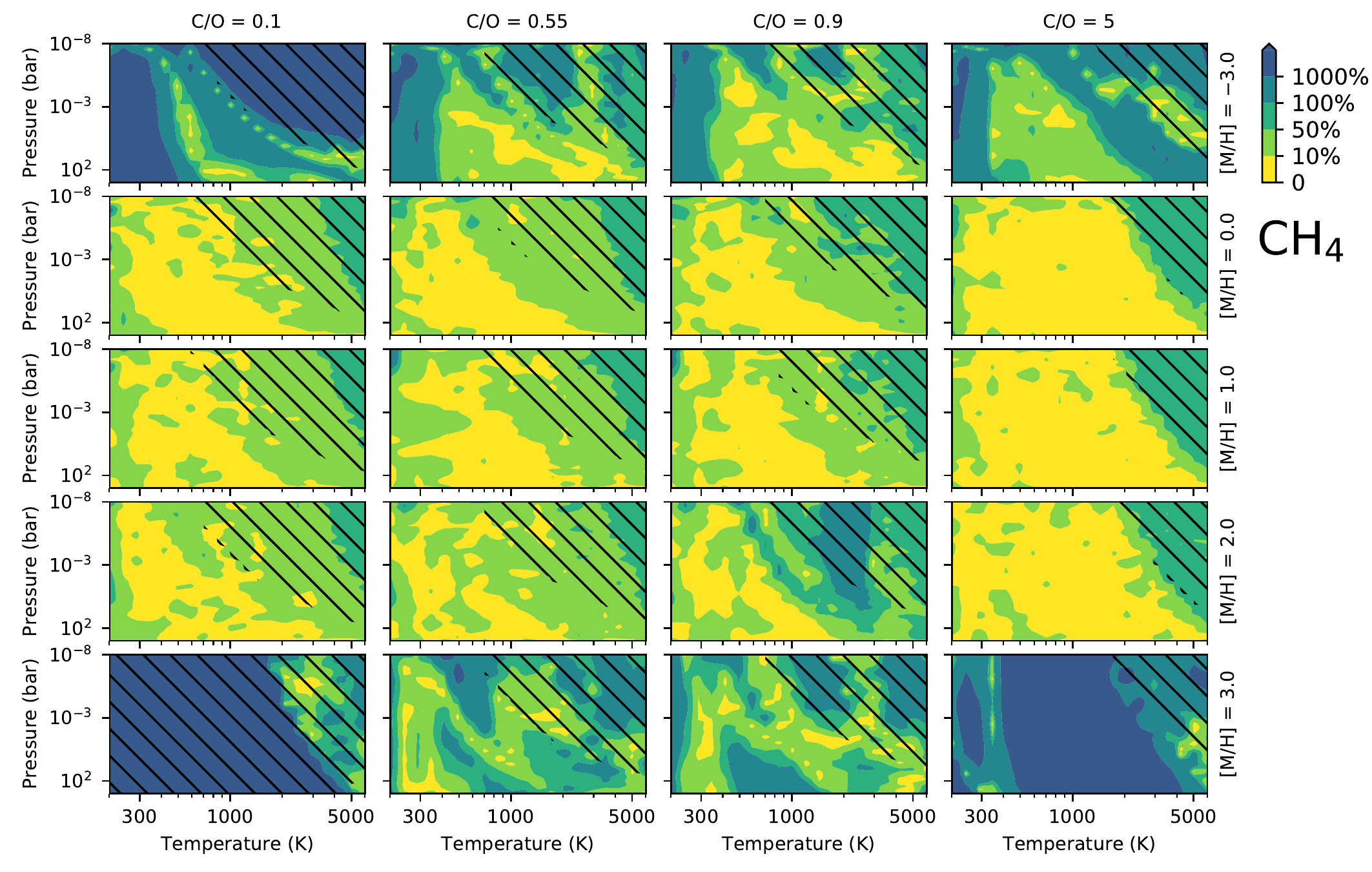}\\
\caption{As in Figure \ref{fig:H2O}, but using the grid of \citet{CubillosEtal2019apjRATE} to illustrate NN inaccuracies at phase space extrema.  While RATE (\textbf{left column}) performs accurately at a metallicity of $-3$, predictions using NN3 (\textbf{right column}) for this metallicity feature significant error in certain pressure--temperature regions.  Note that the RATE plots differ from those in \citet{CubillosEtal2019apjRATE} as we additionally consider helium in this investigation, which adjusts the relative abundances and therefore the relative errors.
}
\label{fig:cubillosgrid}
\end{figure*}

We additionally considered the NN's performance at the edge of the phase space; Figure \ref{fig:cubillosgrid} shows NN2's and RATE's predictions for the grid considered in \citet{CubillosEtal2019apjRATE} for H$_2$O and CH$_4$.  
We find that, unsurprisingly, its accuracy diminishes near the edges of the phase space.
At the metallicity extrema (3 orders of magnitude above and below solar metallicity), the NN's error increases significantly, especially at the C/O extrema.  
By comparison, RATE's accuracy at $Z = -3$ is more or less consistent with less extreme metallicities, though it too becomes inaccurate at $Z = 3$, as reported in \citet{CubillosEtal2019apjRATE}.
We attribute the NN's behavior to its training data set: when making random draws from the phase space, few samples will have multiple parameters at extrema (e.g., both $Z = 3$ and $\mathrm{C/O} = 0.1$), resulting in reduced accuracy in that regime.  
While this could be addressed by preferentially drawing samples that are in these regions, doing so would bias the training data and thus would need to be handled carefully to ensure it does not bias the trained model.  
A more straightforward solution would be to generate data over a larger phase space than needed for the target application, thus ensuring the model is more accurate at the extrema of the target phase space.

\begin{table}[tb]
\caption{Model Performance Comparison}\vskip -.25in
\label{tbl:metrics}
\begin{center}
\begin{tabular}{l c c c c}
 \toprule
Model & RMSE & $R^2$ & Dex & Speedup Factor\\
\hline
RATE   & 1.3386 & 0.9302 & 0.1002 &  891  \\
Linear & 1.1131 & 0.9821 & 0.0037 & 8330  \\
IDW    & 0.8314 & 0.9879 & 0.0696 & 2691  \\
NN1    & 1.4942 & 0.9613 & 0.0316 &  206  \\
NN2    & 0.4872 & 0.9967 & 0.0250 &  207  \\
NN3    & 0.4390 & 0.9973 & 0.0280 &  207  \\
NN4    & 0.6343 & 0.9950 & 0.0258 &  205  \\
\hline
\multicolumn{5}{p{0.9\linewidth}}{\textbf{Notes.}  Here we present only the best-performing linear and IDW interpolation models.  For each model listed, we present only the median RMSE, $R^2$, and absolute dex error for conciseness.  For the full data, see the Reproducible Research Compendium (RRC) download link at the end of Section \ref{sec:conclusions6}.  Note also that the speedup factors for the NNs are calculated for a single atmosphere using the central processing unit (CPU); multiple atmospheric models could be calculated simultaneously and/or a graphics processing unit (GPU) could be used for calculations, increasing the speedup factor.}
\end{tabular}
\end{center}
\end{table}

\begin{table*}[tb]
\caption{Linear Interpolation Model Performance Comparison}
\label{tbl:interp}
\begin{center}
\begin{tabular}{l l c c c c}
 \toprule
Number of & Number of elemental & RMSE & $R^2$ & Dex & Speedup Factor\\
temperatures & abundances & & & & \\
\hline
 40 & 11 & 1.4937 & 0.9623 & 0.0251 & 8434  \\
 80 & 11 & 1.4719 & 0.9641 & 0.0127 & 8518  \\
 80 & 21 & 1.1131 & 0.9821 & 0.0037 & 8332  \\
\hline
\multicolumn{6}{p{0.55\linewidth}}{\textbf{Notes.}  For each model listed, we present only the median RMSE, $R^2$, and absolute dex error for conciseness.  For the full data, see the RRC download link at the end of Section \ref{sec:conclusions6}.}
\end{tabular}
\end{center}
\end{table*}

\begin{table*}[tb]
\caption{IDW Model Performance Comparison}
\label{tbl:idw}
\begin{center}
\begin{tabular}{l l c c c c}
 \toprule
Exponent & Neighbors & RMSE & $R^2$ & Dex & Speedup Factor\\
          & per axis & & & & \\
\hline
 1 &  1 & 1.1341 & 0.9830 & 0.0706 & 15190  \\
 2 &  1 & 1.0780 & 0.9799 & 0.0619 & 15250  \\
 2 &  2 & 1.0795 & 0.9802 & 0.0650 &  2763  \\
 2 &  3 & 5.6737 & 0.3177 & 0.0809 &   886  \\
 3 &  1 & 1.0068 & 0.9803 & 0.0650 & 15000  \\
 5 &  1 & 0.8495 & 0.9840 & 0.0780 & 14920  \\
 5 &  2 & 0.8502 & 0.9839 & 0.0722 &  2716  \\
 5 &  3 & 5.6693 & 0.3243 & 0.0767 &   841  \\
 5 &  4 & 7.5990 &-0.0859 & 0.0993 &   299  \\
10 &  1 & 0.8301 & 0.9884 & 0.0781 & 14930  \\
10 &  2 & 0.8194 & 0.9886 & 0.0721 &  2188  \\
10 &  3 & 5.6713 & 0.3345 & 0.0741 &   835  \\
10 &  4 & 7.6006 &-0.0631 & 0.0996 &   299  \\
20 &  1 & 0.8405 & 0.9877 & 0.0761 & 15160  \\
20 &  2 & 0.8322 & 0.9879 & 0.0687 &  2710  \\
20 &  3 & 5.6721 & 0.3344 & 0.0717 &   871  \\
30 &  1 & 0.8392 & 0.9877 & 0.0761 & 15020  \\
30 &  2 & 0.8314 & 0.9879 & 0.0696 &  2691  \\
30 &  3 & 5.6721 & 0.3343 & 0.0748 &   868  \\
40 &  1 & 0.8380 & 0.9876 & 0.0756 & 13680  \\
40 &  2 & 0.8316 & 0.9878 & 0.0719 &  2689  \\
\hline
\multicolumn{6}{p{0.55\linewidth}}{\textbf{Notes.}  For each model listed, we present only the median RMSE, $R^2$, and absolute dex error for conciseness.  For the full data, see the RRC download link at the end of Section \ref{sec:conclusions6}.}
\end{tabular}
\end{center}
\end{table*}

\section{Conclusions}
\label{sec:conclusions6}

In this study, we presented a comparison between thermochemical equilibrium estimation methods.  
We found that neural-network (NN) surrogate models outperform both interpolation approaches considered here as well as the analytical approximation of \citet{CubillosEtal2019apjRATE}, which is based on the formalism of \citet{HengEtal2016apjAnalyticalThermochemicalEquilibrium},  \citet{HengLyons2016apjAnalyticalThermochemicalEquilibrium}, and 
\citet{HengTsai2016apjAnalyticalThermochemicalEquilibrium}.
However, all approaches offer orders of magnitude reductions in computational cost compared to the Gibbs energy minimization implemented in TEA \citep{BlecicEtal2016apjsTEA}.  
Our results suggest that these thermochemical equilibrium estimation methods can run fast enough and perform accurately enough to be used in place of a Gibbs-minimization method for thermochemical equilibrium during retrievals.  
More importantly, these fast approximation approaches enable computationally expensive chemistry models, such as those considering disequilibrium processes, to be utilized within models requiring thousands of model calculations, such as Bayesian atmospheric parameter retrieval from spectra, global circulation models, etc.

For our application, our results show that NNs trained on random data outperform the considered interpolation methods even when the random data set size is ${\sim}7$\% of the size of the uniform data set used for interpolation.  
We also found that training on a combination of gridded and random data results in a less accurate NN than training on only the random data, even when trained for more iterations. 
This finding is consistent with previous studies in the literature which showed that random sampling outperforms uniformly sampled grids when trying to approximate some forward model \citep{LoyolaEtal2016nnSmartSampling, FisherHeng2022apjOptimalSampling}.
While this suggests that gridded data are less effective at training NNs compared to random data, a future study that more thoroughly investigates this behavior is necessary to determine this definitively.

When considering interpolation within a fixed grid, we found that interpolation error significantly increases in regimes of non-linear behavior, such as C/O $\sim$ 1.  
We thus recommend investigators utilizing a grid-based interpolation approach to verify that the interpolation is sufficiently accurate across the phase space of application.  
While a more finely sampled grid will reduce this error, the increased number of models would similarly benefit the NN approach.
Based on our finding that the NN outperformed grid interpolation when considering a smaller gridded data set than the interpolation approaches, it is likely that the NN approach would continue to outperform the grid approach except at the highest resolutions where the non-linear behavior becomes nearly linear between grid points.  
Future work should determine how finely sampled a grid must be for accurate interpolation across the phase space of interest.

While the NNs performed accurately over most of the phase space, their accuracies significantly decrease near the edge of the phase space.  
Future investigations that seek to train comprehensive NN surrogate models should keep this in mind when generating data, as the trained surrogate model may not be valid at the extrema of the phase space.
To account for this, we recommend generating data over a slightly larger phase space than required. 
For situations where physical limits prevent expanding the phase space, it may be helpful to force some of the random data to occur at the extrema, that is, fixing one or more parameters to extrema and randomly generating the other parameters.  
A future study should investigate this in detail to determine how best to handle this situation.

The Reproducible Research Compendium for this work is available for download.\footnote{\url{https://zenodo.org/record/7783281}}

\acknowledgements

We graciously thank the anonymous reviewers for comments which improved the quality of this manuscript.  We thank contributors to NumPy, SciPy, Matplotlib, Tensorflow, Keras, the Python Programming Language, the free and open-source community, and the NASA Astrophysics Data System for software and services. 
We gratefully acknowledge the support of NVIDIA Corporation with the donation of the Titan Xp GPU used for this research.
This research was supported by the NASA Fellowship Activity under NASA Grant 80NSSC20K0682.

\bibliography{MLTEA}{}
\bibliographystyle{aasjournal}

\end{document}